\documentclass[fleqn,usenatbib]{mnras}

\usepackage{newtxtext,newtxmath,siunitx}

\usepackage[T1]{fontenc}

\DeclareRobustCommand{\VAN}[3]{#2}
\let\VANthebibliography\thebibliography
\def\thebibliography{\DeclareRobustCommand{\VAN}[3]{##3}\VANthebibliography}

\usepackage{amsmath,enumitem,multirow,listings,array}
\usepackage{mathtools,orcidlink,graphicx}

\definecolor{codegreen}{rgb}{0,0.6,0}
\definecolor{codegray}{rgb}{0.3,0.3,0.3}
\definecolor{codepurple}{rgb}{0.58,0,0.82}
\definecolor{backcolour}{rgb}{0.975,0.975,0.975}
\usepackage[T1]{fontenc}
\lstdefinestyle{mystyle}{
    backgroundcolor=\color{backcolour},   
    commentstyle=\color{codegreen},
    keywordstyle=\color{magenta},
    numberstyle=\tiny\color{codegray},
    stringstyle=\color{codepurple},
    columns=fullflexible,
    basicstyle=\ttfamily\footnotesize,
    breakatwhitespace=true,         
    breaklines=true,                 
    captionpos=b,                    
    keepspaces=true,                 
    numbers=left,                    
    numbersep=2pt,              
    tabsize=2
}
\newcommand{\dOmega}{${\rm d}\Omega$}
\newcommand{\kms}{$\rm km~s^{-1}$}
\newcommand{\msy}{$\rm M_\odot~yr^{-1}$}
\newcommand{\lb}{$\ell_{\rm w}$}
\newcommand{\tablepm}[1]{\multicolumn{1}{S[table-format=4.2]@{\hspace*{\tabcolsep}\makebox[0pt]{$\pm$}}}{#1}}

\title[{\tt OutLines}: Modeling Spectral Line Profiles]{{\tt OutLines}: Modeling Spectral Lines from Winds, Bubbles, and Outflows}

\author[S. R. Flury]{
Sophia R. Flury\orcidlink{0000-0002-0159-2613},$^{1}$\thanks{E-mail: sflury@roe.ac.uk}
\\
$^{1}$Institute for Astronomy, University of Edinburgh, Royal Observatory, Edinburgh, EH9 3HJ, UK\\
}

\date{Accepted XXX. Received YYY; in original form ZZZ}

\pubyear{\the\year{}}

\begin{document}
\label{firstpage}
\pagerange{\pageref{firstpage}--\pageref{lastpage}}
\maketitle

\begin{abstract}
Studies of kinematics and geometry of outflowing gas rely on modeling features in integrated spectra using empirical quantiles or fitting multiple Gaussian or Voigt profiles. Such methods can miss key underlying physics and even lead to spurious interpretations of observations. To address this problem, we present the public python code {\tt OutLines}, which predicts spectral emission and absorption line profiles produced by winds, bubbles, and outflows. { By design, {\tt OutLines} is phenomenologically and scale agnostic to bridge the gap between observations and simulations across a broad swath of astrophysics.} The {\tt OutLines} code accounts for differences in velocity field, density profile, and outflow geometries, making OutLines versatile for a wide variety of astrophysical phenomena. We demonstrate the wide applicability of {\tt OutLines} by using the code to model line profiles in an \ion{H}{ii} region knot, super star clusters, a starburst galaxy, and an AGN. In each of these contexts, we illustrate how {\tt OutLines} can illuminate key underlying physics in ways that improve scientific understanding and address important open questions in astronomy, including the key mechanisms in the baryon cycle, the evolution of \ion{H}{ii} regions and galaxies, and even Lyman continuum escape. {\tt OutLines} will be a critical resource as massively multiplexed spectroscopic surveys with facilities like WEAVE and 4MOST come online, providing the means to probe feedback with deeper, higher resolution spectroscopy for unprecedented large samples of objects.
\end{abstract}

\begin{keywords}
ISM: jets and outflows --
ISM: bubbles --
ISM: kinematics and dynamics --
galaxies: kinematics and dynamics 
\end{keywords}

\section{Introduction} \label{sec:intro}

Outflows, winds, and bubbles are ubiquitous astrophysical phenomena occurring on a dynamic range of spatial scales and in a variety of contexts, from young stellar objects (YSOs) to nebulae, active galactic nuclei (AGN), galaxies, and even galaxy clusters. This feedback has the capacity to shape many key astrophysical systems, from YSO winds affecting the formation of exoplanets \citep[e.g.,][]{Suzuki2009,Manara2023} to massive stars redistributing the leftover material from their birth clouds \citep[e.g.,][]{Dopita2005,Menon2024} to AGN and starbursts driving the baryon cycle of galaxies \citep[e.g.,][]{2005ApJ...618..569M,Tumlinson2017}. Many physical mechanisms can accelerate, or drive, winds and bubbles, from cosmic rays \citep[e.g.,][]{Ipavich1975} to hot expanding gas \citep[e.g.,][]{1985Natur.317...44C} to radiation pressure on dust \citep[e.g.,][]{Poynting1904,LyndenBell1971} and ions \citep[so-called line driving, e.g.,][]{Lucy1970}, each with direct implications for how feedback shapes the surrounding environment \citep[see, e.g.,][and references therein]{2020A&ARv..28....2V,Thompson2024}. A common method of detecting and studying feedback is through broad, often asymmetric spectroscopic features \citep[e.g.,][]{Castor1970,1975ApJ...195..157C,1974ApJ...193..651N,1980A&A....81..172P,1981MNRAS.195..787P,1981ApJ...247..403H,Chu1986}, perhaps most famously with P Cygni wind lines in stellar atmospheres \citep[e.g.,][]{1986A&A...164...86P,1987ApJ...314..726L,2001A&A...369..574V}. Physically motivated models for these line profiles is thus of critical importance to understanding how energetic phenomena shape their environment.

While the physics and formalism for spectroscopic features arising from winds, bubbles, and outflows is not new \citep[e.g.,][]{Beals1931,1960mes..book.....S,Castor1970,1975ApJ...195..157C,1986A&A...164...86P,1987ApJ...314..726L}, their application is relatively sparse outside the phenomena of stellar atmospheres \citep[e.g.,][]{2001A&A...369..574V,2021MNRAS.504.2051V,Sander2017,Puls2020}. Instead, fitting multiple Gaussian or Voigt profiles to broad, asymmetric lines has become the standard \citep[e.g.,][]{1980A&A....81..172P,1981MNRAS.195..787P} for a wide variety of objects, including \ion{H}{ii} regions \citep[e.g.,][]{Bresolin2020,Yarovova2025}, AGN \citep[e.g.,][]{Liu2020}, and galactic winds \citep[e.g.,][]{Hogarth2020,Peng2023,Peng2025}, despite early evidence that physics-based profiles are not only more appropriate but also demonstrably different in shape while still providing good descriptions of observations \citep[e.g.,][]{Zheng1990,TenorioTagle1996}. Some studies have skirted the issue by adopting a more empirical approach using velocity quantiles \citep[e.g.,][]{1981ApJ...247..403H,1985MNRAS.213....1W,1991ApJS...75..383V} or by fitting the line profile wings with a log-linear \citep[e.g.,][]{2021ApJ...920L..46K,Komarova2025} or exponential \citep[e.g.,][]{Rusakov2025} functions to characterize the steepness of the non-Gaussian flux distribution. The benefit of such methods is that they require no physical model while still providing insight into kinematics. However, bridging the gap between these approaches and the underlying physics is not always straightforward.

Some effort towards a physically motivated approach has emerged in recent years, including detailed treatment of radiation pressure and conservation effects \citep[e.g.,][]{2017MNRAS.471.4061K}, turbulent mixing layers \citep[e.g.,][]{2009A&A...500..817B}, and { relativistic accretion disk winds \citep[e.g.,][]{DISKWIND1,DISKWIND2,XRADE}} along with more generalized models \citep[e.g.,][]{2010ApJ...717..289S,2011ApJ...734...24P,Martin2013,2015ApJ...801...43S,2016MNRAS.463..541C,Luminari2018,Flury2023}. Despite these exciting advancements, wide-spread adoption of these methods has stalled. A key to furthering the use of physically-motivated line profiles is versatile, well-documented, and user-friendly software tools which make such models readily accessible to the community to generate spectral line profiles linking observations to outflow physics { without forward-modeling simulations}.

In this paper, we present {\tt OutLines}, a {\tt python}-based code for generating model emission, absorption, and P Cygni line profiles arising from astrophysical winds, bubbles, and outflows { on arbitrary spatial scales and independent of the phenomenon}. The input geometry and { parameterizations of wind physics} can be readily customized, and functionality for modeling is provided. We outline spectral line mechanisms, velocity fields, gas density profiles, and geometry in \S\ref{sec:model}. Then, in \S\ref{sec:OutLines} we present the implementation of these components in the {\tt OutLines} code\footnote{available at \url{github.com/sflury/OutLines} and via {\tt pip install}.}. We demonstrate the versatile applicability and scientific usefulness of {\tt OutLines} in \S\ref{sec:application} by modeling broad spectral lines in an \ion{H}{ii} region knot, a super star cluster (SSC), a starbursting galaxy, and an active galactic nucleus (AGN), with profiles generated by {\tt OutLines}. { In the final case, we demonstrate that {\tt OutLines} can successfully recover wind geometry through comparison with observations.} Finally, we summarize {\tt OutLines} and its potential for high scientific impact in \S\ref{sec:conclusion}, providing { details of code setup and installation in Appendix \ref{apx:OutLines} and } worked examples in Appendices \ref{apx:examples} and \ref{apx:modeling}.

\section{Physical Model}\label{sec:model}

Within the framework of the Sobolev approximation \citep{1944SvA.....21..143S,1957SvA.....1..678S,1960mes..book.....S}, small scale Doppler motions of gas within clouds can be ignored in the context of the large-scale motion of the clouds themselves. { While the Sobolev approximation can break down when turbulent motions exceed the bulk motion of the wind \citep[e.g.,][]{Li2026}, the regime in which this violation occurs tends to be high velocity, low density, meaning that such regimes will contribute negligibly to the net profile. Moreover, studies have shown that this assumption is largely valid in a variety of contexts, including stellar \citep{Castor1979,1987ApJ...314..726L} and galactic \citep{2011ApJ...734...24P} winds.} Indeed, the Sobolev approach has a long-standing precedent in the study of winds \citep[e.g.,][]{Castor1970,1974ApJ...193..651N,1975ApJ...202..465M,1987ApJ...314..726L}, with successful application to galactic winds and outflows over the past 15 years \citep[e.g.,][]{2010ApJ...717..289S,2011ApJ...734...24P,Martin2013,2015ApJ...801...43S,2016MNRAS.463..541C,Flury2023}. Under this assumption, the key characteristics of winds, outflows, and bubbles which dictate the line profile are the velocity field, density profile, and geometry of the clouds \citep[e.g.,][]{Zheng1990}. Below, we elaborate how the observed projected velocity $u$ in a spectral line, given by { the longitudinal relativistic Doppler shift \citep[][also \citealt{Sher1968} their Eqn 4]{Einstein1905} such that}
\begin{equation}\label{eqn:vel_obs}
    u = \frac{c}{v_\infty}
    \frac{(\lambda/\lambda_0)^2-1}{(\lambda/\lambda_0)^2+1}
\end{equation}
for a line with rest wavelength $\lambda_0$ and normalized to the wind terminal velocity $v_\infty$ and { reducing to the classical Doppler shift $u=c(\lambda/\lambda_0-1)/v_\infty$ when $v_\infty\ll c$}, relates to the intrinsic wind velocity $w$ given by some radially-symmetric velocity field
\begin{equation}\label{eqn:vel_wind}
    w = \frac{v_{\rm wind}[x]}{v_\infty}
\end{equation}
{ with respect to some radial distance $x=r/R_0$ from the source of the outflow, making $R_0$ effective a launch radius}. In addition to the velocity field, one must also know the radial gas density distribution, given by
\begin{equation}\label{eqn:den_wind}
    n^\prime[r] = n_0 n[x]
\end{equation}
for a dimensionless profile $n$ and maximum density $n_0$, and the azimuthal geometry of the wind. With the velocity field $w$, density profile $n$, and outflow geometry in hand, radiative processes can translate outflows from the frame of the source to the spectral line in the frame of the observer.

\subsection{Radiative Processes}\label{sec:radiativeprocesses}

Below, we outline the formalism for absorption and emission line profiles arising from astrophysical winds, with implicit assumptions about the velocity field $w[x]$, the radial gas distribution $n[x]$, and the gas geometry.

\subsubsection{Line Absorption}

In general, line absorption { optical depth $\tau_\lambda$} at any particular wavelength $\lambda$ in a feature is given by
\begin{equation}\label{eqn:line_abs}
    \tau_\lambda = \frac{N}{v} f \lambda_0\sigma_a \phi_\lambda
\end{equation}
where $v$ is the velocity scale of the line ($b$ in the case of static interstellar medium (ISM) lines, $v_\infty$ in the case of winds), $N$ is the column density of the gas, $f$ is the quantum mechanical oscillator strength, $\lambda_0$ is the rest wavelength of the electronic transition which absorbs incident light, $\sigma_a=10^{-14.8427}\rm~cm^2~s~km^{-1}~\AA$ is the classical absorption cross-section per unit wavelength per unit velocity, and $\phi_\lambda$ is the line profile evaluated at $\lambda$. Traditional treatment of the static interstellar medium (ISM) assumes a Voigt profile for $\phi_\lambda$ (most robustly calculated using the real part of the Faddeeva function, although approximations can also be used, e.g., \citealt{Hjerting1938,TepperGarcia2006}) to account for kinetic and natural broadening. 

Under the Sobolev approximation, instead one must evaluate the relative column density at each velocity $w$ in the wind or bubble that is projected onto the line of sight velocity $u$.
The optical depth profile for the outflow component given by the Sobolev approximation is thus
\begin{equation}\label{eqn:phi_abs}
    \phi_\lambda = K^{-1}\int n[x]\frac{{\rm d}x[w]}{{\rm d}w}{\rm d}\Omega^\prime
\end{equation}
over some observed differential solid angle \dOmega$^\prime$\ given by the geometry \citep[see][their Equation 2.14]{Rybicki1983} and where $K$ is a normalization constant ensuring $\int_\lambda\phi_\lambda{\rm d}\lambda = 1$ \citep[][their Equation 1.65]{Rybicki1979}. The differential term $\frac{{\rm d}x}{{\rm d}w}$ arises from integrating over all column densities associated with the observed velocity $u$ (\citealt{Castor1970}, their Equation 6, \citealt{2016MNRAS.463..541C}, their Equation 4, assuming no stimulated emission). The corresponding integral limits are given by constraints from the geometry (\S\ref{sec:geometry}). { Following the equation of radiative transfer, one must sum up contributions of each velocity slice to the optical depth, not to the emergent flux, or else risk falling victim to Jensen's inequality ($\exp[-\int\tau{\rm d}\Omega^\prime]<\int\exp[-\tau]{\rm d}\Omega^\prime$) by over-estimating the column density and optical depth \citep[e.g.,][]{Jennings2025}.}

The emergent flux, including a static component with a Voigt profile, is therefore given by
\begin{equation}\label{eqn:abs}
    \frac{F_\lambda}{F_{\lambda0}} = \exp\left[\tau_{\rm s}\phi_{\lambda,\rm Voigt}+\tau_{\rm o}\phi_{\lambda,\rm out}\right]
\end{equation}
{ for intrinsic (backlight) flux density $F_{\lambda0}$ and emergent flux density $F_\lambda$ and} where 
\begin{equation}\label{eqn:column_static}
    \tau_{\rm s} = f\lambda_0\sigma_a b^{-1}N_{\rm s}
\end{equation}
for static gas column density $N_s$ and
\begin{equation}\label{eqn:column_outflow}
    \tau_{\rm o} = f\lambda_0\sigma_a v_\infty^{-1}N_{\rm o}
\end{equation}
for outflowing gas column density $N_{\rm o}$.

\subsubsection{Resonant and Fluorescent Line Emission}

The upper state population of resonant and fluorescent lines depends primarily on the absorption of photons { which are subsequently re-emitted in a random direction}. Therefore, the line emission coefficient $j_\lambda$ is proportional to $n$d$x/$d$w$, meaning that the line profile is determined using Equation \ref{eqn:phi_abs} but with integral limits for line emission given by integrating over the entire volume (see \S\ref{sec:geometry}). However, photons emitted in this fashion can be re-absorbed and subsequently re-emitted in an arbitrary direction, effectively a scattering process, which will alter the line profile. Resonant and fluorescent lines re-emit light which has previously been absorbed isotropically, meaning that the line flux is determined by the source function $S_\lambda=1-\exp[-\tau_\lambda]$ \citep[see, e.g.,][ their \S6, \citealt{Rybicki1983}, their \S3]{Castor1970} such that 
\begin{equation}
    \frac{F_\lambda}{F_{\lambda0}} = \epsilon f_{\rm esc,\lambda}\left(1-\exp\left[-\tau_\lambda\right]\right)
\end{equation}
where $\tau_\lambda$ is given by line absorption in Equations \ref{eqn:line_abs} and \ref{eqn:phi_abs} with integral limits consistent with line emission in Equation \ref{eqn:phi_ems}, $\epsilon$ is the filling factor to account for clumpiness because individual clouds do not sample every possible sightline even when they are isotropically distributed, and $f_{\rm esc,\lambda}$ is the fraction of photons which escape radiative trapping at a given wavelength $\lambda$. The escape fraction of photons depends on the fraction $p_i=A_i/\sum A$ of photons which are produced by a transition $i$ given by the transition rate $A_i$ divided by the total radiative depopulation rate $\sum A$, where each $i$ corresponds to a resonant ($p_{res}$) or fluorescent ($p_{fluor}$) channel. Following formalism from \citet[][their Equation A12, see also \citealt{Rybicki1983}, their Equation 3.19]{Castor1970}, the probability that a resonant photon can escape from radiative ``trapping'' through resonant scattering is given by the ratio of emission to absorption $(1-\exp[-\tau_{\rm res,\lambda}])/\tau_{\rm res,\lambda}$. Trapped photons either scatter through emission and re-absorption in the resonant channel or re-emit in the fluorescent channel. After multiple scatterings, the resonant photons are increasingly trapped relative to the fluorescent photons, meaning that the escape fraction is
\begin{equation}
    f_{\rm esc,\lambda}^{\rm res} = p_{\rm res} \frac{1-\exp[-\tau_{\rm res,\lambda}]}{\tau_{\rm res,\lambda}} \left(1-p_{\rm res}\frac{1-\exp[-\tau_{\rm res,\lambda}]}{\tau_{\rm res,\lambda}}\right)^{-1}
\end{equation}
for resonant photons and 
\begin{equation}
    f_{\rm esc,\lambda}^{\rm fluor} = p_{\rm fluor} \left(1-p_{\rm res}\frac{1-\exp[-\tau_{\rm res,\lambda}]}{\tau_{\rm res,\lambda}}\right)^{-1}
\end{equation}
for fluorescent photons under the infinite scattering limit given by \citet[][their Equations 20a,b]{2015ApJ...801...43S}.

\subsubsection{Nebular Line Emission}

For nebular emission lines, the emergent line { flux density $F_\lambda$} is simply the profile $\phi_\lambda$ scaled by the integrated line flux $F$ such that
\begin{equation}\label{eqn:flux_ems}
    F_\lambda = F\phi_\lambda
\end{equation}
for any given wavelength. ``Static'' gas undergoes Doppler broadening from thermal, orbital, and turbulent motion which produces the traditional Gaussian form of $\phi_\lambda$. In the case of an outflow, we must first consider how the line emission depends on density and velocity. The line emission coefficient $j_\lambda$ is typically defined as proportional to the density of the populated upper state producing the transition \citep{Rybicki1979,source:osterbrock2006}. Since nebular lines emerge from upper electronic states populated by recombination or collisional excitation, $j\propto n[x(w)]^2$ \citep[e.g.,][]{source:osterbrock2006,Peimbert_2017}. Thus, we can simply express the line profile as
\begin{equation}\label{eqn:phi_ems}
    \phi_\lambda = K^{-1} \int n\left[x[w]\right]^{2} {\rm d}\Omega^\prime
\end{equation}
where $K$ again is a normalization constant ensuring $\int_\lambda\phi_\lambda{\rm d}\lambda = 1$ \citep[see][]{Flury2023}. As with absorption, limits on the integral are given by the geometry in \S\ref{sec:geometry}.

The emergent flux, including a static component with a Gaussian profile, is then given by
\begin{equation}\label{eqn:ems}
    F_\lambda = F_{\rm s}\phi_{\lambda,\rm Gauss}+F_{\rm o}\phi_{\lambda,\rm out}
\end{equation}
for nebular emission.

\subsection{Velocity Field Parameterizations from Wind Physics}\label{sec:velocityfields}

Velocity fields for winds can take on a variety of shapes arising from different treatments of the wind equations of motion \citep[e.g., CAK theory][]{1975ApJ...195..157C,1976ApJS...32..715L,1977PhDT.........3A,1977ApJ...213..737B,Castor1979}. The simple $\beta$ law approximation
\begin{equation}\label{eqn:betaCAK}
    w[x] = \left(1-x^{-1}\right)^\beta{\rm ,\hspace{3em} BetaCAK}
\end{equation}
first appears in early treatment of Wolf-Rayet P Cygni lines for $\beta=0.5$ \citep[Equation 23][]{Castor1970} and arises from a generalization of CAK theory solutions \citep[see, e.g., Equation 47 of ][]{1975ApJ...195..157C} without loss of physical insight \citep[see, e.g.,][]{1977ApJ...213..737B,Castor1979,1986A&A...164...86P}.
The CAK $\beta$ law approximation is ubiquitous in the study of stellar winds \citep[e.g.,][]{2001A&A...369..574V,2021MNRAS.504.2051V,Telford2024} and has even seen application to galactic outflows \citep[e.g.,][]{2016MNRAS.463..541C,Flury2023}. Given its widespread adoption and underlying physical motivation, the CAK $\beta$ law is our default and fiducial velocity field.

In addition to the $\beta$ law approximation to CAK theory, {\tt OutLines} includes two alternate solutions to the wind equations proposed in the literature. The first, from \citet{2010ApJ...717..289S}, assumes a power law approximation to the radial acceleration field to obtain
\begin{equation}\label{eqn:accplaw}
    w[x] = \left(1-x^{1-\beta}\right)^{1/2}{\rm ,\hspace{3em} AccPlaw}
\end{equation}
where the normalization terms are implicit in $w$ and $x$ and $r_{min}=1$. While those authors do not provide a corresponding physical model, their acceleration power law has a velocity field qualitatively similar to those of \citet{Ipavich1975}, which assumed cosmic ray driving, and \citet{1985Natur.317...44C}, which assumed constant momentum and energy deposition into a steady-state wind from a fluid in which it is entrained \citep[see also, e.g.,][]{Owocki2025}, and formally akin to radiative acceleration of expanding clouds \citep[see Equations 16-17][]{Mathews1974}. Note that the acceleration power law field with $\beta=2$ is identical to that of the CAK $\beta$ law when $\beta=0.5$, { both of which correspond to a $r^{-2}$ driving force}.

The second velocity field alternative, from various sources \citep[e.g.,][]{FaucherGiguere2012}, constitutes a power law approximation to the radial velocity field such that
\begin{equation}\label{eqn:velplaw}
    w[x] = A(x-1)^\beta{\rm ,\hspace{3em} VelPlaw}
\end{equation}
where the wind is assumed to accelerate from rest at $R_0$ and $A$ scales the field. We set $A=0.5$ to allow clouds to reach 50\% terminal velocity in roughly the same range of radii as the CAK theory $\beta$ law and generally overlap with the same region in $w$-$x$ space as both the CAK $\beta$ law and acceleration power law fields. Single-shell models of stellar wind bubbles can give rise to power law velocity fields from the shared time dependence of $r\propto t^{3/5}$ and $v\propto t^{-2/5}$ \citep[$v\propto r^{-2/3}$, e.g.,][]{Weaver1977,Dyson1977}. Building on this framework, \citet{FaucherGiguere2012} arrive at a velocity power law solution assuming continuous production of ``shells'' by an AGN to produce a wind.

While the {\tt OutLines} default assumes that gas is launched from rest, we have allowed for the possibility that the gas starts from some launch velocity $v_0$, which we include as an optional free parameter. The velocity limits are accordingly updated so that the minimum velocity $w_\ell$ which contributes to an observed velocity $|u|$ is never below $v_0/v_\infty$. The velocity fields in Table \ref{tab:velocity_terms} and Equations \ref{eqn:betaCAK}-\ref{eqn:velplaw} are modified by a factor of $1-v_0/v_\infty$ with the addition of a constant $v_0/v_\infty$ such that the relative velocity not from rest $w_{\rm nfr}$ is
\begin{equation}\label{eqn:nonrest}
    w_{\rm nfr} = w[x|\beta]\left(1-\frac{v_0}{v_\infty}\right)+\frac{v_0}{v_\infty}
\end{equation}
\citep[e.g.,][]{Castor1979}.
Emergent profiles are similar to those presented in \citet{TenorioTagle1996}, which included distributions of non-zero shell ages, and to those presented in \citet{2015ApJ...801...43S}, which account for non-zero launch velocities. { For visualization purposes, we assume in example profiles in \S\ref{sec:model} that $v_0=0.1v_\infty$.}

\begin{figure}
    \centering
    \includegraphics[width=\linewidth,clip=True,trim={0.35in 0.2in 0in 0in}]{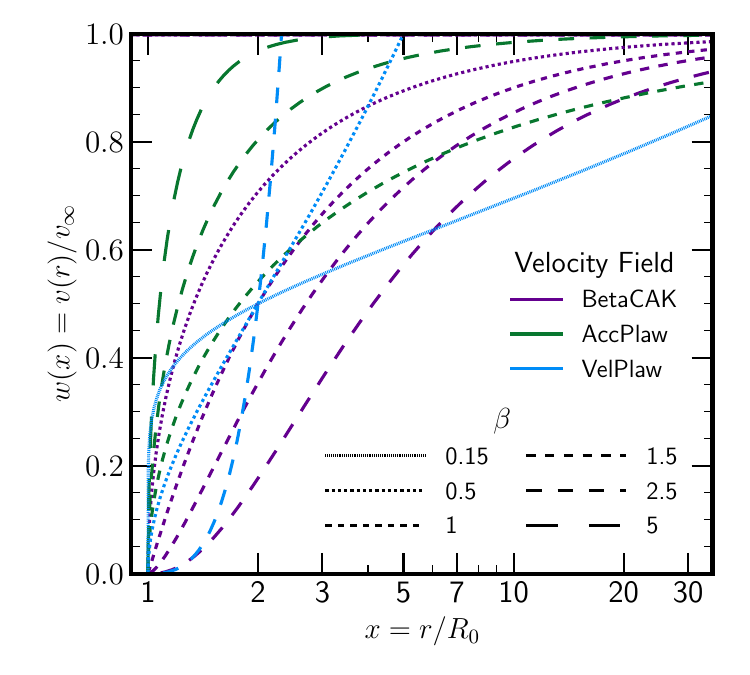}
    \caption{Radial velocity fields included in {\tt OutLines} and given by Equations \ref{eqn:betaCAK}-\ref{eqn:velplaw}. Acceleration from rest at the base $R_0$ of the wind ($x=1$) proceeds until the wind reachs the terminal velocity $v_\infty$ ($w[x]=1$). Velocity fields include the \citet{1986A&A...164...86P} $\beta$ law approximation to CAK theory \citep{1975ApJ...195..157C} (blue, Equation \ref{eqn:betaCAK}), the \citet{2010ApJ...717..289S} power law acceleration (green, Equation \ref{eqn:accplaw}), and the power law velocity field (purple, Equation \ref{eqn:velplaw}). Dash length increases with increasing $\beta$.}
    \label{fig:velocity_fields}
\end{figure}

Figure \ref{fig:velocity_fields} depicts the different velocity fields for a range of $\beta$. From this comparison, as suggested by the formalism, the power law velocity field approaches $v_\infty$ more rapidly than the power law acceleration field or the $\beta$-law approximation to CAK theory, reaching $v_\infty$ at distances far closer to the launch radius. Moreover, the power law velocity field does not approach $v_\infty$ asymptotically as do the other velocity fields, meaning that the stopping condition at $v_\infty$ is imposed solely by integral limits rather than the formalism for the velocity field. We note that physical wind models almost always exhibit an acceleration which gradually decreases as the outflow asymptotically approaches some terminal velocity \citep[e.g.,][]{1975ApJ...195..157C,1977ApJ...213..737B,1985Natur.317...44C,Thompson2015,2017MNRAS.471.4061K,2018MNRAS.476..512I,Quataert2022a}, indicating that the underlying wind physics strongly favors a CAK $\beta$ law or acceleration power law field over a velocity power law field. While individual wind bubble models indicate a velocity power law field \citep[e.g.,][]{Weaver1977,Dyson1977}, in ensembles of shells,
the distribution of shell ages will dominate the net velocity field \citep[e.g.,][]{TenorioTagle1996} and is likely determined by the physical driving mechanism \citep[e.g.,][]{Chu1994}, suggesting that the CAK theory or acceleration power law velocity fields provide a more physically appropriate description for wind scenarios. That being said, the velocity power law field prescription can be consistent with turbulent streaming including the classic Kolmogorov cascade, with the case of $\beta=1$ indicative of Couette flow \citep[e.g.,][]{2009A&A...500..817B}, although such scenarios may not capture the underlying driving mechanism.

\begin{figure}
    \centering
    \includegraphics[width=\linewidth,clip=True,trim={0.35in 0.2in 0in 0in}]{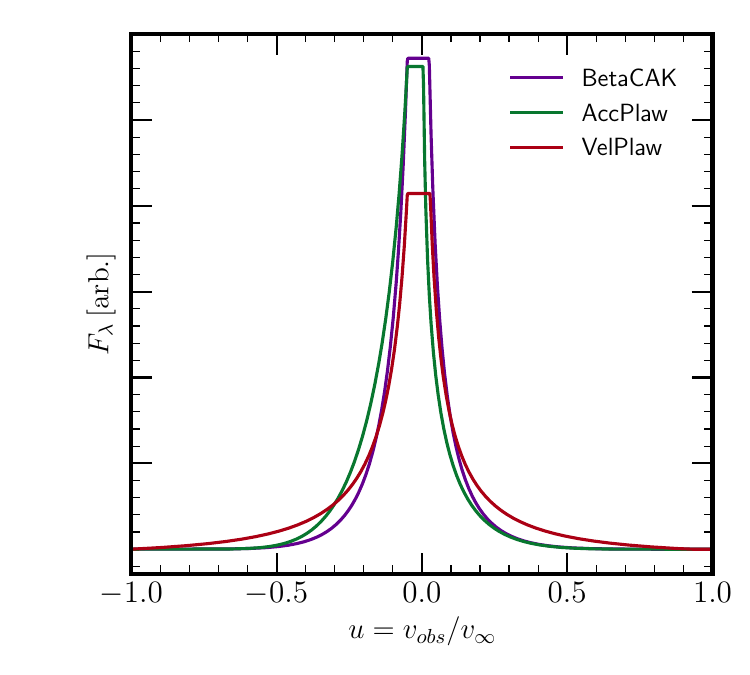}
    \includegraphics[width=\linewidth,clip=True,trim={0.35in 0.2in 0in 0in}]{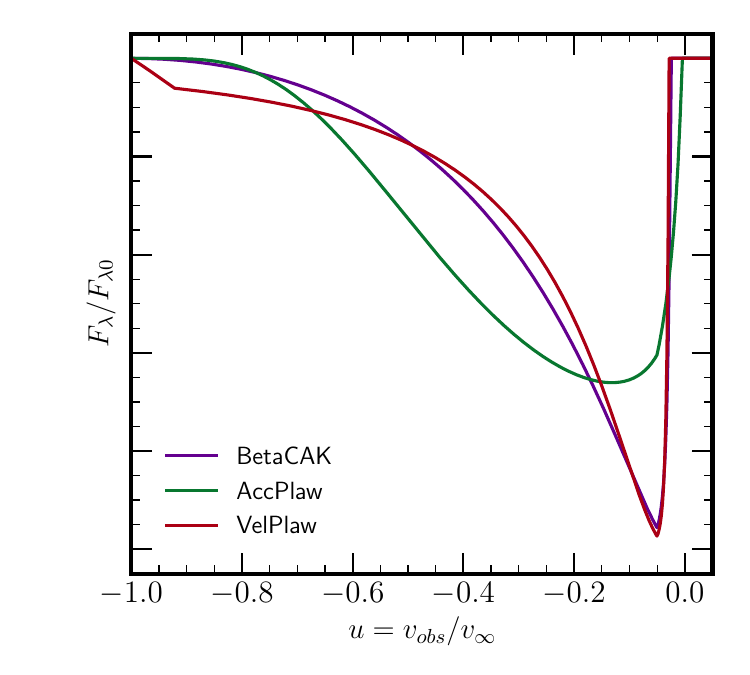}
    \caption{Nebular emission ({\it top}) and resonant absorption ({\it bottom}) line profiles for each of the velocity fields included in {\tt OutLines} assuming $\beta=1.5$ (CAK theory), $\beta=1$ (velocity power law), and $\beta=2$ (acceleration power law) for the {\tt OutLines} defaults.}
    \label{fig:velocity_profiles}
\end{figure}

We illustrate the effect of varying the velocity field for fixed $\beta$ and the default {\tt OutLines} geometry and density profiles in Figure \ref{fig:velocity_profiles}. As suggested by the velocity fields in Figure \ref{fig:velocity_fields}, the acceleration power law causes $w\to1$ more rapidly than the other two scenarios, resulting in more emergent flux at higher projected velocities and thus the most pronounced intermediate wings. The $\beta$ CAK law field more rapidly populates moderate velocities than the velocity power law field, resulting in a slightly brighter core. However, the acceleration of the velocity power law field overtakes that of the $\beta$ CAK at high velocities, producing more prominent broad wings in the former than the latter.

\subsection{Radial Density Profiles}\label{sec:densityprofiles}

The gas / clouds within a wind likely vary with projected radial distance from the launching source due to a variety of factors, including conservation of cloud surface area \citep[e.g.,][]{2017MNRAS.471.4061K}, fountains \citep[e.g.,][]{2015ApJ...814...83L}, or an isothermal mass-conserving gas \citep[e.g.,][]{2005ApJS..160..115R,2011ApJ...734...24P}.

We adopt a smooth function to describe a continuous distribution of delta functions, with each delta function ``pulse'' representing an instantaneous burst or ejection of material with gas density $n$ from the launch site and projected radially. For simplicity, $n$ can be assumed to follow a power law such that
\begin{equation}\label{eqn:denplaw}
    n[x] = x^{-\alpha}{\rm ,\hspace{3em} PowerLaw}
\end{equation}
a reasonable approximation for stellar and galactic winds \citep[e.g.,][respectively]{Castor1979,1985Natur.317...44C} and a common choice in the literature \citep[e.g.,][]{FaucherGiguere2012,2015ApJ...801...43S,2016MNRAS.463..541C,Flury2023,Luminari2024A&A} arising from power law solutions to differential equations (as also the case for the velocity profile) and can be indicative of momentum or energy conservation \citep[$\beta=(\alpha-2)/2$ and $\beta=(\alpha-2)/3$, respectively][for the power law velocity field $\beta$]{FaucherGiguere2012}. Perhaps intuitively, $\alpha=2$ represents a mass-conserving scenario with density { for a fixed velocity}, changing only as a $r^{-2}$ projection effect, { and for an arbitrary velocity field represents an isothermal gas \citep{Parker1958}}. As an alternative, we can adopt an exponential profile where
\begin{equation}\label{eqn:expon}
    n[x] = \exp\left[-\gamma (x-1)\right]{\rm ,\hspace{3em} Exponential}
\end{equation}
which is also readily invoked from solutions to differential equations and may correspond to the relaxation or dissipation of wind material as it projects out from the launch point. For cases where a continuous flow may flatten or precipitously decline beyond some radius, we also implement a broken power law
\begin{equation}\label{eqn:dplaw}
    n[x] = 
    \begin{cases}
    x^{-a_1}, & x<x_1 \\
    x_1^{a_2-a_1} x^{-a_2}, & x>=x_1 
    \end{cases}{\rm ,\hspace{3em} PowerLaw2}
\end{equation}
with a break-point or ``knee'' at radius $x_1$, inner slope $a_1$ and outer slope $a_2$. Examples of a broken power law include the shock-driven outflow launched by the interaction of winds from massive stars in proximity to Sag A$^*$ \citep[$a_1<a_2$, drop off in density profile, e.g.,][]{Quataert2004} and the relativistic outflow triggered by the waking/activating supermassive black hole in Swift J164449.3+57345 \citep[$a_1>a_2$, flattening in density profile, e.g.,][]{Berger2012}.

\begin{figure}
    \centering
    \includegraphics[width=\columnwidth,trim={0cm 2.5cm 0cm 1cm},clip=True]{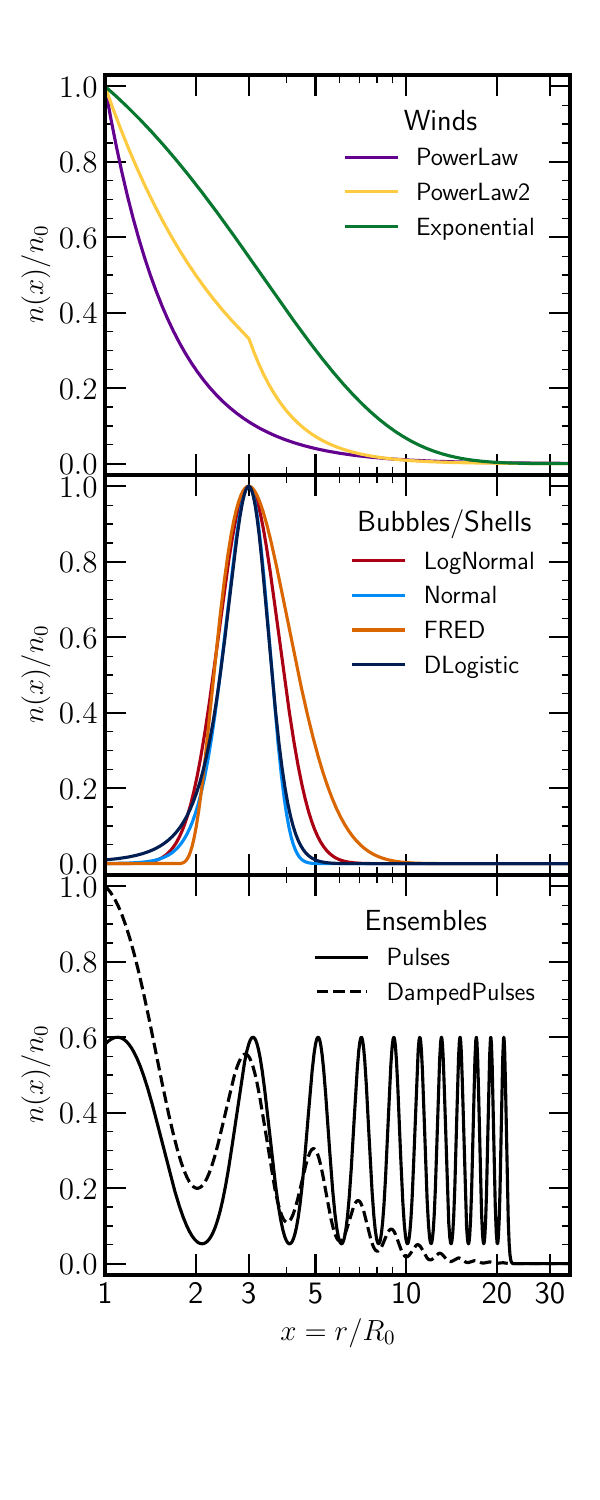}
    \caption{ Radial density profiles included in {\tt OutLines} and given by Equations \ref{eqn:denplaw}-\ref{eqn:dampedpulses}. {\it Top}: Wind-like density profiles, including power law (`PowerLaw', purple, Eqn \ref{eqn:denplaw}), double power law (`DoublePowerLaw', yellow, Eqn \ref{eqn:dplaw}), and exponential (`Exponential', green, Eqn \ref{eqn:expon}) distributions. {\it Center}: Pulse-like density profiles, including the normal (`Normal', blue, Eqn \ref{eqn:norm}), log normal (`LogNormal', red, Eqn \ref{eqn:lognorm}), fast-rise exponential decay (`FRED', orange, Eqn \ref{eqn:fred}), and derivative of the logistic (`DLogic', indigo, Eqn \ref{eqn:dlogic}) distributions, represent physical scenarios such as blast waves, expanding shells, and single burst events.
    {\it Bottom}: Ensembles of pulse-like profiles without (black solid, Eqn \ref{eqn:pulses}) and with (black dashed, Eqn \ref{eqn:dampedpulses}) damping. These ensembles represent the resolved elements of a radially clumpy wind or a series of outflowing episodes.}
    \label{fig:dens_dists}
\end{figure}

For a variety of physical scenarios, including swept up shells and blast waves, we also allow in {\tt Outlines} the possibility of pulse-like radial density profiles consistent with a shell or bubble with a scaled radius $x_1=R_{\rm shell}/R_0$ and characteristic width $\sigma_x=\sigma_r/R_0$, including a shell with gas in a normal distribution
\begin{equation}
    n[x] = \left(2\pi\sigma_x^2\right)^{-1/2}\exp\left[-\frac{(x - x_1)^2} {2\sigma_x^2} \right] {\rm ,\hspace{1em} Normal,}\label{eqn:norm}
\end{equation}
with a log normal distribution 
\begin{equation}
    n[x] = \left(2\pi(\log\sigma_x)^2\right)^{-1/2}\exp\left[-\frac{(\log x - \log x_1)^2} {2(\log\sigma_x)^2} \right]{\rm ,\hspace{1em}  LogNormal,}\label{eqn:lognorm}
\end{equation}
a fast-rise exponential decay (or FRED, Equation \ref{eqn:fred}, see \citealt{Norris2005}) distribution
\begin{equation}
    \begin{split}
    n[x] = & \exp[2\mu]\exp\left[-\frac{\tau_1}{x+\mu-x_1}-\frac{x+\mu-x_1}{\tau_2}\right],\\ &\mu=\sqrt{\tau_1\tau_2}
    \end{split}{\rm\hspace{1em}  FRED,}\label{eqn:fred}
\end{equation}
a logistic column density distribution
\begin{equation}
    n[x] = \frac{k\exp[-k(x-x0)]}{(1+\exp[-k(x-x0)])^2}{\rm ,\hspace{1em}  DLogistic,}\label{eqn:dlogic}
\end{equation}
and a uniform distribution
\begin{equation}
    n[x] = \begin{cases}
        \frac{1}{2\sigma_x},& |x-x_1|<\sigma_x\\
        0,\text{otherwise}
    \end{cases}{\rm ,\hspace{1em}  Uniform Shell}
    \label{eqn:shell}.
\end{equation}
These pulse-like density profiles are consistent with simple single expanding shells or bubbles which are symmetric (normal or uniform), leading (FRED), or trailing (log normal) material with respect to the outward direction of the gas. We illustrate the continuous and pulse-like density profiles available in {\tt OutLines} in Figure \ref{fig:dens_dists}.

\begin{figure}
    \centering
    \includegraphics[width=\linewidth,clip=True,trim={0.35in 0.2in 0in 0in}]{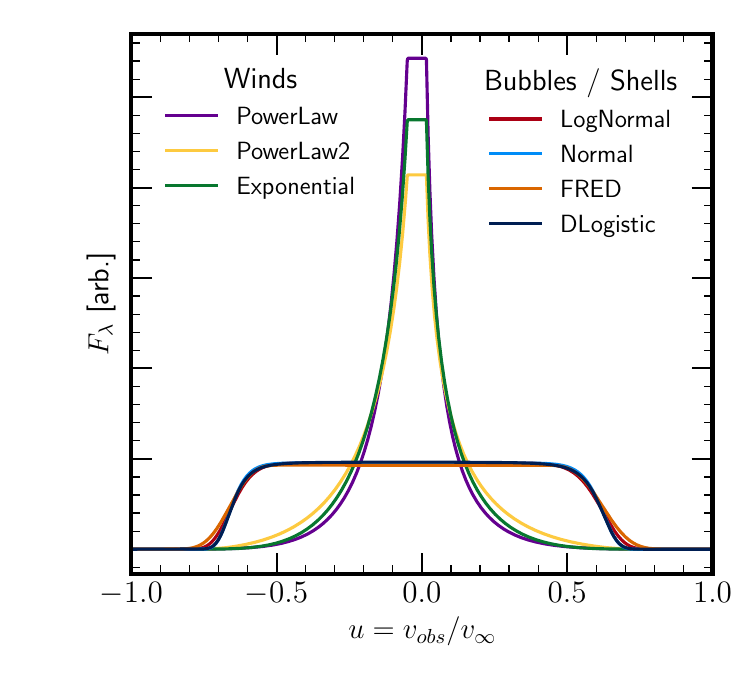}
    \includegraphics[width=\linewidth,clip=True,trim={0.35in 0.2in 0in 0in}]{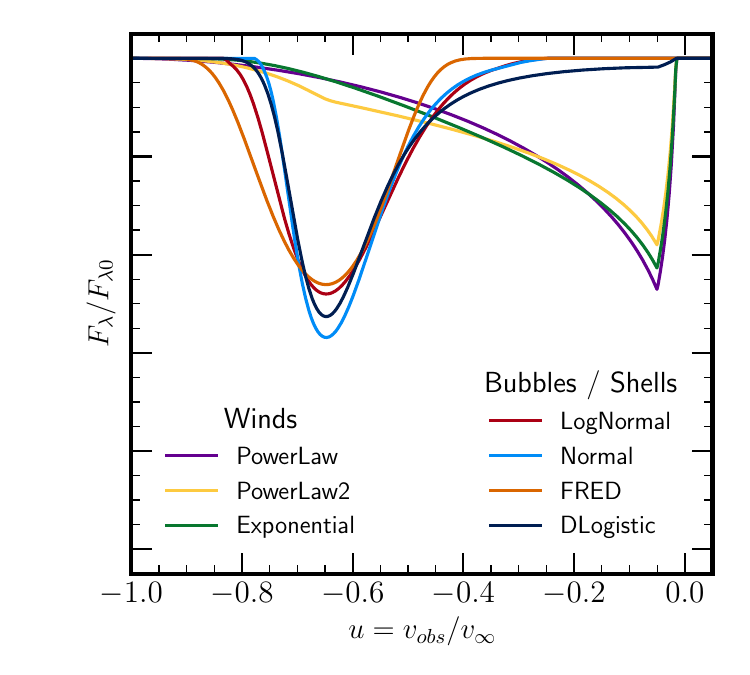}
    \caption{Nebular emission ({\it top}) and absorption ({\it bottom}) line profiles predicted by {\tt OutLines} for fixed flux, column density, and velocity field for different density distributions assuming a spherical geometry. Colors matched to the corresponding profiles in Figure \ref{fig:dens_dists}.
    \label{fig:dens_profs}}
\end{figure}

We show the effects of changing the density distribution for a fixed density ($n_0$), total column density ($N$), and velocity field ($\beta$, $v_\infty$) in Figure \ref{fig:dens_profs} assuming a spherical geometry. The emission line profile wings for the shell or pulse-like density distributions are distinct from the continuous distributions; however, distinctions among the various subsets are less clear, likely owing to the geometric dilution of the density profile by integrating over the entire volume of the line-emitting gas. The effect of pulse profiles on the emission line profile is largely consistent with early fixed-velocity single expanding shell models for Wolf-Rayet features and stellar wind bubbles wherein no line core emerges \citep[e.g.,][]{Beals1931,TenorioTagle1996}. Such marked differences in radial gas distribution may explain the line profiles of ionized gas around pre- and post-SNe stellar populations as in the different clusters in the Sunburst Arc \citep[e.g.,][]{Mainali2022}. 

{ A hallmark of the pulse-like density profiles which distinguishes them from continuous distributions is their relatively flat/uniform flux distributions across velocity space in emission and their concentration at the shell velocity in absorption. In emission, bubbles always have a distinctive box-like shape \citep[e.g.,][]{Beals1931} while winds always exhibit wings.} The absorption line profiles are more distinct for each individual density profile, likely owing to the direct tracing of the line-of-sight gas distribution without any geometric dilution. { In absorption, bubbles will largely distinguish themselves from winds with the deepest trough in the profile corresponding to the mean bubble velocity. Wind absorption profiles, on the other hand, will be deepest at systemic rest since the density by design is maximal at the base of the flow.} Deviations from spherical geometry may affect this result, particularly in the case of a narrow bicone normal to the plane of the sky and aligned with the observer, which directly traces the density distribution in emission and absorption. { Additionally, a wind not launched from rest will exhibit some flattening of the profile in emission and a shift of the profile trough in absorption due to the effect of the inner edge behaving itself as an interior shell. This behavior highlights the sum-of-shells nature of the continuous profile.}

\begin{figure}
    \centering
    \includegraphics[width=\linewidth,clip=True,trim={0.32in 0.2in 0in 0in}]{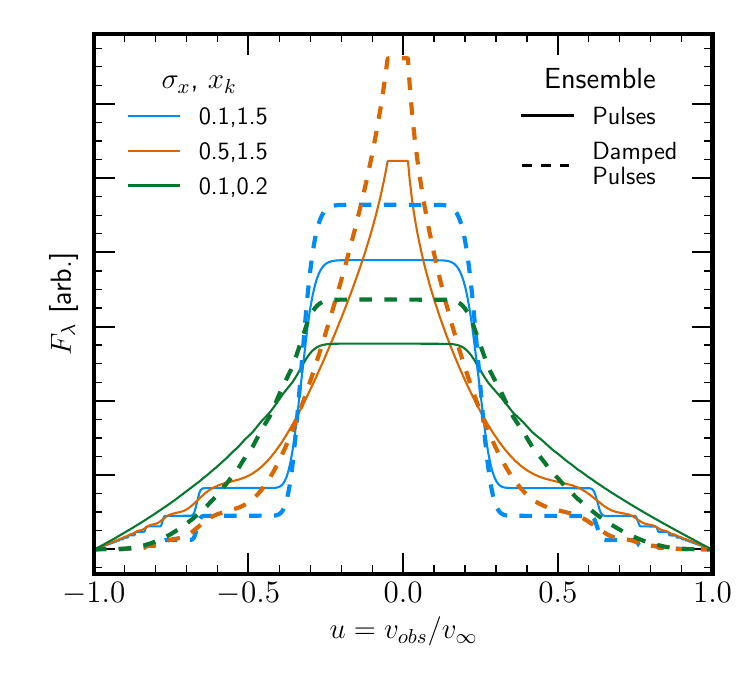}
    \includegraphics[width=\linewidth,clip=True,trim={0.32in 0.2in 0in 0in}]{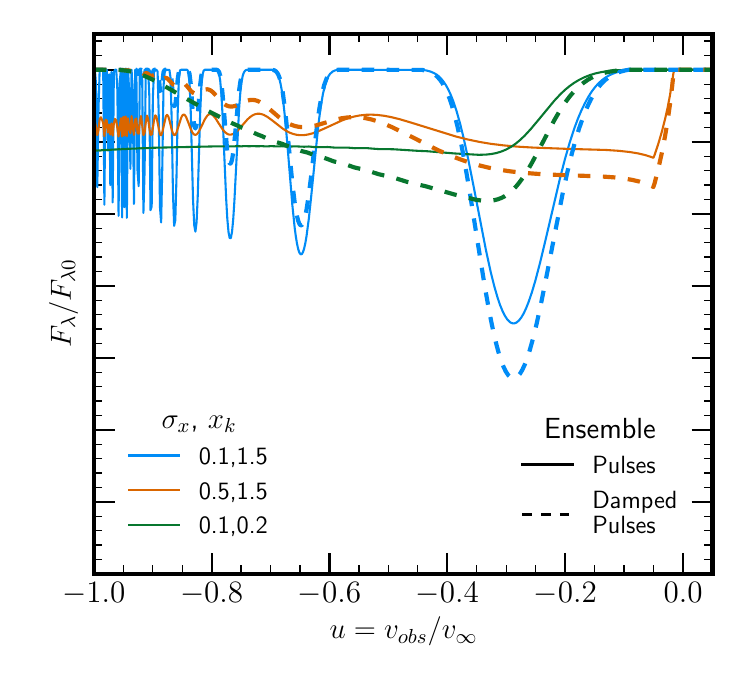}
    \caption{Nebular emission ({\it top}) and absorption ({\it bottom}) line profiles predicted by {\tt OutLines} for an ensemble of expanding bubbles, shells, or other pulse-like episodes in cases of thick (orange, $\sigma_x=0.5$) or thin (blue, $\sigma_x=0.1$) widely-spaced shells and thin closely-spaced shells (green, $x_k=0.2$) with (dashed) and without (solid) damping to mimic the effects of dissipation. In all cases, density, column density, velocity field, and terminal velocity are fixed, and a spherical geometry is assumed.}
    \label{fig:pulse_profiles}
\end{figure}

The exponential and power law profiles can be considered a combination of many instantaneous wind ``pulses'' or multiple expanding shells, together resulting in a nearly continuous flow which in reality consists of multiple bursts. A prime example of this multi-episode flow phenomenon in nature is the consecutive shells observed in Wolf-Rayet nebulae like WR140, where traces of the individual pulses dissipate on the timescales much shorter than that of the outflow, producing a decline in flux comparable to an exponential or power law density profile \citep{Lieb2025}. Similar evidence is found from spectroscopy of the base or launch site of outflows, where successive episodes begin to resemble a continuous density distribution, as in the case of the Ly$\alpha$ and \ion{C}{iv} lines in NGC 3783 \citep[e.g.,][]{Mehdipour2017,Mehdipour2025}, and may even indicate an outflow or wind in formation. Alternatively, a system with multiple expanding shells might appear as a series of pulses even when not spatially coincident (i.e., not nested), as in the case of starburst galaxies \citep[e.g.,][]{RiveraThorsen2015}. To that end, we also include density profiles for a series of pulse-like episodes (shells, bubbles, etc) such that, for $N_p$ total pulses,
\begin{equation}\label{eqn:pulses}
    n[x] = \left(2\pi\sigma_x^2\right)^{-1/2}\sum_{i=0}^{N_p}\exp\left[-\frac{(x - x_1 - ix_k)^2} {2\sigma_x^2} \right]{\rm,\hspace{1em} Pulses}
\end{equation}
where $\sigma_x$ is the width of each pulse, $x_1$ is the radius of the smallest pulse (akin to a phase shift) and $x_k$ is the interval spacing between pulses. We assume $N_p$ such that pulses span the observationally relevant velocity field from $x=1$ to $x[0.95]$ for any given $x_1$ and $x_k$. To allow pulses to relax or dissipate over time, we also include a damped pulse series
\begin{equation}\label{eqn:dampedpulses}
    n[x] = \frac{\exp\left[-\gamma x\right]}{\sqrt{2\pi\sigma_x^2}}\sum_{i=0}^{N_p}\exp\left[-\frac{(x - x_1 - ix_k)^2} {2\sigma_x^2} \right]{\rm,\hspace{1em} DampedPulses}
\end{equation}
which is likely a more physical representation of outflows in formation. We show the effects of a series of pulse-like events or damped pulses on the emergent line profiles in Figure \ref{fig:pulse_profiles} for shells resolved and unresolved in velocity. { Figure \ref{fig:pulse_profiles} illustrates how infinitely thin shells with infinitely small separations ($\sigma_x,x_k\to0$) or relatively thick shells ($\sigma_k\gtrsim x_k$) produce a wind-like profile.}

One might invoke the central limit theorem to argue that a system of many shells or bubbles will result in a Gaussian line profile; however, we find that neither the damped nor undamped pulses behave in this manner.  Indeed, \citet{TenorioTagle1996} find that \emph{only} a Gaussian velocity field will produce an intrinsically Gaussian line profile for a sum of many shells, a scenario which does not appear to be consistent with the bulk motion of gas predicted by theory. They further demonstrate that profiles might only appear Gaussian due to smearing by a line spread function. These results are consistent with comparisons of integrated and spatially resolved observations of the ensemble of expanding shells in 30 Doradus where the bulk velocity field produces a line profile with a Gaussian-like core but distinctly non-Gaussian wings \citep{Chu1994}.

In all cases, simple algebra gives the density in the wind as a function of velocity. As the density term $n_0$ simply scales the total line profiles in our formalism below, the $n_0$ is absorbed into column density $N_o$ and flux $F_0$ terms, respectively.

\subsection{Geometry}\label{sec:geometry}

{ For any given radial velocity $w$, line emission or absorption is distributed evenly across a given projection angle $\varrho$ (the polar coordinate), which we illustrate here under full special relativistic treatment. The special relativistic velocity projection of $w$ in an arbitrary polar direction $\varrho^\prime$ in the observer's frame incorporates both longitudinal and transverse effects such that, following \citet[][their \S7]{Einstein1905},
\begin{equation}
    \lambda = \lambda_0\gamma\left(1+\omega\cos\varrho^\prime\right)
\end{equation}
where $\omega=w\:v_\infty/c$ and $\gamma^{-1}=\sqrt{1-\omega^2}$ is the Lorenzt factor. Additionally from \S7 of \citet[][]{Einstein1905}, the angle $\varrho$ in the source frame requires a relativistic aberration correction to the observer's frame $\varrho^\prime$, with a differential which reduces to 
\begin{equation}
    \sin\varrho^\prime{\rm d}\varrho^\prime = \left(\frac{\lambda_0}{\lambda}\right)^2 \sin\varrho{\rm d}\varrho .
\end{equation} 
Combining these identities and respective differentials with Equation \ref{eqn:vel_obs} provides a convenient substitution for the $\sin\varrho{\rm d} \varrho$ term in ${\rm d}\Omega$ such that
\begin{equation}
    {\rm d}\Omega^\prime = 2\pi\frac{\sqrt{1-\omega^2}}{\omega(1-\omega)^2}\sqrt{\frac{1-\omega}{1+\omega}} {\rm d}\omega
\end{equation}
where $2\pi$ is the integrated azimuthal angle assuming any given deprojected velocity $\omega$ is fully inscribed by the outflow (a spherical geometry) -- we return to adjust this assumption later. In the classical limit of $\omega\ll1$, the differential solid angle reduces to ${\rm d}\Omega = 2\pi/w\:{\rm d}w$, the classical solution given by \citet{Beals1931}. This ${\rm d}w$ substitution for d$\Omega$ is key to solving Equations \ref{eqn:phi_abs} and \ref{eqn:phi_ems} as it allows the line profile to be expressed solely in terms of differential velocity \citep[e.g.,][]{Castor1979}. The implication, here, is that a shell with smaller velocity (and thus smaller radius) contributes more to the differential surface area.}

\begin{figure}
    \centering
    \includegraphics[width=\linewidth,clip=True,trim={0in 0.5in 0in 0.15in}]{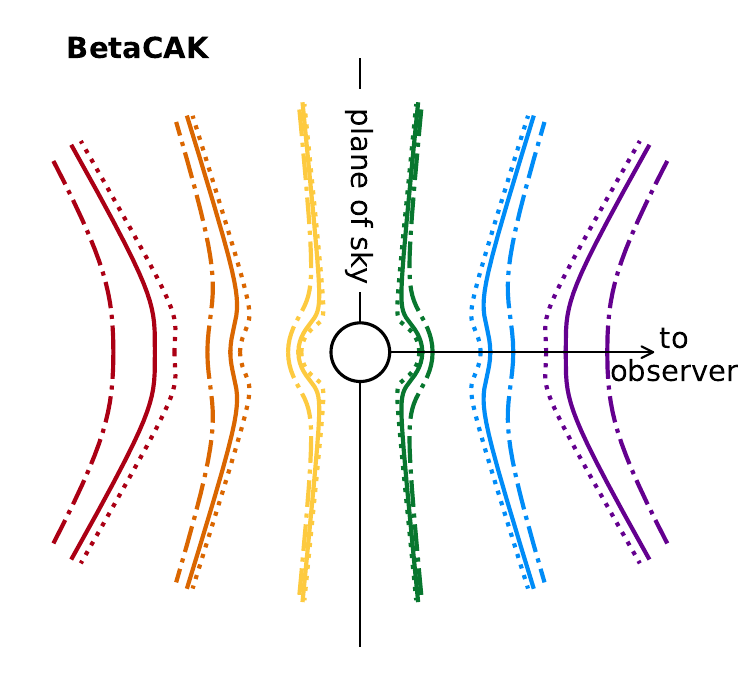}
    \includegraphics[width=\linewidth,clip=True,trim={0in 0.5in 0in 0.15in}]{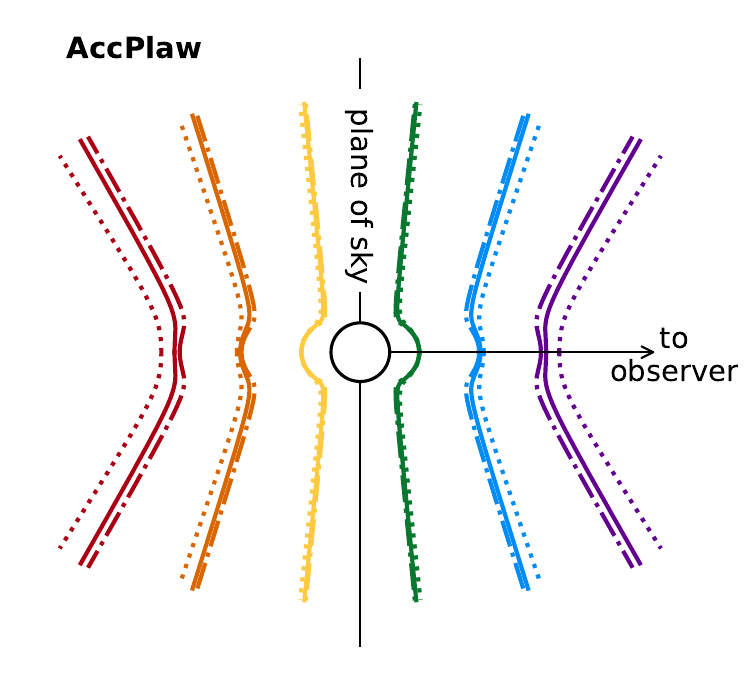}
    \includegraphics[width=\linewidth,clip=True,trim={0in 0.5in 0in 0.15in}]{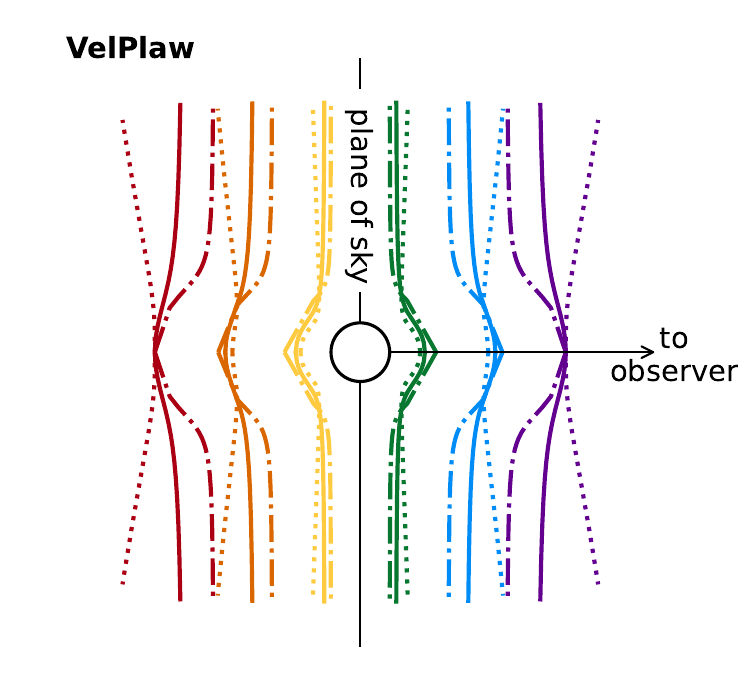}
    
    \caption{Velocity isocontours for the different velocity fields available in {\tt OutLines}: the $\beta$ law approximation to CAK theory ({\it top}), an acceleration power law ({\it middle}), and a velocity power law ({\it bottom}). Colors indicate fixed observed velocities of $v_{obs}/v_\infty=u=0.1$ (yellow/green), 0.3 (orange/blue), and 0.5 (red/purple). Line styles indicate low (dotted), intermediate (solid), and high (dash-dotted) values of $\beta$ for each velocity field. Arrow indicates the direction towards the observer, with cool colors representing blue Doppler shifts and warm colors representing red Doppler shift.}
    \label{fig:isocontours}
\end{figure}

To illustrate the effects of assumed velocity field and related $\cos\varrho$ projection onto the line of sight, we illustrate contours of constant observed velocity (isocontours) in Figure \ref{fig:isocontours} \citep[see also \citealt{Castor1970}, their Figure 1,][their Figure 2]{Lucy1971}. Sometimes referred to as velocity surfaces \citep[as in][]{Castor1970}, these isocontours represent the velocities which are cosine projected to a single observed velocity, assuming a particular velocity field and varying the power law index.

\subsubsection{Velocity Limits}

\begin{figure}
    \centering
    \includegraphics[width=\columnwidth]{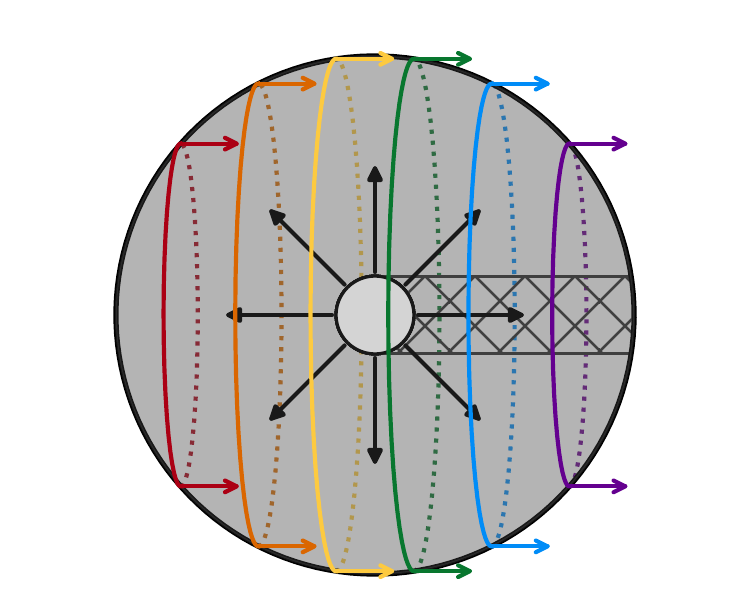}
    \caption{Spherically symmetric outflow presented in the text with a constant-velocity surface with a radius of $u$. Black arrows indicate the motion of the outflow. Central circle indicates the launching source of the outflow. Filled grey region indicates line-emitting material while the hatched region indicated line-absorbing material. Colored arrows indicate the direction of light propagation towards the observer. Colored circles indicate different deprojected velocities $w$ which can contribute to a single observed velocity surface $u$ with values of $u/w=0.15,0.5,0.85$ corresponding to yellow/green, orange/blue, and red/purple, respectively.
    }
    \label{fig:sphere_geometry}
\end{figure}

The weighted contributions of each deprojected velocity to the absorption or emission allows us to reduce the geometry to the velocity limits $w_\ell$ and $w_u$ of the integrals in Equations \ref{eqn:phi_abs} and \ref{eqn:phi_ems}. 
Boundary conditions for velocity in the case of a uniform spherical outflow (illustrated in Figure \ref{fig:sphere_geometry}) are given by some minimum and maximum projected velocities $w_\ell,\ w_u$, respectively. The lower limit $w_\ell$ is simply the observed velocity $|u|$ for a uniform sphere if no obstruction occurs. From Figure \ref{fig:sphere_geometry}, the line-emitting gas posterior to the source (redshifted in the case of an outflow) cannot be seen by the observer, meaning that $w_\ell$ must account for obstruction by the launching source at $x=1$.
Solutions for the source obstruction velocity limit are given in \citet{Flury2023}, which we summarize here. From the Pythagorean theorem, the minimum velocity $w_\ell$ must satisfy
\begin{equation}\label{eqn:w0source}
    w_0 = |u|\frac{x[w_0]}{\sqrt{x^2[w_0]-1}}
\end{equation}
giving a lower velocity limit of
\begin{equation}\label{eqn:w_lower}
    w_\ell = \begin{cases}
        |u| & {\rm no\ source\ effect} \\
        w_0 & {\rm source\ effect}\\        
    \end{cases}
\end{equation}
where the `source effect' applies only for the line-emitting gas posterior to the launch site relative to the observer (left half of the isocontours in Figure \ref{fig:isocontours}, of the sphere in Figure \ref{fig:sphere_geometry}, and of the cartoon in Figure \ref{fig:cone_geometry}). 

\begin{figure*}
    \centering
    \includegraphics[width=0.75\linewidth]{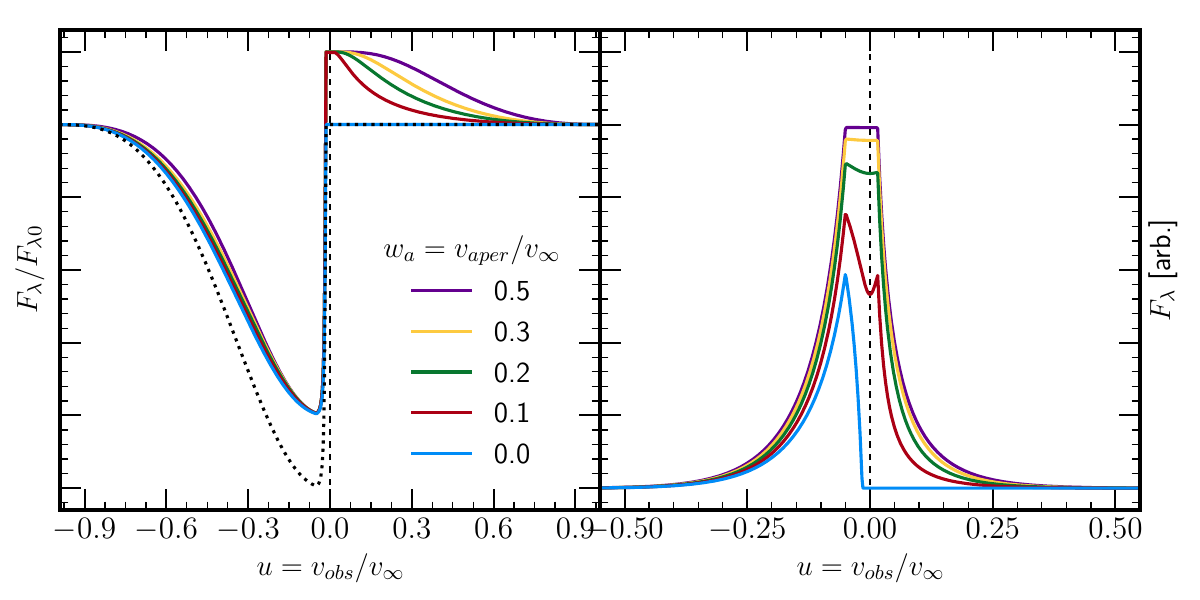}
    \caption{Effect of aperture clipping on the observed line profile for a P Cygni line consisting of absorption and resonant emission with filling factor $\epsilon=0.2$ and escape fraction $p_{res}=1$ (left panel) and for nebular emission (right panel). For visualization, lines are normalized to the case of no aperture clipping ($w_a=1$). Dotted black line in the left panel indicates pure absorption, demonstrating that infilling can still occur in the fully aperture-clipped limit but that only an absorption profile is observed \citep[cf.][]{2015ApJ...801...43S}. Note that the effect of radiative trapping on the resonant line core decreases as $w_a$ decreases.}
    \label{fig:aper_profiles}
\end{figure*}

\begin{figure}
    \centering
    \includegraphics[width=\linewidth]{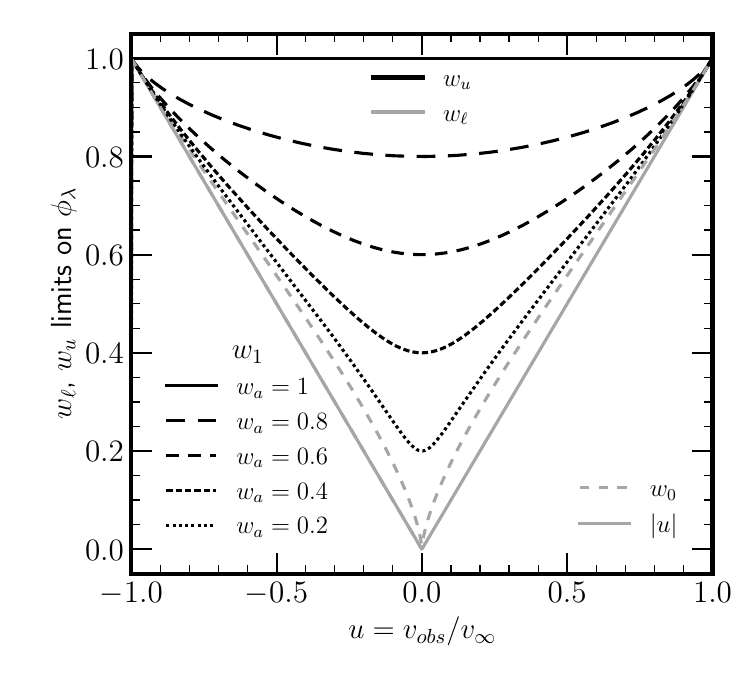}
    \caption{Integral velocity limits for the emission line profile $\phi_\lambda$. Lower limit $w_\ell$ is shown in grey with (dashed, $w_0$, Equation \ref{eqn:w0source}) and without (solid, $|u|$) the outflow source. Upper limit $w_u$ is shown in black with (dashed, $w_1$, Equation \ref{eqn:wAper}) and without (solid, unity) aperture clipping.}
    \label{fig:lims}
\end{figure}

In principle, the upper velocity limit is simply the terminal velocity, i.e., $w_u=1$. However, a spectroscopic aperture might not capture the full extent of the outflow. This aperture clipping effect is most likely to affect emission line profiles as the core arises from extended material. {\tt OutLines} incorporates aperture clipping into the geometry as some maximum velocity $v_{a}$ captured by the spectroscopic aperture on the transverse axis (plane of the sky) assuming that (i) the aperture is circular, as is the case for fiber spectrographs, and (ii) the outflow source is perfectly centered within the aperture. This effect can be imposed via the integral limits relative to the terminal velocity as $w_{a}=v_{a}/v_\infty$. The corresponding limiting velocity can be written as
\begin{equation}\label{eqn:wAper}
    w_1 = |u|\frac{x[w_1]}{\sqrt{x^2[w_1]-x^2[w_a]}}
\end{equation}
following the same approach as with $w_0$ in Equation \ref{eqn:w0source}. We show the change in $w_1$ for different $w_a$ in Figure \ref{fig:lims} and illustrate the effect of aperture clipping for different $w_a$ in Figure \ref{fig:aper_profiles}. 

The limit of $w_a$ most dramatically affects the projected velocities close to the transverse axis. As a result, aperture clipping will predominantly truncate the core of the profile close to the central wavelength \citep[see also][]{2015ApJ...801...43S}; i.e., the broadest extent of the wings remains \emph{relatively} unaffected.
Moreover, absorption lines arise from gas anterior to the launch site that is backlit by the source, resulting in a ``pencil beam'' effect where the maximum velocity contributing to the profile is $w_0$ (hatches in Figure \ref{fig:sphere_geometry}). Combining $w_1$ and $w_0$ gives a final upper velocity limit of 
\begin{equation}
w_u = 
    \begin{cases}
        w_0 & \textrm{absorption} \\
        w_1 & w_1 < 1 \\
        1 & \textrm{otherwise}\\
    \end{cases}.
\end{equation}
We visualize the integral limits $w_\ell$ and $w_u$ in Figure \ref{fig:lims}.

{ To accommodate a variety of scenarios, both the source and the aperture velocity limit effects can be toggled on or off in our models. For instance, if the source is sufficiently small so as not to obstruct the outflow, then source obstruction may be negligible, resulting in a symmetric profile. Alternatively, the assumed spherical geometry of the source may also be substituted with an obstructing disk following discussion below. For absorption lines, removing the source constraint changes the extent of background lighting of the proceeding gas rather than the obstruction of the receding gas. Removing the source constraint but allowing for an ``aperture'' allows for a variable-size backlight in cases where starlight may be more extended.}

\subsubsection{Directed Outflows}
\begin{figure}
    \centering
    \includegraphics[width=0.9\columnwidth]
    {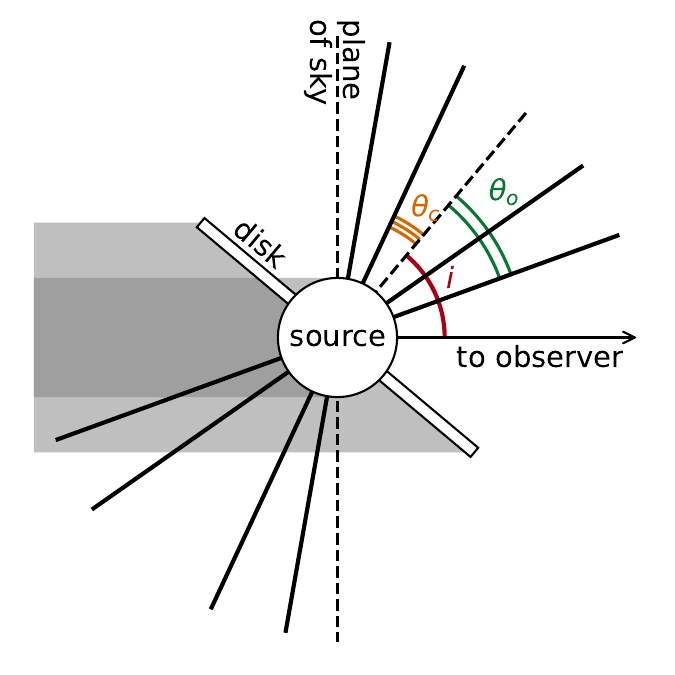}
    \caption{Outflow geometry accounting for obstruction by the outflow source (white circle with dark grey shaded region) and by an optional disk (rectangle with light grey shaded region). Red arc indicates the inclination angle $i$ of the cone axis with respect to the observer. Green double arc indicates the opening angle $\theta_o$ of the cone between the axis and the edge. Orange triple arc indicates the optional cavity angle $\theta_c$ between the cone axis and the inner edge of the outflow.}
    \label{fig:cone_geometry}
\end{figure}

Many outflows are not spherically uniform but rather are observed as bi-directional cones on a wide variety of scales \citep[e.g.,][]{Fukui1986,Clegg1987,Heckman1990,Colina1991,Hirano1992,Marlow1995,Cameron2021}. We consider a bi-directional conical geometry with a half opening angle $\theta_o$ and an inclination angle $i$ with respect to the observer, which we depict in Figure \ref{fig:cone_geometry} (see also, e.g., Figure 5 in \citealt{Hjelm1996}). {\tt OutLines} includes a tool {\tt PlotGeometry} for producing such figures as a visual aid.

\begin{figure*}
    \centering
    \includegraphics[width=0.75\linewidth]{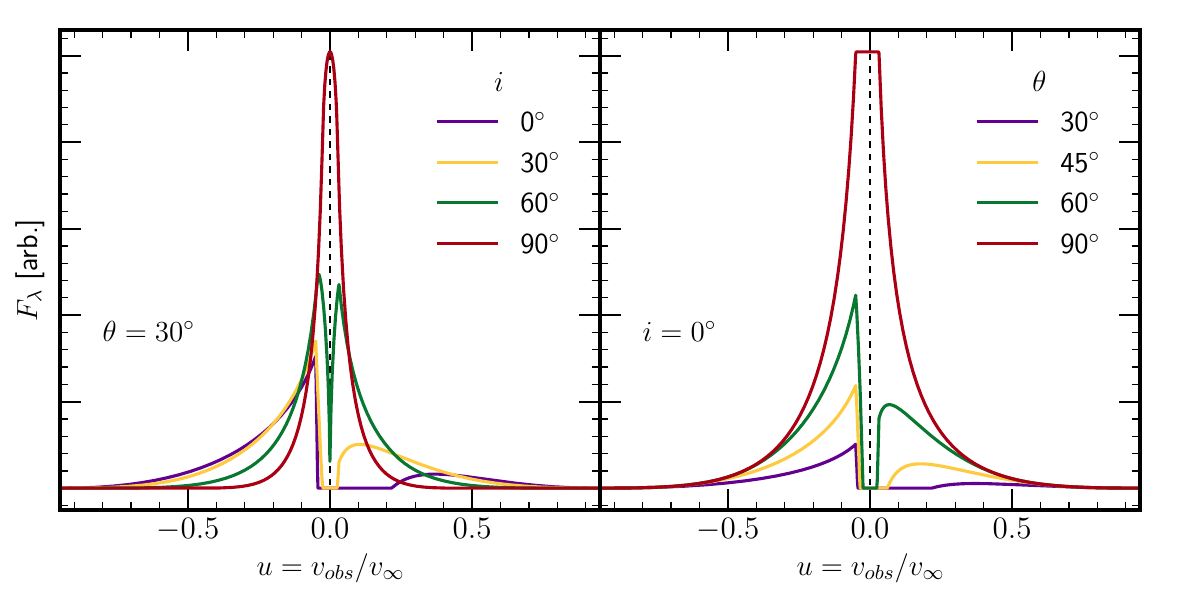}
    \includegraphics[width=0.75\linewidth]{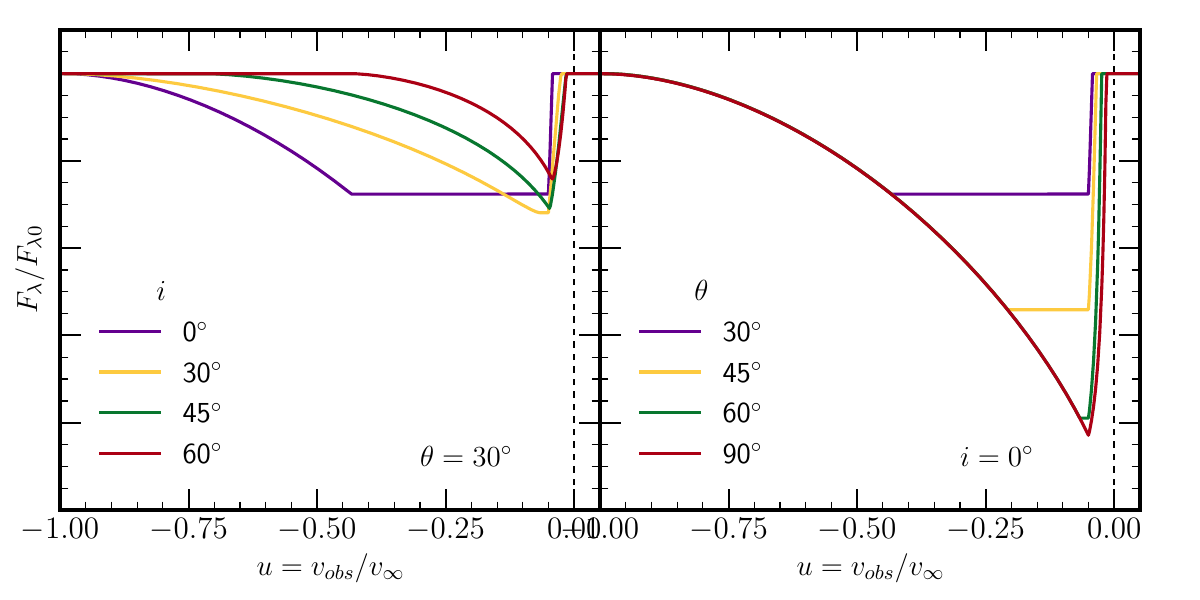}
    \caption{Effects of bi-directional conical geometry on the emergent line profiles of emission (top) and absorption (bottom) profiles for fixed density (power law with $\alpha=2$) and velocity ($\beta=1$) fields. To illustrate the effects of geometry, we either vary the inclination with respect to the observer (left) from face-on ($i=0^\circ$) to edge-on ($i=90^\circ$) with fixed opening angle of $\theta=30^\circ$ or vary opening angle of the cones (right) from narrow ($\theta_o=30^\circ$) to a full sphere ($\theta=90^\circ$) with face-on inclination. The case of $i=0$ and $\theta=90^\circ$ (red line, right) is consistent with a filled sphere. For visualization, line profiles are all normalized to the spherical case ($i=0^\circ$, $\theta=90^\circ$). The effects of inclination on the emergent line profile shown in the upper left are similar to the 3D Monte Carlo simulation predictions for X-ray fluorescent lines from the {\sc skirt} code \citep{SKIRT2024,XRISM2026_NGC1068}, with face-on inclination yielding broader and flatter profiles and edge-on inclination yielding narrower and peakier profiles.}
    \label{fig:filled_cones}
\end{figure*}

The major difference with conical geometry is that the polar coordinate of the differential solid angle \dOmega\ now depends on $i$ and $\theta$; we cannot simply assume each deprojected velocity contributes equally to the integral due partial coverage. Therefore, we must account for azimuthal angle of the deprojected velocity which falls within the region of the cone projected onto the plane of the sky. The azimuthal angle inscribed by the projected cone simply scales the total solid angle by a factor \lb$(u,w,i,\theta)\in[0,1]$ to give the relative contribution of a de-projected velocity $w$ to the observed velocity $u$. {\tt OutLines} computes the conical projection from $i,\theta_o$ and then calculates \lb\ by determining the angular length(s) of each deprojected velocity inscribed by the projected cone. For details on this method, we refer the reader to Appendix \ref{apx:projections} \citep[see, e.g.,][for earlier approaches to solving this problem]{Zheng1990,Carr2018,Luminari2018}. We illustrate the effects of changing the opening angle $\theta_o$ and inclination angle $i$ in Figure \ref{fig:filled_cones}.

\subsubsection{Directed Outflows with Cavities}

\begin{figure*}
    \centering
    \includegraphics[width=0.75\linewidth]{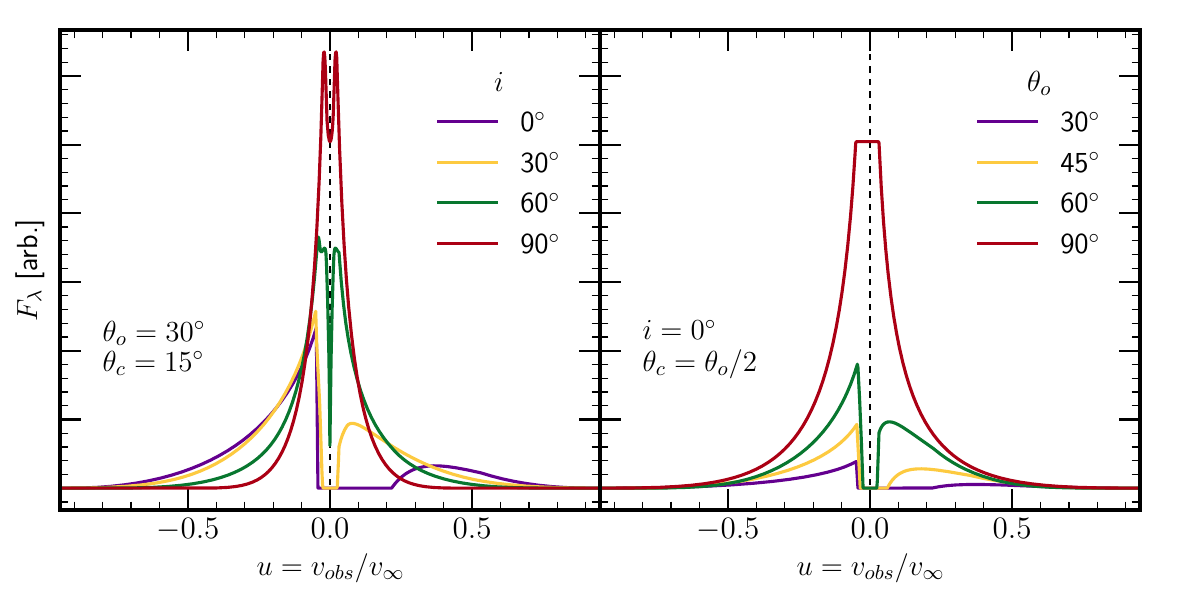}
    \includegraphics[width=0.75\linewidth]{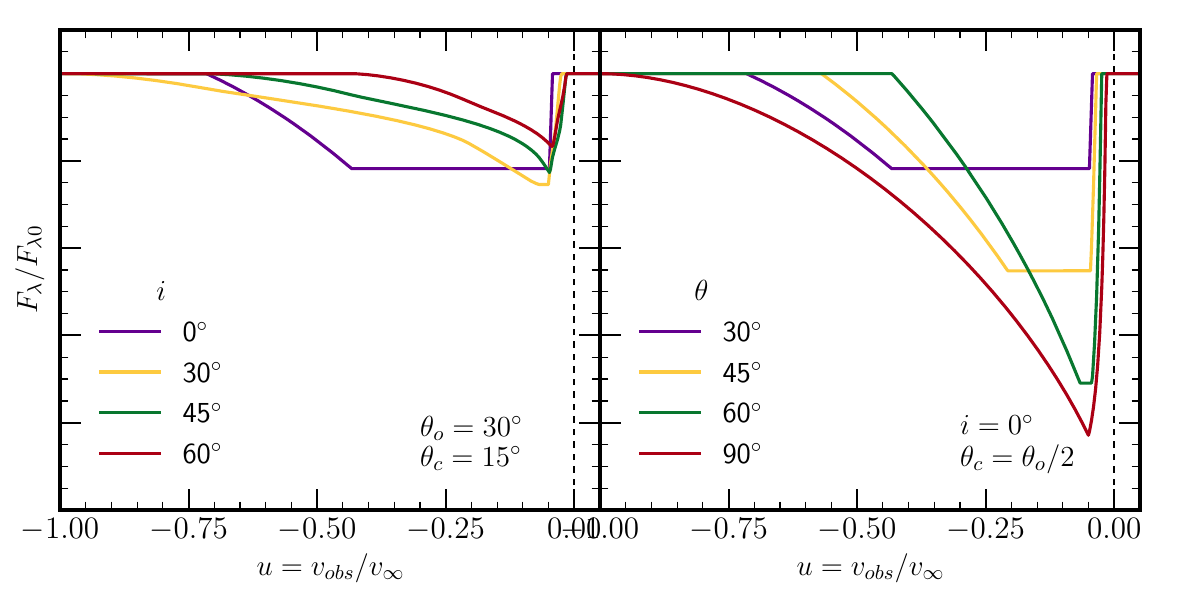}
    \caption{Same as Figure \ref{fig:filled_cones} but for hollow bi-directional cones with cavity angle $\theta_c=\theta_o/2$.}
    \label{fig:hollow_cones}
\end{figure*}

An additional geometry observed in young stellar objects \citep[e.g.,][]{Padgett1999,Louvet2018,deValon2020}, stellar winds \citep[e.g.][]{Sirianni1998}, starbursts \citep[e.g.,][]{Axon1978,Marlow1995,Heckman1995,Rich2010,Yoshida2010}, and AGN \citep[e.g.,][]{Phillips1983,Hjelm1996,Crenshaw2000a,Crenshaw2000b,Muller2011,Venturi2018} is that of hollow cones of outflowing material. We include this geometry, assuming some cavity angle $\theta_c$ and opening angle $\theta_o$ \citep[see also models by][]{Das2006,Muller2011,Bae2016,Luminari2024A&A}. {\tt OutLines} implements this geometry by subtracting the fractional angular length $\ell_{\rm cavity}$ of each deprojected velocity inscribed by the cavity from the fractional angular length inscribed by the total cone (see Appendix \ref{apx:projections}). The difference will always be positive and greater than zero since $\theta_o>\theta_c$ by definition. We illustrate the impact of a hollowing of the outflow cone on the emergent line profile in Figure \ref{fig:hollow_cones}. Somewhat intuitively, an edge-on hollow-cone outflow loses flux from the least Doppler-shifted gas (truncating the core) while a face-on hollow-cone outflow loses flux from the most Doppler-shifted gas (truncating the wings).

\subsubsection{An Obstructing Disk}

The outflow may be launched from a region embedded in a disk of optically thick material, a phenomenon observed in a variety of astrophysical scenarios: expelled debris in planetary nebulae \citep[e.g.,][]{Icke1981}, protoplanetary material around young stellar objects \citep[e.g.,][]{Padgett1999,Ilee2016}, the dusty material concentrated around an AGN \citep[e.g.,][]{Veilleux2001}, or even the plane of a galaxy \citep[e.g.,][]{Hjelm1996,Cronin2025}. In such cases, the disk will obstruct some of the observer's view of the outflow. {\tt OutLines} optionally includes obstruction by a circular disk which is orthogonal to the central axis of the cones and which has some radius $x_d=r_{\rm disk}/R_0$. As with the cones, the disk must be projected onto the plane of the sky and subsequently compared to the deprojected velocity inscribed by the projected cone to determine the fractional arc length(s) \lb\ inscribed by the cone but not the projected disk. We elaborate on this formalism in Appendix \ref{apx:projections}. {\tt OutLines} implements a disk either in tandem with the cone geometries or alone as a hemisphere geometry.

\begin{figure*}
    \centering
    \includegraphics[width=0.75\linewidth,clip=True]{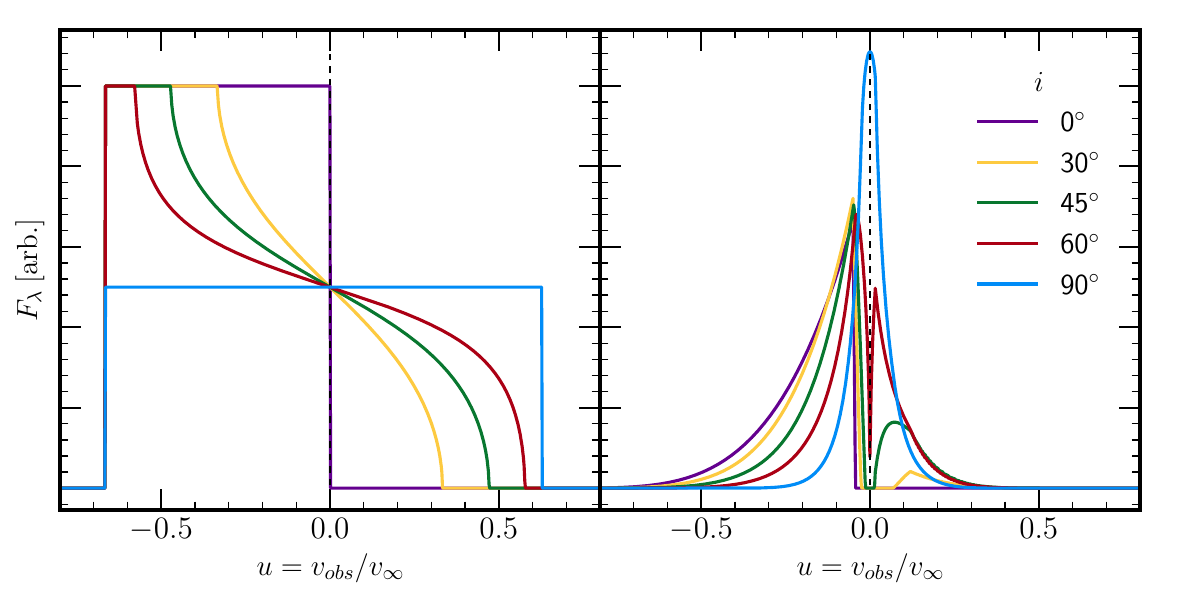}
    \caption{Same as Figure \ref{fig:filled_cones} but for the effects of an obstructing disk as a function of inclination with respect to the observer in ({\it left}) a thin single-shell hemisphere geometry where all material posterior to the disk is obstructed \citep[cf.][their Fig 12]{Xrism2025_PDS456} and ({\it right}) a biconical wind with $\theta_0=30^\circ$ and $x_{\rm disk}=5$. Colors indicate changes to the inclination angle $i$, illustrating the redistribution of flux from entirely blue-shifted ($i=0^\circ$) to evenly blue- and red-shifted ($i=90^\circ$).}
    \label{fig:hemisphere}
\end{figure*}

\begin{figure*}
    \centering
    \includegraphics[width=0.75\linewidth]{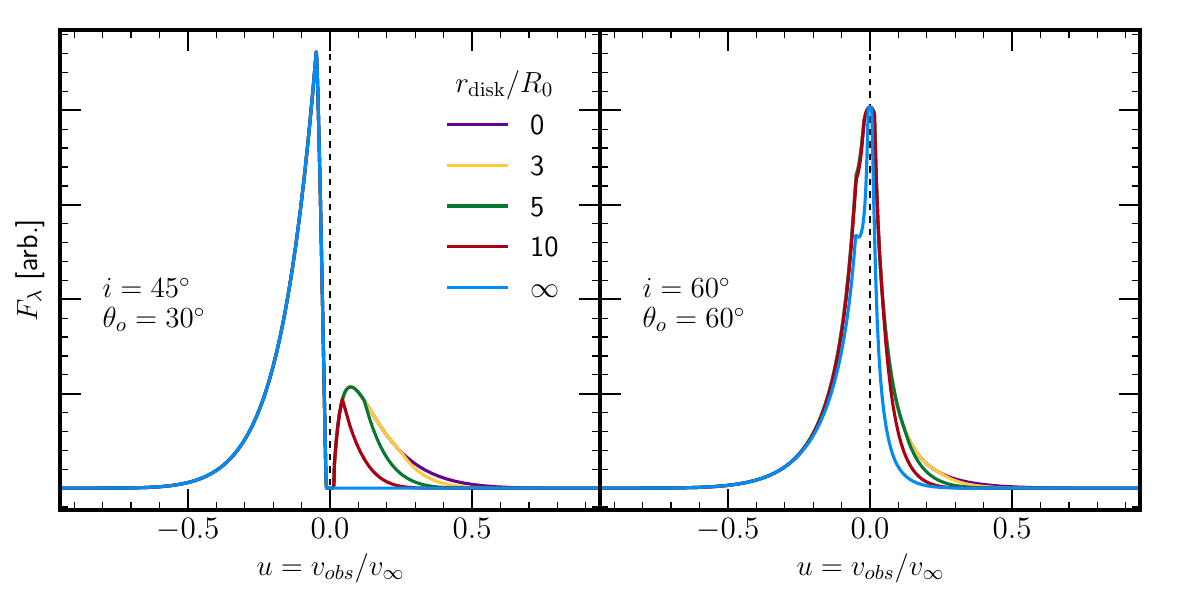}
    \caption{Same as Figure \ref{fig:filled_cones} but for bidirectional cones with fixed inclination and opening angle and an obscuring disk with varying radii, where a radius of 0 indicates a disk-free line profile. All profiles are normalized to the disk-free case for illustrative purposes. On the left, $i+\theta_o<90^\circ$, resulting in loss of the redshifted component. On the right, $i+\theta_o>90^\circ$, resulting in losses of both red- and blueshifted components when the disk radius is sufficiently high. For visualization, line profiles are normalized to the disk-free case in each panel.}
    \label{fig:cone_disk_profiles}
\end{figure*}

{ Disk inclination largely serves to determine how evenly emission is distributed across the red- and blue-shifted components, which we  illustrate for a hemisphere and for bidirectional cones in Figure \ref{fig:hemisphere}. As the disk radius changes relative to the size of the outflow, the cone posterior to the disk and projected away from the observer is increasingly obstructed, which we demonstrate in Figure \ref{fig:cone_disk_profiles}.}
By definition, the disk will only obstruct the posterior cone, so if $i+\theta_o < 90^\circ$, the disk will only obstruct the receding outflow, solely obstructing the redshifted component (left of Figure \ref{fig:cone_disk_profiles}). However, when $i+\theta_o>90^\circ$, the disk obstructs the lune projected by the posterior cone toward the observer, resulting in loss of the blueshifted component of the line profile as well (right of Figure \ref{fig:cone_disk_profiles}).

\subsection{Outflow Properties}\label{sec:properties}

Following \citet[][see their Appendix B]{Flury2023}, the characteristic outflow radius $x_{\rm out}$ corresponds to the maximum momentum density $\dot{\rho}_p = \mu m_{\rm H}n_0v_\infty n[x]w[x]$, i.e., where the wind effects are strongest. In the case of pulse-like density distributions, the center of the gas distribution $x_1$ will be close to, but not always the same as, $x_{\rm out}$.
The radius $x_{\rm out}$ of maximum momentum density scales all other outflow properties, first and foremost the characteristic outflow velocity $v_{\rm out}$ by determining the velocity at $x_{\rm out}$ using the velocity field $w$ such that
\begin{equation}\label{eqn:vout}
    v_{\rm out} = v_\infty w[x_{\rm out}].
\end{equation}
The effective column density $N_{\rm out}$ at the characteristic radius can be determined by integrating over the density profile from the base of the outflow ($x=1$) to the peak momentum density $x=x_{\rm out}$ \citep[e.g.,][their Equation 13 for the case of a CAK velocity field and power law density profile]{Flury2023}. Note that $N_{\rm out}$ is not to be confused with $N$ from the absorption line profile, which is the total column density along the line of sight. $N_{\rm out}$ necessarily depends on the space density $n_0$ and the radius at the base of the outflow $R_0$, which are unconstrained parameters in {\tt OutLines}. As such, {\tt OutLines} computes the relative column density $\mathcal{R}$ where
\begin{equation}\label{eqn:Rcal}
    \frac{N_{\rm out}}{n_0 R_0} = \mathcal{R} = \int\limits_1^{x_{\rm out}} n[x] {\rm d}x.
\end{equation}
This quantity $\mathcal{R}$ becomes important for computing the strength of the outflow expressed via $\dot{M}$, $\dot{p}$, and $\dot{E}$.

The mass outflow rate $\dot{M}$ is given by the momentum flux $\mu m_{\rm H} n_0 v_{\rm out}$ with a geometric projection $4\pi R_0^2 (\cos\theta_c-\cos\theta_o)$ such that
\begin{equation}\label{eqn:mdot}
    \dot{M} = 4\pi(\cos\theta_c-\cos\theta_o) n_0 R_0^2 \mathcal{R} \mu m_{\rm H} v_{\rm out},
\end{equation}
a more general case of Equation 12 from \citet{Flury2023} \citep[see also, e.g.,][their Equation 4]{Martin2005}. Such formalism is frequent in the literature with differences in implementation of the geometric projection \citep[e.g.][]{Martin2005,2005ApJS..160...87R,2005ApJS..160..115R,Crenshaw2012,Maiolino2012,Fiore2017,Luminari2018,Luminari2024A&A,Flury2023}.
The outflow momentum rate $\dot{p}$ expresses the rate at which momentum is injected into the outflow. Typically, $\dot{p}$ is defined as the product of outflow velocity and mass outflow rate \citep[e.g.,][their Equation 9]{2005ApJS..160..115R} such that
\begin{equation}\label{eqn:pdot}
    \dot{p} = v_{\rm out} \dot{M} = 4\pi(\cos\theta_c-\cos\theta_o) n_0 R_0^2 \mathcal{R} \mu m_{\rm H} v_{\rm out}^2 .
\end{equation}
The outflow mechanical luminosity $\dot{E}$ expresses the injection of kinetic energy into the outflow. Typically, $\dot{E}$ is defined as the product of outflow velocity and mass outflow rate \citep[e.g.,][their Equation 11]{2005ApJS..160..115R} such that
\begin{equation}\label{eqn:edot}
    \dot{E}_{out}
    \begin{split}
        =& \frac{1}{2}\dot{M}\left(v_{\rm out}^2 + 3\sigma_v^2\right)\\
        =& 2\pi(\cos\theta_c-\cos\theta_o) n_0 R_0^2 \mathcal{R} \mu m_{\rm H}\left(v_{\rm out}^3 + 3v_{\rm out}\sigma_v^2\right).
    \end{split}
\end{equation}
These values can be readily computed using constraints on $x_{\rm out}$ and $v_\infty$ and possible geometric terms $\theta_o$ and $\theta_c$, all obtained from the line profile. The only two unknowns are $n_0$ and $R_0$, which cannot be directly informed by the line profile and require additional measurements (spatial for $R_0$, spectral for $n_0$) or assumptions based on prior knowledge. For instance, $n_0=10\rm~cm^{-3}$ and $R_0=0.5$ kpc may be reasonable for galactic outflows (e.g., \citealt{1985Natur.317...44C,Flury2023}).

As implemented here, we compute $\dot{M}$, $\dot{p}$, and $\dot{E}$ normalized to $n_0R_0^2$ with $n_0$ in cm$^{-3}$ and $R_0$ in kpc, meaning the characteristics of the outwardly moving gas are written as 
\begin{align}
        \frac{\dot{M}}{\rm M_\odot~yr^{-1}}\left[\frac{n_0R_0^2}{\rm cm^{-3}~kpc^{2}}\right]^{-1} & =0.191\mathcal{R}[\cos\theta_c-\cos\theta_o]\left[\frac{v_{\rm out}}{\rm km~s^{-1}}\right]\label{eqn:MdotNR2}\\
        \frac{\dot{p}}{\rm dyne}\left[\frac{n_0R_0^2}{\rm cm^{-3}~kpc^{2}}\right]^{-1} & = 10^{-3.2}\left[\frac{v_{\rm out}}{\rm km~s^{-1}}\right]\left[\frac{\dot{M}}{\rm M_\odot~yr^{-1}}\right]\label{eqn:pdotNR2}\\
        \frac{\dot{E}}{\rm erg~s^{-1}}\left[\frac{n_0R_0^2}{\rm cm^{-3}~kpc^{2}}\right]^{-1} & = 10^{-6.5}\left[\frac{v_{\rm out}}{\rm km~s^{-1}}\right]^2\left[\frac{\dot{M}}{\rm M_\odot~yr^{-1}}\right]\label{eqn:EdotNR2}
\end{align}
Typically, $n_0R_0^2$ is of order unity for the case studies in \S\ref{sec:application}, with values ranging from about 1 to 5 cm$^{-3}$ kpc$^{2}$, although for Mrk 462, \citet{Flury2023} find $n_0R_0^2=12$ cm$^{-3}$ kpc$^{2}$. 

\section{The {\tt OutLines} Code}\label{sec:OutLines}

{\tt OutLines}, a portmanteau of outflows and spectral lines, implements all of the above model components in a single code for emission and absorption lines. The code is designed to be simultaneously efficient and user-friendly, with a straightforward user interface and ready scientific application. Below, we assess {\tt OutLines} in the contexts of standard techniques, including empirical percentiles and multi-Gaussian fitting. Details of the architecture and use of the {\tt OutLines} code can be found in Appendix \ref{apx:OutLines} { with additional examples available in Jupyter notebooks online at \url{https://github.com/sflury/OutLines}}.

\subsection{Empirical Percentiles}

\begin{figure}
    \centering
    \includegraphics[width=\linewidth]{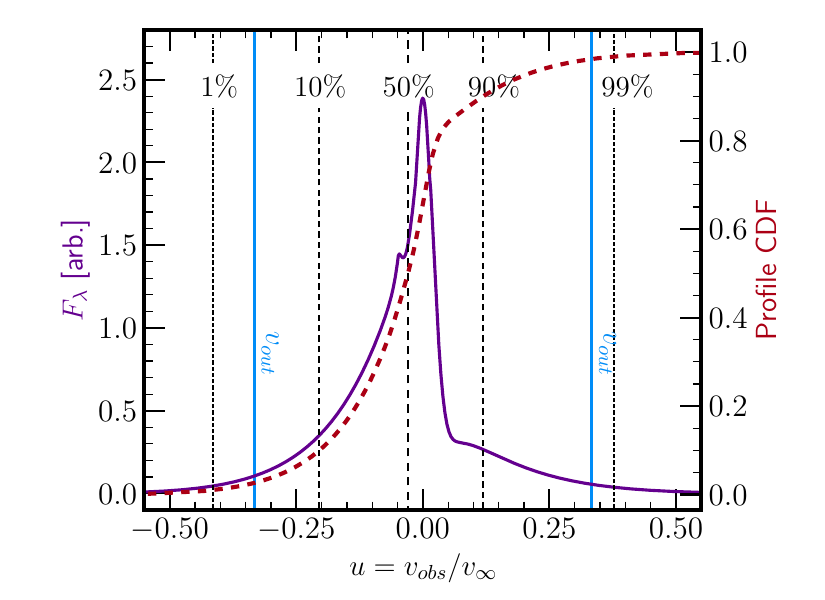}
    \includegraphics[width=\linewidth]{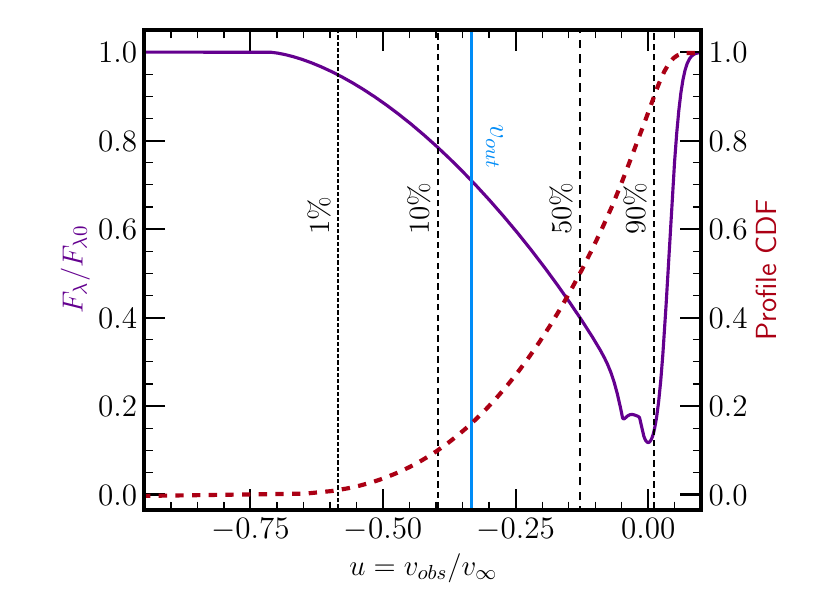}
    \caption{Velocity percentiles (black dashed lines) for an arbitrary emission ({\it top}) and absorption ({\it bottom}) line (purple) generated by {\tt OutLines} for a directed, filled cone geometry ({\tt Geometry='FilledCones'}) for $i=45^\circ$ with $v_0=0.1v_\infty$ ({\tt FromRest=False}), default parameters, and otherwise default settings. The cumulative distribution function (fraction of integrated flux) is shown in red. For reference, we show the characteristic outflow velocity $v_{\rm out}$ (Equation \ref{eqn:vout}) in blue.}
    \label{fig:quantiles}
\end{figure}

The empirical approach of velocity percentiles \citep[e.g.,][]{1981ApJ...247..403H,1985MNRAS.213....1W,1991ApJS...75..383V,2013ApJ...768...75R,2013MNRAS.436.2576L,2020A&ARv..28....2V} measures the velocity at which the cumulative flux or equivalent distribution reaches some percentile. {\tt OutLines} provides quantiles of individual model line profiles through the {\tt velocity\_quantiles} method bound to the line profile object class and a convenience function {\tt get\_W80}, a common interval in the literature \citep[e.g.,][]{2014MNRAS.441.3306H,Fiore2017}, which returns the $W80$ value from the 10-90 percentiles. While {\tt OutLines} includes a {\tt get\_FWHF} function similar to the full-width at half-maximum (FWHM), this function is the velocity width of the profile containing the inner 50\% of the flux ($W_{50}$) and can be thought of as the full-width at half-flux (FWHF) centered on the first moment (median) of the flux distribution. We note that the FWHM can be somewhat misleading, particularly in cases of a high velocity outflow where the flux density is low relative to the maximum and would not be represented by the FWHM velocity (see Fig \ref{fig:quantiles}) or could yield multiple values in cases where the outflow component is multi-peaked. Illustrations of the 1$\sigma$ and 2$\sigma$ quantiles for an absorption and emission line with the {\tt OutLines} defaults for filled cone geometry and a static component are shown in Figure \ref{fig:quantiles}. 

\subsection{Modeling Line Profiles}\label{sec:modeling}

Given that {\tt OutLines} will sometimes require a large number of parameters (anywhere from four to nine for a given line depending on the geometry and density profile), we strongly recommend the use of Markov-chain Monte Carlo (MCMC) or other Gaussian-process samplers of the posterior on the parameters rather than the various least squares minimizers. Not only can this improve fit results but also allow users to incorporate informative priors, plus marginalization of particular nuisance parameters, which the trust-reflective algorithm, Levenberg-Marquardt, and other nonlinear least squares routines cannot. E.g., in the hollow cone scenario, $\theta_o>\theta_c$ necessarily and should be enforced.

To facilitate modeling of observed line profiles, the {\tt Profile} classes contain several attributes and functions. Users can access the model line profile function directly via the {\tt Profile} attribute, which receives rest wavelengths and parameters from the user and returns the total line profile predicted by the {\tt OutLines} code. When instantiated, the {\tt Profile} classes also contain initial parameter guesses and boundaries accessible via the {\tt get\_params()} and {\tt get\_bounds()} methods. For convenience, these two classes also contain the {\tt init\_params} function which will sample walkers within the current parameter bounds given the current parameter values. These initial parameter guesses can be readily seeded into an MCMC sampler while the boundaries can be folded into a prior. For this purpose, the {\tt log\_probability()} bound method includes both the likelihood and a uniform bounded prior incorporating the boundary values contained in the object. Users are able to update any or all of these bound methods as needed. An example of modeling with {\tt OutLines} is provided in Appendix \ref{apx:modeling}.

While we encourage simultaneous fitting of multiple features from the same or very similar species, we caution against modeling lines from different species which may not necessarily share, say, the same density profile parameters due to ionization structure within a multiphase outflow.

{\tt OutLines} does not presently include line-spread function (LSF) corrections, { in part because not all instrumental LSFs are well-behaved (e.g., \emph{HST}/COS). Such corrections can be readily implemented by generating a line profile with {\tt OutLines} at the sampling resolution of the detector and convolving with the LSF before comparing with the data.} To assess the impact of the LSF on the recovery of line profiles, we generate $10^4$ mock observations varying $R$ from 1 000 to 10 000, $v_\infty$ from 100 to 1 000 \kms, and $\beta$ from 0 to 2.5 assuming the {\tt OutLines} defaults. We convolve the predicted line profiles with a Gaussian LSF and add noise consistent with S/N between 3 and 10 in the continuum. We fit the mock observations using the {\tt scipy.least\_squares} trust reflective algorithm to attempt to recover the input parameters.
Our simulation indicates that $R\sim1500$ is sufficient to recover outflow properties to within $\la15$\% with a simple Gaussian LSF correction and $R\sim3000$ to within $10$\%, suggesting that facilities such as \emph{JWST}/NIRSpec can be used with {\tt OutLines} to recover key information about winds, bubbles, and outflows. Moreover, we find that $R\sim6000$ can recover parameters within $\approx5$\% of the input values even without an LSF correction, suggesting LSF corrections in this regime are negligible.

\begin{figure*}
    \centering
    \includegraphics[width=0.9\linewidth]{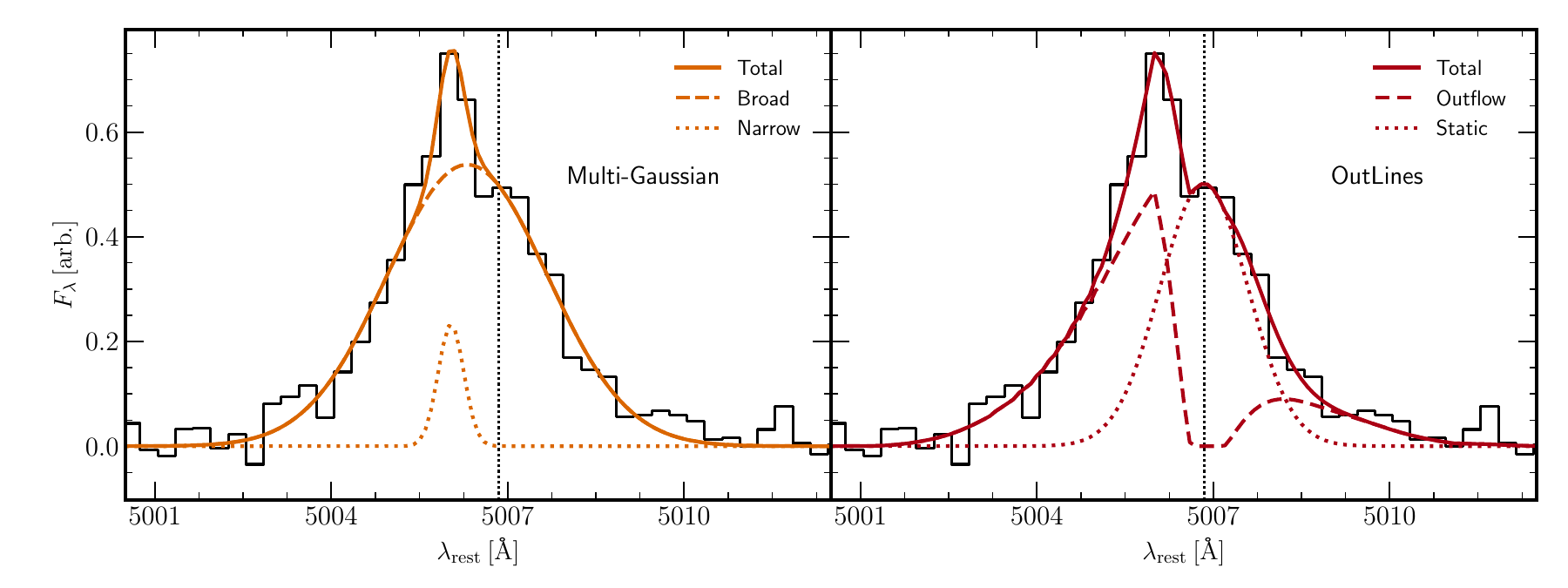}
    \caption{Fit to a mock [\ion{O}{iii}] $\lambda5007$ outflow profile (see \S\ref{sec:modeling}) with a double Gaussian (left) and with {\tt OutLines} (right). While the fit quality is comparable in terms of $\chi^2$, the inferred fluxes and kinematics differ dramatically, leading to disparate scientific conclusions.}
    \label{fig:mock-fit}
\end{figure*}

We illustrate the importance of using {\tt OutLines} over traditional methods by comparing a fit to a mock [\ion{O}{iii}] $\lambda5007$ line profile profile at $R=5\ 000$ with Nyquist sampling and S/N$=10$ at the peak flux with {\tt OutLines} and with a double Gaussian, both of which we show in Figure \ref{fig:mock-fit}. For this purpose, we generate a mock line profile using {\tt OutLines} for an acceleration power law velocity field with $\beta=1.5$ and $v_\infty=500$ \kms, power law density profile with $\alpha=1$, and filled cone geometry with $i=45^\circ$ and $\theta_o=30^\circ$. We add a static component with $\sigma_v=50$ \kms. Both the static and outflow components have equivalent integrated fluxes normalized to unity in the mock profile.

Overall quality of the best-fit profile we obtain is comparable 
($\chi^2=2.057$ and $2.942$ for {\tt OutLines} and a double Gaussian, respectively).
The double Gaussian model does have fewer parameters (6) than the {\tt OutLines} model (9). Given the similarity in $\chi^2$, the difference in Akaike information criterion (AIC) between the {\tt OutLines} and double-Gaussian model is dominated by the number of parameters, leading to $\Delta$AIC$=6.7$. At face value, the AIC might favor the multi-Gaussian approach; however, the multi-Gaussian approach has no strong physical argument and, as we elaborate below, leads to spurious conclusions about the physical nature of the observed phenomenon. The AIC results are therefore more indicative of the shortfalls of relying solely on statistical metrics rather than physical motivation. In other words, all free parameters are equal, but some are more equal than others.

With this caveat in mind, we compare the results between the standard method and our new approach with {\tt OutLines} in modeling an outflow. {\tt OutLines} is able to recover all input parameters within the imposed 10\% uncertainty due to noise. The double Gaussian model over-predicts the outflow flux by a factor of two and under-predicts the static gas flux by a factor of three while characterizing the outflow as red-shifted relative to the systemic velocity of the static gas. The double Gaussian model results lead to vastly different physical interpretations of the observations than those from {\tt OutLines}: one would conclude the presence of a more massive outflow that is systematically traveling away from the static gas at $15$ \kms, perhaps merger-like activity.
One might even be inclined to interpret the double-Gaussian result as evidence for an inflow rather than an outflow, which is not the case. The {\tt OutLines} best-fit model indicates an outflow launched from the same systemic velocity as the static gas with a bidirectional cone geometry and with characteristic velocities of $v_{\rm out}=220$ \kms. Following Equation \ref{eqn:mdot}, accounting for the geometry (a factor of $1/(1-\cos\pi/6)\approx7.5$), differences in line flux ($n_0\propto\sqrt{F_0}$ gives a factor $\sqrt{2}$), and difference in $v_{\rm out}$ (factor of 0.5), the double Gaussian model yields a mass loss rate higher than that of {\tt OutLines} by a factor of five. Thus, while the velocity inferred from the double Gaussian model is half that from {\tt OutLines}, the double Gaussian model implies a substantially higher impact of the outflow on its environment than the {\tt OutLines} model.

\subsection{Absorption vs Emission}

A key result of the development and applications of {\tt OutLines} is that absorption and emission lines optimize constraints on different properties of outflows. Absorption lines are more sensitive to the line-of-sight, as evidenced in the line profiles (Figures \ref{fig:velocity_profiles} and \ref{fig:dens_profs}) and the narrow range in velocities contributing to each wavelength bin due to the ``down-the-barrel'' effect (Figures \ref{fig:sphere_geometry} and \ref{fig:lims}), making them good probes of the density profile and velocity field but poor tracers of the geometry. Emission lines trace the global gas distribution since velocities from across the majority of the gas contribute to each velocity bin (Figure \ref{fig:lims}), making them good probes of the geometry (Figures \ref{fig:filled_cones}, \ref{fig:hollow_cones}, \ref{fig:cone_disk_profiles}) but, as a result, lose some constraining power on the density profile and velocity field. As such, combined modeling of both emission and absorption provides optimal constraints on the outflow properties. 

\section{A High Capacity for Science}\label{sec:application}

Below, we demonstrate application of {\tt OutLines} to a variety of phenomena on different spatial and temporal scales, all involving astrophysical winds: stellar wind bubbles in the brightest cluster of the giant extragalactic \ion{H}{ii} region NGC 5471, radiative feedback and Lyman continuum escape from a super star cluster in Green Pea J1044+4012, blowout and the formation of a directed outflow in starburst Mrk 1486, and thermal/cosmic ray pressure from AGN-shocked bubbles in the Seyfert 2 nucleus of NGC 2992. Results from modeling these examples with {\tt OutLines} are given in Table \ref{tab:examp_fits} and are discussed on a cases-by-case basis in the subsections below.

All modeling with {\tt OutLines} is done here using {\tt emcee} to sample the posterior on the parameters. The MCMC chain consists of a number of walkers equal to twice the number of free parameters plus one (the {\tt OutLines} default), with each walker taking $10^3$ steps with an additional $200$ burn-in steps given the number of walkers (minimum of 10 walkers for at least 2000 individual burn-in steps), which we find to be sufficient to ``forget'' the initial guess parameters. The initial guess parameters, unless otherwise stated, are obtained using the {\tt scipy.optimize.least\_squares} trust-reflective algorithm fit. The prior is assumed to be uniform bounded unless otherwise stated. We show posterior sampling for each case in Appendix \ref{apx:corners}.


\begin{table*}
    \centering
    \caption{Best-fit parameters and properties of test-case objects: Knot A of giant extragalactic \ion{H}{ii} region NGC 5471, super star cluster in the center of Green Pea J1044+3053, star-forming galaxy Mrk 1486, and Seyfert 2 NGC 2992. As discussed in \S\ref{sec:properties}, reported properties of $\dot{M}$, $\dot{p}$, and $\dot{E}$ are normalized to $n_0R_0^2$ with $n_0$ in units of cm$^{-3}$ and $R_0$ in kpc (see Equations \ref{eqn:MdotNR2}-\ref{eqn:EdotNR2}).
    \label{tab:examp_fits}}
    \begin{tabular}{l *{10}{ S[table-format=2.2]} }
    Object      & \multicolumn{2}{c}{NGC 5471 Knot A} & \multicolumn{2}{c}{J1044+0353 SSC} & \multicolumn{2}{c}{Mrk 1486} & \multicolumn{2}{c}{NGC 2992 Sy2 nucl.} \\
    Scenario      & \multicolumn{2}{c}{\ion{H}{ii} region winds} & \multicolumn{2}{c}{Lyman continuum escape} & \multicolumn{2}{c}{galaxy-scale outflows} & \multicolumn{2}{c}{AGN disk winds} \\
    Feature(s)      & \multicolumn{2}{c}{H$\alpha$, [\ion{N}{ii}] $\lambda\lambda6548,6583$} & \multicolumn{2}{c}{\ion{C}{ii} $\lambda1334$} & \multicolumn{2}{c}{\ion{Si}{ii} $\lambda\lambda1191,1193$} & \multicolumn{2}{c}{[\ion{O}{iii}] $\lambda\lambda4959,5007$} \\
    {\tt OutLines} model      & \multicolumn{2}{>{\centering\arraybackslash}p{2.5cm}}{BetaCAK, PowerLaw, Sphere} & \multicolumn{2}{>{\centering\arraybackslash}p{2.5cm}}{BetaCAK, PowerLaw, Sphere, Launched} & \multicolumn{2}{>{\centering\arraybackslash}p{3cm}}{BetaCAK, DampedPulses, Sphere} & \multicolumn{2}{>{\centering\arraybackslash}p{2.5cm}}{AccPlaw, PowerLaw, HollowCones, Disk} \\

    \hline
    \hline

\multicolumn{9}{c}{Parameters from MCMC Posterior Samples}\\
\hline
$\alpha$                                & \tablepm{\ 0.23} & 0.05       & \tablepm{0.58} & 0.50       &  &       & \tablepm{0.03} & 0.06\\
$\beta$                                 & \tablepm{7} & {2}      & \tablepm{1.87} & 1.09      & \tablepm{\ 0.75 } & 0.27      & \tablepm{\ 2.02 } &   0.17  \\
$v_\infty$ [\kms]                       & \tablepm{309} & 14      & \tablepm{195} & 17      & \tablepm{567} & 42     & \tablepm{925} &  65 \\
$b$ or $\sigma_v$ [\kms]                & \tablepm{30} & 0.5     & \tablepm{5.0} & 2.4      & \tablepm{64} & 20      & \tablepm{82} & 1 \\
\hline
\multicolumn{9}{c}{Properties from MCMC Posterior Samples}\\
\hline

$W_{50}$ (``FWHF'') [\kms]                             & \tablepm{47} & 1      & \tablepm{88} & 4      & \tablepm{184} & 27      & \tablepm{235} &   28 \\
$W_{80}$ [\kms]                            & \tablepm{105} & 2      & \tablepm{150} &  6     & \tablepm{318} & 77      & \tablepm{541} &   35 \\
$x_{\rm out}$                               & \tablepm{43} & 23      & \tablepm{4.22} & $^{11.83}_{~~2.52}$       & \tablepm{1.24 } &  0.17     & \tablepm{20   } &   0.02 \\
$v_{\rm out}$                               & \tablepm{246} & 6      & \tablepm{118} & 32      & \tablepm{188} & 78      & \tablepm{883} &  46 \\
$\dot{M}/n_0R_0^2$ [\msy]                        & \tablepm{777} &  295     & \tablepm{44.5} & $^{268.0}_{~~32.7}$       & \tablepm{4.8} & 3.2      & \tablepm{721} &   321 \\
$\dot{p}/n_0R_0^2$ [$\rm 10^{34}~dyne$]          & \tablepm{120} & 47      & \tablepm{3.21} & $^{27.00}_{~~2.67}$       & \tablepm{0.57} & 0.54      & \tablepm{430} &   168 \\
$\dot{E}/n_0R_0^2$ [$\rm 10^{42}~erg~s^{-1}$]    & \tablepm{15} & 6      & \tablepm{0.19} & $^{0.16}_{2.09}$       & \tablepm{0.05} & 0.05      & \tablepm{195} &   70 \\

    \end{tabular}
\end{table*}

\subsection{Evolution of \ion{H}{ii} Regions}\label{sec:hbNGC5471}

\ion{H}{ii} regions are dynamic environments with radiation and winds from O and B stars shaping and enriching the proximal ISM. Winds launched from stars carry out recently formed metals dredged up to the surface and subsequently sweep up surrounding material. Radiation heats gas, often through photoionization, and imparts momentum to the ISM, which can result in additional launching or driving of winds, bubbles, and outflowing gas \citep[e.g.,][]{Castor1975b,Weaver1977,Dyson1977,Dyson1979,Oey1997,Dopita2005}. While stellar population models often contain predictions for these feedback effects \citep[e.g.,][]{1999ApJS..123....3L,starburst99,Eldrige2017,Byrne2022,Hawcroft2025}, empirically constraining the kinematics and related mechanical feedback in \ion{H}{ii} regions remains an active area of research \citep[e.g., models and simulations by][]{Geen2020,Geen2023,Lancaster2025a,Lancaster2025b} with implications for understanding chemical enrichment \citep[e.g.,][]{Esteban1995,LopezSanchez2007,Stock2011} and dispersal timescales \citep[e.g.,][]{Dopita2005,Menon2024}.

\begin{figure*}
    \centering
    \includegraphics[width=0.9\linewidth,clip=True,trim={0in 0.2in 0in 0in}]{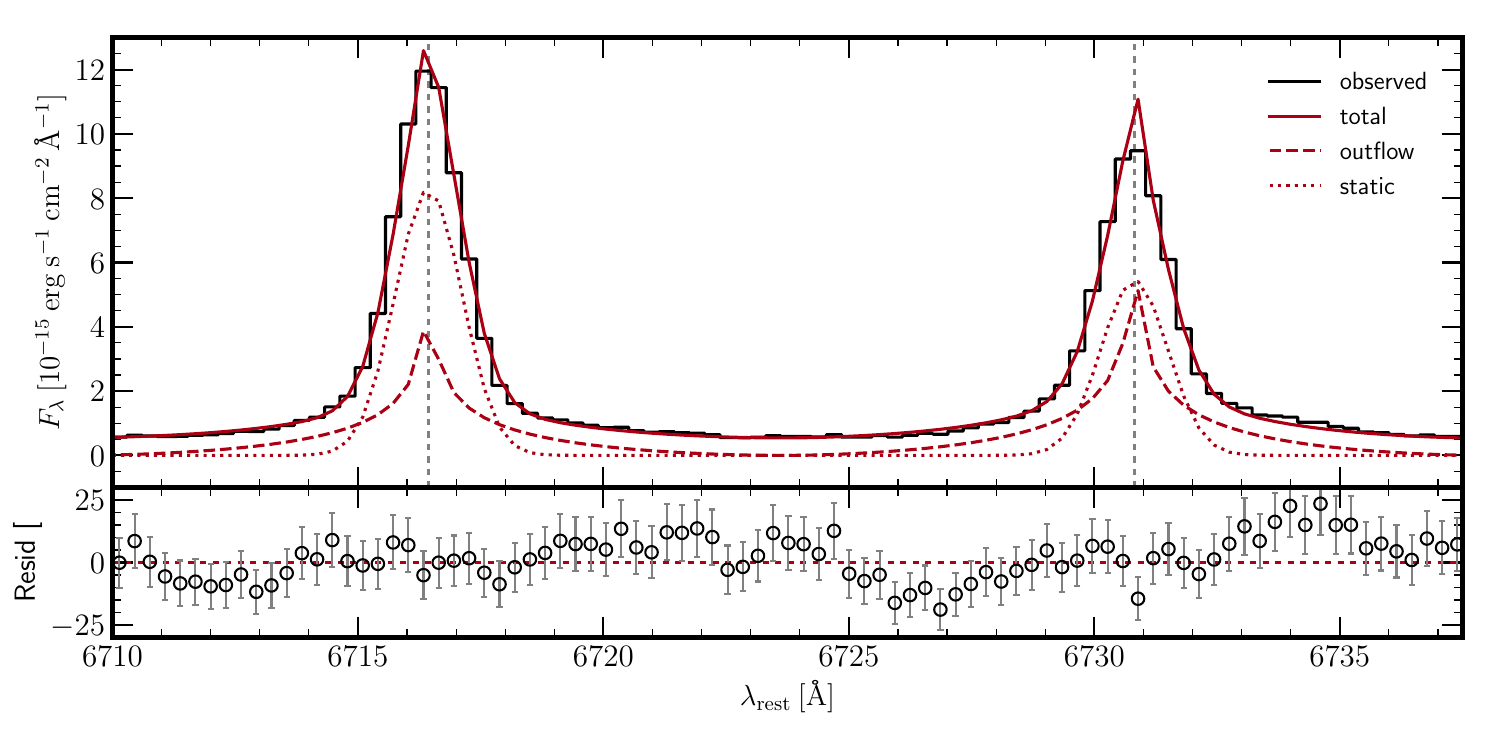}
    \caption{{\it Top}: [\ion{S}{ii}] $\lambda\lambda6716,6731$ doublet in Knot A of NGC 5471. Grey dashed vertical line indicates the line center given by atomic data. The best-fit {\tt OutLines.Nebular} model is shown in red with the outflow (dashes), static (dotted), and total (solid) line profiles. {\it Bottom}: residual difference between the best-fit {\tt OutLines} profile and the observed spectrum.}
    \label{fig:NGC5471_SII}
\end{figure*}

One such kinematically rich and complex system is the giant extragalactic \ion{H}{ii} region NGC 5471 located in M101. Many knots within this region exhibit broad and sometimes asymmetric nebular lines which cannot be described by a single Gaussian \citep[e.g.,][]{Chu1986,Bresolin2020}. 
To investigate the evolution of \ion{H}{ii} regions in this case-study, we apply {\tt OutLines} to the 
[\ion{S}{ii}] $\lambda\lambda6716,6731$ doublet, which traces the density and kinematics of low ionization gas,
in Knot A
of NGC 5471, the brightest knot identified by \citet{Skillman1985}, as observed with the $R=6000$ MEGARA IFU spectrograph on the Gran Telescopia Canarias (Arellano-Cordova in prep, PI Esteban). For our modeling, we use a 2.5$^{\prime\prime}$ aperture extraction of the Knot A spectrum from the IFU cube. We obtain atomic data from {\tt NIST} \citep{NIST_ASD} for wavelengths \citep{Kaufman1993} and spontaneous emission coefficients (transition probabilities) \citep{Podobedova2009}. We use collision strengths from \citet{Tayal2010}. First, we fit the adjacent featureless continuum with a third-order polynomial, an appropriate approximation in the case of extended \ion{H}{ii} regions which do not suffer appreciably from contamination such as stellar absorption except near permitted lines from, e.g., \ion{H}{i} and \ion{He}{i}. We subtract this continuum before modeling the line profile with {\tt OutLines}. { For our tailored {\tt OutLines} model, we adopt the following: a spherically symmetric geometry given the relatively symmetric line profile, a CAK $\beta$ velocity field, and a wind profile since the broad profile component exhibits wings rather than a box-like edge. To obtain the best-fit profile, we fit both lines simultaneously to maximize statistical constraints on the parameters.} Then, we sample the posterior on the {\tt OutLines} profile parameters to constrain the outflow and static gas properties. No LSF corrections are necessary as $R=6000$ is sufficiently high to constrain the kinematics (see \S\ref{sec:modeling}). { To verify that the best-fit fluxes are a global rather than a local solution, we rerun our fit with 11 different seeded flux ratios assuming that the wind and static components have the same density and find that the best-fit parameters do not change.} We compare the best-fit model to the observed line profiles in Figure \ref{fig:NGC5471_SII}.

From the best-fit line fluxes, we determine the electron density of both the nebula and outflow components from the [\ion{S}{ii}] ${6716}/{6731}$ flux ratio using {\tt PyNeb} \citep{2015A&A...573A..42L}. The static gas flux ratio of 1.459$\pm0.165$ is in the low-density limit, requiring $n_e<10\rm\ cm^{-3}$. 
For the outflowing gas component, the doublet ratio is 0.831$\pm0.070$, which is where [\ion{S}{ii}] is most sensitive simultaneously to density and to temperature. For a range in $T_e$ from 10 to 15 $\rm kK$, the inferred outflow density is anywhere from $970$ to $1140\rm~cm^{-3}$, with density values increasing with the assumed temperature. The implication from these densities is that the outflow is roughly two orders of magnitude more dense than the gas within the static nebular gas. While this increase in density may be due to a sweeping up of material in the wind, this dramatic change in density coupled with the supersonic flow implied by $v_{\rm out}$ could instead arise from compression by shocked gas as it rapidly cools via radiation \citep[e.g.,][]{Flury2025b}.

Though the posterior on $\beta$ is not well constrained, the slope of the velocity field strongly favors $\beta>4$, indicating gradual acceleration likely caused by optically thick radiation pressure from early type stars \citep[e.g.,][]{1977ApJ...213..737B}. While \citet{Bresolin2020} argue that the non-thermal radio SEDs of \citet{Sramek1986} are clear-cut evidence for supernova-driven bubbles, the line profile shape indicates a continuous wind rather than shells or bubbles { due to the presence of wings rather than a box-like flux density distribution} and, from modeling with {\tt OutLines}, disfavors the supernova scenario. Indeed, the {\tt OutLines} results are very much in agreement with the fact that Knot A features the a very young ($<3\rm~Myr$) cluster--in fact the youngest in the entirety of NGC 5471 \citep[e.g.,][based on HST photometric color-magnitude diagrams]{GarciaBenito2011}. Knot A is also very massive ($>10^4\rm~M_\odot$) and requires a significant population of O stars to produce the $Q=4\times10^{51}\rm~phot~s^{-1}$ associated with the star cluster \citep{GarciaBenito2011}. The age and mass of the stellar population indicate sufficient radiation available in the UV to radiatively drive a supersonic wind. Shocks triggered by this wind could explain the non-thermal emission in the radio in addition to the dense gas. Inhomogeneities in the gas distribution can readily dissipate the outward force due to pressure from the hot shocked gas \citep[e.g.,][]{Dyson1977}, allowing radiation pressure to dominate the acceleration of material out from the \ion{H}{ii} region. Our results indicating escape of hot gas are consistent with the \citet{HarperClark2009} model for the Carina Nebula and the \citet{Lopez2011} multiwavelength study of the massive extragalactic \ion{H}{ii} region 30 Doradus, both of which found strong evidence for radiation pressure driving and a deficit of X-ray flux associated with shocked gas.

With the velocity field and line fluxes inferred by modeling nebular lines with {\tt OutLines}, we have obtained key insight into the nature of the outflowing material: (i) that it is a wind, not a bubble, (ii) that the wind exhibits gas densities far higher than that of the static nebula, and (iii) that the wind is driven by substantial radiation pressure from O stars. These results are consistent with the observed stellar population ages. Moreover, this wind is supersonic, which could explain the non-thermal emission and high gas densities via shocks.

\subsection{Lyman Continuum Escape from a Super Star Cluster}\label{sec:c2J1044}

Starburst galaxy J1044+0353 is a low mass ($M_\star\approx10^{6.8}\rm~M_\odot$) Lyman continuum emitter (LCE) candidate  with $f_{esc}^{\rm LyC}$ as high as 60\% \citep{Parker2025} and, as a nearby ($z=0.01287$) Green Pea, is a local analog for star-forming galaxies at cosmic dawn that are actively involved in re-ionizing the Universe \citep{Martin2024}. Recent IFU observations of J1044+0353 reveal expanding shells of gas at the periphery of the nebular gas flux distribution \citep{Martin2024}, indicative of blowout due to SNe from an aging stellar population 10-20 Myr after a burst of star formation { in one region of the galaxy} \citep{Peng2023}. { An adjacent star-forming region} of the galaxy hosts a super star cluster (SSC) with several young ($<2$ Myr) stellar populations, however, which may be responsible for driving additional feedback on shorter timescales \citep{Olivier2022}. { \citet{Martin2024} argue that the stellar population ages, combined with kinematics, demonstrates how bursty star formation can drive Lyman continuum escape.} As such, J1044+0353 provides a unique laboratory to explore the role of feedback on LyC escape using {\tt OutLines}.

To undertake this investigation, we draw on archival \emph{HST}/COS G130M FUV spectroscopy ($R=10\ 000$) of the central starburst of J1044+0353 from the COS LegAcy Spectroscopic SurveY \citep[CLASSY,][GO 15840]{CLASSY,CLASSY2}, which allows for detailed modeling of outflow properties along the line of sight without significant effects from the LSF even after downsampling to 0.073 {\AA} resolution \citep[see][]{CLASSY,CLASSY2,Parker2024}. The \ion{C}{ii} $\lambda1334$ absorption line is one of the strongest low ionization lines in the FUV due to the transition oscillator strength and comparatively high ionic abundance, making it an excellent tracer of outflows. Visual inspection of the \ion{C}{ii} line profile suggests a classic blue-shifted continuous outflow plus a static ISM component \citep[see, e.g., Figure 1 of][]{2015ApJ...809..147H}. To obtain the normalized flux, we fit the starlight continuum within 3 {\AA} of the feature with a third-order polynomial and divide the local spectrum by this fit \citep[see, e.g., the {\tt VoigtFit} software,][which takes a similar approach using Chebyshev polynomials]{VoigtFit}.

For modeling the \ion{C}{ii} line with {\tt OutLines}, we use atomic data from \citet{Morton2003}. No infilling is apparent in the observed line profile, suggesting either maximal aperture clipping where pure absorption is an appropriate description of the emergent feature (see Figure \ref{fig:aper_profiles}) or an optically thick wind. For simplicity, we assume the {\tt OutLines} default geometry and density profile, although we note that applying more sophisticated geometry or density profiles negligibly impacts the quality of the fit. Indeed, a filled cone model yields best-fit inclinations of $i\to0^\circ$ and opening angles of $\theta_o\to90^\circ$, consistent with the filled sphere geometry. Additional consideration of pulse-like density profiles consistently favors a peak location at the base of the outflow with wide dispersion, which combined are consistent with the power law approximation. { Thus, while we have considered the possibilities of bubbles or directed outflows, we must conclude that the line profile is consistent with an isotropic wind.}

{ In their analysis of spatially resolved optical emission line flux maps (as in their Figure 9), \citet{Martin2024} argue that the UV absorption lines must coincide with anisotropic breakout of bubbles rather than the wind interpretation asserted by \citet{Xu2022}. However, the break-out bubbles traced by nebular maps are not spatially coincident with the young stellar populations which backlight the \ion{C}{ii} absorption feature. Moreover, the shape of the broad [\ion{O}{iii}] line profile associated with the SSCs in J1044+0353 \citep[][again, their Figure 9]{Martin2024} can be accounted for with a wind-like density distribution identical to that of the \ion{C}{ii} absorption line. While the $R\sim3600$ resolution of KCWI cannot distinguish between a bubble and a wind, the velocity width of a bubble necessary to account for the convolved [\ion{O}{iii}] line profile would be only 100 \kms, which, while consistent with estimates by \citet{Martin2024}, is only half of the terminal velocity indicated by the UV absorption line and would not account for the lower or higher velocity line absorption even after accounting for the COS instrumental resolution. As such, we must conclude that the neutral and ionized gas flows in J1044+0353 likely correspond to a spatially coincident wind rather than a bubble or anisotropic breakout.} 

The sudden drop in flux immediately blueward of the systemic velocity of the system suggests a non-zero launch velocity. We therefore allow for non-zero launch velocities. The {\tt OutLines} default top hat (uniform-bounded) prior is assumed.
Best-fit parameters are in Table \ref{tab:examp_fits}. While the COS G130M grating resolution is in the LSF-insensitive regime, COS LSFs contain prominent broad non-Gaussian wings which necessititate folding in LSF corrections. As such, we convolve the {\tt OutLines} profile with the appropriate COS G130M LSF \citep[see \hyperlink{STScI}{https://www.stsci.edu/hst/instrumentation/cos/performance/spectral-resolution} and method in Appendix B of][]{Flury2025a} before comparing it to the observations in our fit. We find that the results from our initial fit are comparable to those from the fit containing the LSF correction, with each set of parameters well within the 1$\sigma$ confidences of both posteriors. However, we proceed with the LSF-corrected result for robustness.

\begin{figure}
    \centering
    \includegraphics[width=\linewidth]{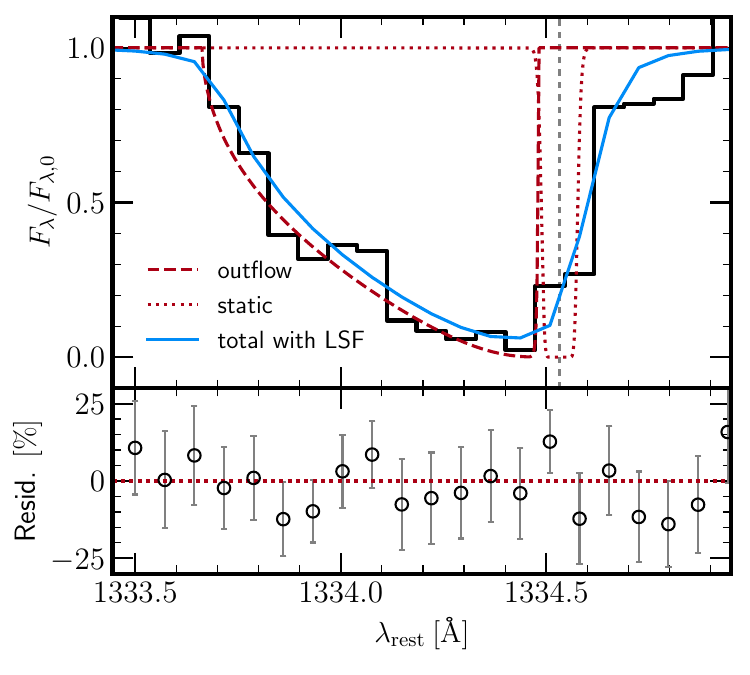}
    \caption{MCMC fit of an {\tt OutLines.Absorption} model to the \ion{C}{ii} $\lambda1334$ absorption line observed by \emph{HST}/COS G130M (black, \citealt{CLASSY}). Line styles and colors as in Figure \ref{fig:NGC5471_SII}. Blue line indicates the intrinsic profile convolved with the \emph{HST}/COS G130M line spread function (LSF).}
    \label{fig:J1044+0353_CII}
\end{figure}

{ The best-fit parameters indicate that the starbursting core of J1044+0353 contains a spherical expanding wind with a velocity field characteristic of radiatively-driven winds most likely due to dusty radiation pressure \citep[e.g.,][]{Flury2023}, although line-driving \citep[][]{1975ApJ...195..157C} and optically thick \citep{1977ApJ...213..737B} mechanisms are consistent with the range of $\beta$. Critically, the posterior on $\beta$ strongly disfavors ram pressure driving, suggesting that supernovae do not accelerate the observed wind.} However, the relatively low wind velocity ($v_\infty=195$ \kms, comparable to Voigt modeling of the silicon lines, \citealt{Parker2024}, and the ``broad'' component of nebular lines, \citealt{Peng2023},\citealt{Martin2024}) and small values of $\dot{p}$ and $\dot{E}$ (marginally non-zero) suggests that mechanical feedback from massive stars is relatively weak or even suppressed. { The similar kinematics across many ion species (\ion{C}{ii} and [\ion{O}{iii}] in this study, but also \ion{Si}{ii}, \ion{Si}{iii}, \ion{Si}{iv} in \citealt{Xu2022}) which in turn trace different ISM phases (\ion{C}{ii} and [\ion{O}{iii}] trace $10^4\rm\:K$ and $10^5\rm\:K$ gas, respectively, e.g., \citealt{Gray2019}) further points to weak feedback since radiative suppression can link dynamics across a wide range of ISM phases \citep{Silich2020}. UV absorption lines of several Si species in samples of Green Pea galaxies indicate that similar wind suppression occurs ubiquitously in this population of galaxies \citep{2017ApJ...851L...9J}, suggesting that J1044+0353 is representative rather than unique.}

{ Wind suppression can occur in cases where strong or catastrophic radiative cooling of swept up dense material prevents acceleration to more ``traditional'' velocities closer to 500 \kms\ and subsequent breakout \citep{Silich2004,Silich2018,Silich2020}. Radiative cooling can even exacerbate the Rayleigh-Taylor breakup of the warm/cool wind front by increasing gas clumping \citep[e.g.,][]{Scannapieco2017}, clearing out escape channels for hot gas (as in NGC 5471-A and 30 Doradus, see discussion and references in \S\ref{sec:hbNGC5471}) and perhaps even for LyC photons \citep{Jaskot2019}. { Similar behavior is observed in the Green Pea analog \ion{H}{ii} region Mrk 71, where catastrophic cooling inhibits energy injection, leading to radiation pressure driving of the observed superwind \citep{Oey2017,2021ApJ...920L..46K}.}

The dominant radiative driving implied by shape of the velocity field further suggests catastrophic cooling for one or both of two possible mechanisms responsible for loss of ram pressure. First, hot gas associated with the shocked interior of the SSC wind could escape due to radiatively augmented clumping, which serves as an opening ``valve'' releasing the ram pressure and allowing radiation pressure to take over \citep{MartinezGonzalez2014}. Power law emission line profiles have been attributed to wind escape through a clumpy medium in past work \citep[e.g.,][]{2017MNRAS.471.4061K,2021ApJ...920L..46K} and appears to be accompanied by LyC escape in Green Peas where ionizing photons behave as a critical radiative accelerant \citep{Komarova2025}. Second, catastrophic cooling can suppress the formation of the thermalized hot shocked gas entirely \citep[as seen in SSC wind models by][for low metallicities and terminal velocities of 250 \kms]{Gray2019,Danehkar2022}, leaving radiation pressure alone to drive the wind. In either case, catastrophic cooling naturally facilitates radiatively driven superwinds by releasing or suppressing ram pressure.} 
Combined with evidence for recent SNe \citep{Peng2023}, the suppressed, radiatively-driven winds inferred from the \ion{C}{ii} line profile using {\tt OutLines} supports the episodic star formation scenario for LyC escape proposed by \citet{Flury2022b} in their assessment of a large sample of star-forming galaxies with LyC detections and later demonstrated by \citet{Flury2025a} with measurement of stellar populations and ISM conditions in stacks of UV spectra of LCEs. In the \citet{Flury2022b,Flury2025a} framework, SNe clear out channels in the ISM via mechanical feedback and a subsequent generation of stars later provides ionizing--but suppressed mechanical--feedback to establish the density-bounded conditions necessary for Lyman continuum photons to escape into the IGM. { However, localized feedback within the young SSCs may still play a pivotal role in linking the LyC production sites with the channels cleared by intense mechanical feedback from SNe.}

{ \citet{Martin2024} present a similar framework for feedback in J1044+0353, identifying anisotropic Lyman continuum escape pathways traced by O$_{32}=$[\ion{O}{iii}]$\lambda5007$/[\ion{O}{ii}]$\lambda\lambda3726,29$ near the SNe-driven blowout beyond the extent of the COS aperture. From their spatially resolved kinematic and flux ratio maps, they claim that the ongoing feedback from the youngest SSCs cannot account for Lyman continuum escape. Their interpretation arises from fundamentally flawed analysis, though. First, they conflate bubbles apparent in \emph{HST} images with wind signatures in the line profiles, which is demonstrably not the case through careful examination of the UV absorption lines. Furthermore, in their assessment of suppressed feedback, they assume that [\ion{O}{iii}] traces $10^4\rm\:K$ gas when the emissivity is well known to peak at $10^5\rm\:K$ \citep[e.g.,][]{Sutherland1993,Gray2019} and, like H$\alpha$, can probe deeper into the hot phase in suppressed winds than in more extreme scenarios where both \ion{H}{i} recombination and [\ion{O}{iii}] are confined to a cooler shell \citep{Danehkar2022}. The consistency of wind kinematics between the \ion{C}{ii} absorption and [\ion{O}{iii}] emission lines, along with the likelihood of [\ion{O}{iii}] tracing the hot gas phase vs \ion{C}{ii} tracing the warm gas phase, is a key result of our own analysis which distinguishes our work from theirs. Nevertheless, they do arrive at a similar conclusion : feedback within the first 2-3 Myrs alone cannot clear LyC escape pathways to the IGM.} 

In our analysis with {\tt OutLines}, we independently conclude that feedback from the youngest stellar populations is likely too weak to drive the escape of ionizing radiation. However, with {\tt OutLines}, we have uniquely identified the driving mechanisms and geometry associated with the feedback in proximity to the SSC : efficient radiative wind cooling and gas clumping facilitate hot gas escape, resulting in isotropic acceleration via radiation pressure and LyC escape. Previous approaches such as \citet{Martin2024} miss these critical details which are fundamental to understanding feedback in young, low-metallicity SSCs and why SNe are necessary to facilitate LyC escape.
Here, inferring underlying physics and geometry of the SSC wind from a single-aperture spectrum demonstrates the power and effectiveness of {\tt OutLines} for exploring the underlying physics of feedback in extreme ionization, low metallicity systems like the Green Peas, particularly highlighting radiative suppression of winds and the need for mechanical feedback from older stellar populations to facilitate LyC escape.

\subsection{Outflow Formation and Blowout in a Starburst Galaxy}\label{sec:Mrk1486}

Mrk 1486 is a starburst galaxy exhibiting a tentative \ion{He}{ii} $\lambda4686$ Wolf-Rayet feature \citep[e.g.,][]{Izotov1997}, metal rich extraplanar biconical outflows beyond the galactic plane \citep[e.g.,][]{Duval2016,Cameron2021}, and $>14$ Myr stellar populations (\citealt{Parker2025}, although see \citealt{Peng2025}, who find from the same UV spectra that these populations are even older, $\approx100$ Myr). The archival \emph{HST}/COS spectrum from \citet[][GO 12583, PI Hayes]{RiveraThorsen2015} exhibits the classic blue-shifted asymmetric feature associated with outflows, albeit with at least two troughs present in the outflow feature across several low ionization lines \citep[\ion{Si}{ii} $\lambda\lambda1191,1193,1304$, \ion{O}{i} $\lambda1302$, and \ion{Al}{ii} $\lambda1670$, see \citealt{RiveraThorsen2015} their Figure 3,][their Figure 7]{Parker2024}. A multi-troughed absorption feature can indicate a ensemble of shells produced by a series of episodes or a system of many star clusters (see Figure \ref{fig:pulse_profiles}, also \citealt{RiveraThorsen2015}). 

For our analysis, we use the \citet{RiveraThorsen2015} spectrum reprocessed by \citet{CLASSY,CLASSY2}, down-sampled to 0.073 {\AA} resolution and normalized following the polynomial approach in \S\ref{sec:c2J1044}. 
To determine the origin of this multi-episode outflow, we can use {\tt OutLines} to model the absorption lines. The \ion{Si}{ii} $\lambda\lambda1191,1193$ doublet provides an excellent probe as both features arise from the same ion, thus providing additional constraining power. The stronger \ion{Si}{ii} $\lambda1260$ line is contaminated by telluric \ion{O}{i} $\lambda1302$ and cannot be used to constrain the fit. We note the presence of weak \ion{Si}{ii} $\lambda1194,1197$ fluorescent emission arising from the re-emission of \ion{Si}{ii} $\lambda\lambda1191,1193$ photon absorption, respectively. These lines suffer from apparent Milky Way contamination in the blue wings due to absorption by Galactic \ion{O}{iv} $\lambda\lambda1234,1237$, \ion{Si}{iii} $\lambda1235$, and \ion{Mn}{ii} $\lambda\lambda1234,1236$. We mask these features in order to include the fluorescent lines in our fit.
Using atomic data from \citet{Kelleher2008} obtained from NIST \citep{NIST_ASD} and the appropriate COS G130M LSF, we model the \ion{Si}{ii} doublet complex of $\lambda\lambda1191,1193$ resonant absorption and emission and $\lambda\lambda1194,1197$ fluorescent emission using {\tt OutLines} assuming a damped-pulses density profile to account for the multiple absorption troughs and otherwise adopting the default options of a spherical geometry and the $\beta$ CAK velocity field. 
The choice of spherical geometry is supported by the presence of relatively symmetric fluorescent \ion{Si}{ii} $\lambda1194,1197$ lines, although the COS LSF may complicate any inference of the geometry. As in \S\ref{sec:c2J1044}, we incorporate the appropriate LSF corrections.

The two \ion{Si}{ii} absorption lines have similar depths. Given the oscillator strengths of the two transitions, the \ion{Si}{ii} $\lambda1193$ line should be deeper than \ion{Si}{ii} $\lambda1191$. For the lines to have similar depths, the \ion{Si}{ii} $\lambda1193$ line must be saturated. Whether the \ion{Si}{ii} $\lambda1191$ line is saturated is not ascertainable solely from the relative line depths; however, the prominent $\lambda\lambda1194,1197$ lines suggest optically thick lines consistent with trapping of the resonant emission, with many scattering events allowing photon escape solely through fluorescent channels, i.e., no resonant emission. Because $\lambda1193$ is saturated and has a depth similar to that of $\lambda1193$ (minimal infilling by the $\lambda1194$ line), any transmitted flux must be due to starlight within the spectroscopic aperture that is not associated with the foreground gas. Traditionally, such residual flux is explained by a partial gas covering fraction, with the radiative transfer rewritten as
\begin{equation}
    \frac{F_\lambda}{F_{\lambda0}} = 1 - C_f\left[1-\exp(-\tau_0\phi_\lambda)\right]
\end{equation}
where $C_f$ expresses the fraction of background light which is not obstructed by the foreground gas \citep[e.g.,][]{Netzer1982,Arav1999,Arav2005,RiveraThorsen2015}. The COS LSF can affect the partial covering fractions of saturated absorption lines \citep[see, e.g., assessment by][their Appendix D, also Figure 3 of \citealt{Jennings2025}]{Flury2025a}. Incorporating both $C_f$ and $\epsilon$ as free parameters into the initial fit, we find that while the COS LSF and infilling by the $1194$ fluorescent line can account for some of the line profile depths, $C_f=0.82\pm0.03$ is necessary to provide an optimal description to the absorption features \citep[comparable to the $C_f=0.8$ obtained by][]{RiveraThorsen2015}. We also find $\epsilon=0.21\pm0.03$ necessary to describe the fluorescent line profile.

\begin{figure*}
    \centering
    \includegraphics[width=0.65\linewidth]{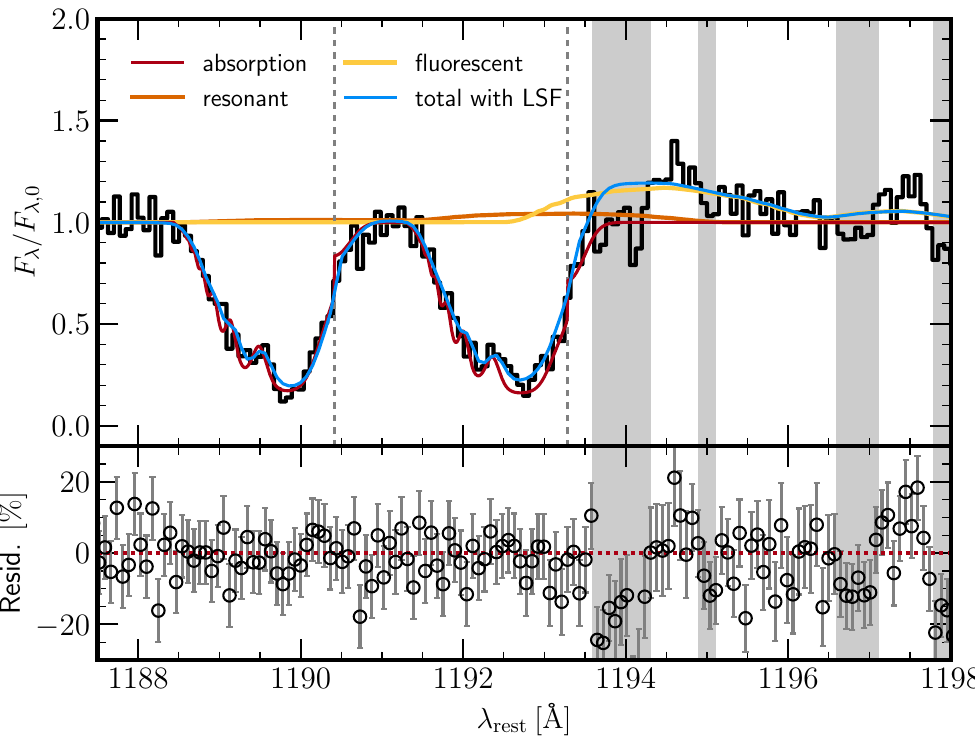}
    \caption{MCMC fit of an {\tt OutLines} model to the \ion{Si}{ii} $\lambda\lambda1191,1193,1194,1997$ doublet complex observed by \emph{HST}/COS G130M (black, \citealt{RiveraThorsen2015}, GO 12583, PI Hayes). Shaded regions indicate suspected Milkyway contamination from Galactic \ion{Si}{iii}, \ion{Mn}{ii}, and and \ion{O}{iv}. \ion{Si}{ii} $\lambda\lambda1191,1193$ resonant emission lines computed using {\tt OutLines.Resonant} are shown in orange, demonstrating strong radiative trapping. \ion{Si}{ii} $\lambda\lambda1194,1197$ fluorescent lines computed using {\tt OutLines.Fluorescent} are shown in yellow. Total line profile corrected for the line spread function (LSF) is shown in blue. Other line styles, symbols, and colors as in Figure \ref{fig:NGC5471_SII}.}
    \label{fig:mrk1486_si2}
\end{figure*}

We show the best-fit line profile in Figure \ref{fig:mrk1486_si2} and list the results in Table \ref{tab:examp_fits}. As a test of our assumption of spherical geometry, we predict the emergent fluorescent \ion{Si}{ii} $\lambda\lambda1194,1197$ lines using {\tt OutLines.Fluorescent} and the best-fit parameters from the posterior, which we show in Figure \ref{fig:mrk1486_si2}.

The posterior sampling indicates relatively small uncertainties on the order of 10\% for each of the expanding shells, with the most prominent and lowest velocity shell having $v_{\rm shell}=188\pm20$ \kms\ (obtained from $v_{\rm shell}=v_\infty w[x_1]$ and propagating the uncertainties in $x_1$, $\beta$, and $v_\infty$). The system of shells suggest an outflow in formation, with regular events forming thin expanding superbubbles. Notably, the less pronounced absorption troughs in \ion{C}{ii} $\lambda1334$, which traces a higher ionization state than \ion{Si}{ii}, \ion{O}{i}, and \ion{Al}{ii}, suggests that the ionized phase of the outflow is established more rapidly than the neutral phase and may indicate that the expanding bubbles are entrained in or even driven by a hotter phase of the ISM. 
The accumulation of superbubbles close to the terminal velocity suggests that an outflow forms from these successive events with sufficient speed to exceed the escape velocity  ($v_{esc}\approx$100 \kms as implied by the $M_\star=10^{9.3}\rm~M_\odot$ from \citealt{2018MNRAS.481.1690C}). The recurrance of many such expanding shells could produce the biconical outflow observed in emission once the superbubbles escape the gravitational potential and blow out from the galactic disk \citep[see spatially resolved cases in, e.g.,][and \citealt{Duval2016} specifically regarding Mrk 1486]{Martin1998,McQuinn2019}. Moreover, the successive events are removing gas at mass outflow rates of $\dot{M}=4.8\rm~M_\odot~yr^{-1}$, which may temporarily rob the galaxy of baryonic material necessary to sustain the production of stars and could explain the 30-40 Myr lapse between bursts of star formation.

Given that the outflows in Mrk 1486 are oxygen rich relative to their host galaxy \citep{Cameron2021}, it is useful to consider whether the outflow component of \ion{Si}{ii} is metal-enhanced since, like oxygen, silicon is an $\alpha$ element and is predominantly produced by core-collapse supernovae \citep[CCSNe, e.g.,][]{Kobayashi2020}.
From the best-fit line profiles from {\tt OutLines}, the outflow $N_{o\ \rm Si^+}=10^{15.5}\rm~cm^{-2}$ while the static $N_{s\ \rm Si^+}=10^{13.5}\rm~cm^{-2}$. Ionization structure may account for some of the difference in Si$^+$; nevertheless, the 2 dex increase in the outflow likely suggests substantial metal loading of the outflow \citep[see, e.g.,][]{2018MNRAS.481.1690C} due to enrichment by recent CCSNe.

It may be tempting to conclude that hot gas or cosmic rays from these supernovae are driving the shells' expansion, especially since, for the lowest velocity shell at 188 \kms, the dynamical time to traverse the inner 500 pc is $\la2.5$ Myr, well after the onset of CCSNe for a burst at least 10 Myr old. That being said, the age reported by \citet{Parker2025} from fits to the UV continuum that Mrk 1486 corresponds to a combination of a substantial young ($\leq5$ Myr) stellar population and a far older (40 Myr) population. In light of the more detailed star formation history, while CCSNe may be the progenitors of the material in these expanding shells, they may not necessarily be responsible for their acceleration. Here, the shape of the velocity field from {\tt OutLines} can provide insight into the driving mechanism. Indeed, the velocity index of $\beta=0.7\pm0.2$ disfavors (but cannot fully not reject) the rapid acceleration associated with ram pressure driving \citep[$\beta\la0.5$, e.g.,][]{2005ApJ...618..569M,2016MNRAS.463..541C}. However, these $\alpha$-rich shells are more likely continuously accelerated by optically thin line-driven radiation pressure associated with $\beta=0.8$ \citep[e.g.,][]{1986A&A...164...86P}. Moreover, the total energy injection rate of $10^{40}\rm~erg~s^{-1}$ for the series of shells is roughly consistent with stellar winds, which have $L_{mech}\approx10^{39}\rm~erg~s^{-1}$ for a given stellar population \citep[see, e.g., {\tt Starburst99} or {\tt BPASS},][respectively]{starburst99,Byrne2022}. 
Radiative driving by young stellar populations may also account for the less pronounced \ion{C}{ii} troughs via ionizing feedback. Furthermore, given that the travel time { from the nucleus to a 1 kpc radius (roughly the COS aperture at the 139.1 Mpc distance of Mrk 1486)} is $\la5$ Myr, the $>40$ Myr stellar populations are unlikely to be responsible for these superbubbles. Corroborating evidence for optically thick radiation pressure is the presence of the fluorescent $\lambda1194,1197$ lines but the absence of resonant line infilling. Such a scenario occurs in the optically thick line limit where photon escape primarily occurs through the fluorescent channel \citep[e.g.,][their Figures 5-6]{2015ApJ...801...43S}.

The bursy star formation in Mrk 1486 could optimize the production of $\alpha$ elements and energetic expanding shells via CCSNe from earlier generations of stars while the most recent stellar populations provide the momentum injection via radiation pressure to form and accelerate optically thick enriched superbubbles, driving them beyond the galactic disk to form an outflow. {\tt OutLines} is necessary here to infer the velocities of the shells as they accelerate under a collective driving mechanism, determine the relative importance of SNe and young stars in producing and accelerating these shells, and whether the shells are capable of eventually exceeding the escape velocity to form an outflow.

\subsection{Thermal Winds in AGN Radio Lobes}\label{sec:o3NGC2992}

The Seyfert 2 galaxy NGC 2992 has long been known to exhibit conical outflows \citep{Allen1999,Veilleux2001}. Using archival observations from the Siding Spring Seyfert Spectroscopic Snapshot Survey \citep[S7,][]{S7DR2} made using WiFES \citep{WiFES}, we have confirmed broad, asymmetric [\ion{O}{iii}] $\lambda\lambda4959,5007$ lines after summing over all the spaxels to boost signal to noise in the line profile and to mitigate any aperture effects. The low resolution ($R\approx$ 3 000 or FWHM of 105 \kms) requires corrections for the LSF by convolving with a Gaussian kernel with $\sigma_{\rm LSF}=$45 \kms; however, we note that the LSF predominantly impacts the narrow Gaussian core. The high signal in the wings ($S/N\approx3$ in the blue broad extent of [\ion{O}{iii}] $\lambda4959$) might allow for tight constraints on the wind properties even without LSF corrections -- indeed fitting with and without LSF corrections yields similar results for the {\tt OutLines} parameters. Nevertheless, incorporating the LSF into the forward modeling with {\tt OutLines} is essential to remain consistent with the observations. As a part of the LSF correction, we fold in the summed best-fit stellar continuum included in S7 to account for stellar contributions to the continuum, including any contamination of the line profile.

\begin{figure*}
    \centering
    \includegraphics[width=0.9\linewidth]{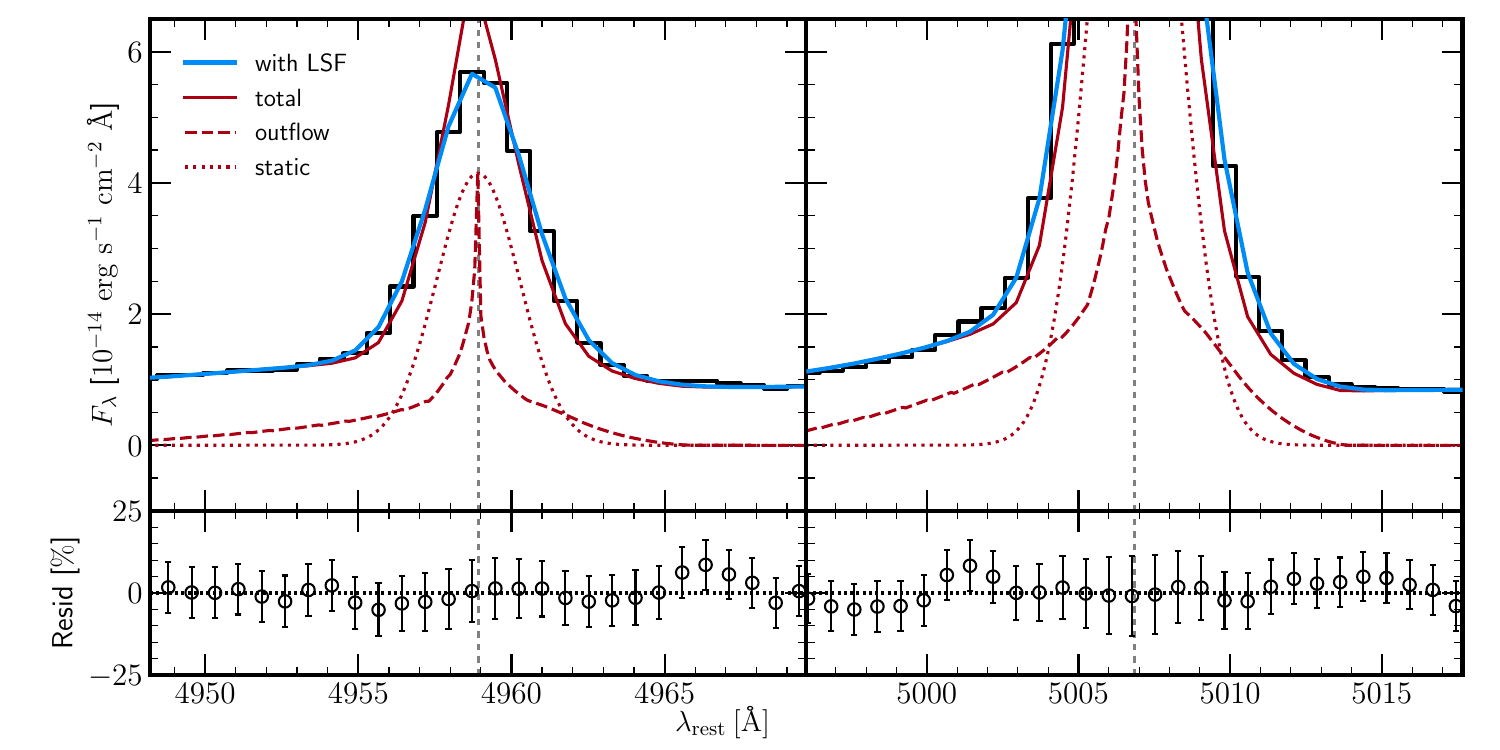}
    \caption{MCMC fit of an {\tt OutLines.Nebular} model to the [\ion{O}{iii}] $\lambda\lambda4959,5007$ doublet observed in the nucleus of Seyfert galaxy NGC 2992 as part of the Siding Spring Seyfert Spectroscopic Snapshot Survey \citep[S7,][]{S7DR2} DR2, assuming a hollow cone plus disk outflow geometry. Line styles and colors as in Figure \ref{fig:NGC5471_SII}. Blue line indicates the profile after LSF correction.}
    \label{fig:NGC2992_OIII}
\end{figure*}

The bright intermediate portion of the wings suggests the acceleration power law for the velocity field. Visual assessment of the highly asymmetric line profile in the spectrum of NGC 2992 and comparison with models in Figures \ref{fig:filled_cones}, \ref{fig:hollow_cones}, and \ref{fig:cone_disk_profiles} also suggests a hollow cone geometry with an obstructing disk. As such, we subsequently model the [\ion{O}{iii}] doublet using {\tt OutLines} assuming an acceleration power law, an open cone geometry, and an obstructing disk. We impose the following into the prior: 
line fluxes within 5\% of the [\ion{O}{iii}] 5007/4959 flux ratio of 2.98 given by atomic data \citep{Storey2000} and $\theta_o-\theta_c>5^\circ$.
With line profiles predicted by {\tt OutLines} added to the continuum and subsequently convolved with the WiFES LSF \citep{WiFES}, we use {\tt emcee} to sample the posterior on the outflow parameters. Due to the large number of parameters, potential degeneracies, and sensitivity to initial conditions, we run the sampler four times longer, for a total of 40 000 samples post burn-in. Implementation of our modeling is demonstrated for reference in Appendix \ref{apx:examples} with the best-fit profile 
shown in Figure \ref{fig:NGC2992_OIII} and results listed in Table \ref{tab:examp_fits}. We note that the posterior sampling ultimately rejects the priors on the angles as the best-guess parameters fell within a local minimum.

From the posterior on the profile parameters, we find terminal velocities in excess of $1\ 000\rm~km~s^{-1}$, indicative of a strong outflow. While this value is far greater than those previously reported for this object, the full width at half flux (FWHF) given by the 25-75 percentiles is  km s$^{-1}$, consistent with the flux-weighted average found by \citet{Veilleux2001}. Additionally, \citet{Allen1999} suggest that velocities closer to 500 \kms\ are necessary to account for possible shock excitation of gas in the outflow. The $v_{\rm out} = 596\rm~km~s^{-1}$ from the best-fit {\tt OutLines} profile could readily provide the conditions necessary for such shock excitation.

The slope $\beta=2.02$ indicates, for the acceleration power law velocity field, 
rapid acceleration to the terminal velocity (see Figure \ref{fig:velocity_profiles}),
which can be interpreted as thermal (ram pressure) or cosmic ray (magnetic pressure)
driving as each can produce similar rapid-acceleration velocity fields based on the shape of the velocity field \citep[see Figure \ref{fig:velocity_fields}, also][]{2005ApJ...618..569M,2016MNRAS.463..541C,Flury2023}. Thermal driving occurs when ``cool" ($10^4\rm~K$) clouds are 
embedded in an adiabatically expanding hot ($>10^6\rm~K$) gas \citep[][]{2005ApJ...618..569M}. 
Cosmic ray occurs driving where clouds are accelerated by the diffusion \citep{Breitschwerdt1993,Quataert2022a} 
or streaming \citep{Ipavich1975,Breitschwerdt1991,Quataert2022b} of relativistic charged particles--launched by hot ionized media such as shocks or supernovae \citep{Colgate1979,Blandford1980}--as they interact with 
magnetic fields in the ISM, resulting in a radially outward pressure gradient.
NGC 2992 is known to contain $\approx0.5$ kpc radio lobes forming an hourglass 
or ``figure eight'' shape, indicative of bi-directional bubbles of hot gas driven 
by nuclear activity \citep{Wehrle1988,Allen1999}. Combining optical and radio 
measurements indicates the presence of hot cavities which are heated (likely 
through shocks) by jets or accretion disk winds and within which are entrained 
clouds of cooler gas \citep{Allen1999,Veilleux2001}. The presence of such lobes 
and the associated hollow cavities suggests either or both \citep[a plausible scenario, e.g.,][]{Breitschwerdt2002,Everett2008} mechanisms are involved
in launching the outflow in NGC  2992. 

\begin{figure}
    \centering
    \includegraphics[width=0.9\linewidth,clip=True,trim={1.6in 0.6in 1in 0.3in}]{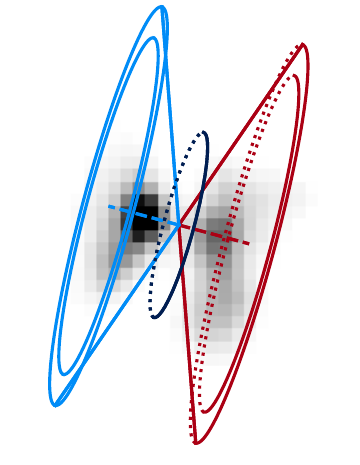}
    \caption{Greyscale map across the central kiloparsec of NGC 2992 of the continuum-subtracted, integrated [\ion{O}{iii}] doublet flux in the S7 DR2 \citep{S7DR2}. Drawn cones indicate the geometry inferred from the best-fit {\tt OutLines} profile to the \emph{integrated} emission line profile. No spatial information was used to determine the shape of the projected cones. Blue indicates the anterior cone, red indicates the posterior cone. Indigo indicates the disk, here with arbitrary radius scaled for visualization. A PA of 75$^\circ$ was assumed to rotate the projected geometry to the PA of the observation.}
    \label{fig:ifu_cones}
\end{figure}

Spatial analysis of the outflow in \citet{Veilleux2001} found an inclination of $i=68\pm3^\circ$ and an opening angle of $\theta_o=62.5-67.5^\circ$. {\tt OutLines} finds $i=53.19\pm1.32^\circ$ and $\theta_o=68.60\pm1.26^\circ$. While $\theta_o$ is in apparent agreement, $i$ is significantly discrepant, requiring further assessment of the fit results. To test the reliability of the geometry inferred from the {\tt OutLines} model, we project the best-fit bidirectional cones onto the plane of the sky and compare with the spatially resolved continuum-subtracted [\ion{O}{iii}] doublet flux in the WiFES IFU in Figure \ref{fig:ifu_cones}. We find good agreement between the cone geometry implied by the integrated line profile using {\tt OutLines} and the observed flux distribution, indicating that modeling with {\tt OutLines} can recover the spatial geometry without any knowledge of that geometry apriori. Not only have we accurately recovered the spatial flux distribution with {\tt OutLines}, but we have also inferred the presence of a cavity comprising the inner two thirds of the outflow cone ($\theta_c=48.64\pm1.82^\circ$), suggesting that cooler line-emitting clouds can only survive towards the edges of the hot cavity where the thermal wind may be somewhat cooler as it expands and contacts the surrounding ISM.

With the best-fit parameters, {\tt OutLines} computes high mass loss rates of at least 40 $\rm M_\odot~yr^{-1}$. From the $v_0\sim300$ \kms\ rotational velocity of the galaxy \citep{Veilleux2001}, the outflow must be capable of ejecting material from the host galaxy with the characteristic velocity exceeding the escape velocity by a factor of 3. Together, these results suggest NGC 2992 is in a state of AGN-triggered inside-out quenching with the combined effects of hot cavities and the evacuation of cool gas away from the plane of the galaxy, both effects preventing future star formation. Interestingly, this quenching feedback is not sufficiently columnated to be driven by a jet but instead is most likely due to expanding gas heated by winds launched from the accretion disk. While the implications of quenching are clear for galaxy evolution, the association of hot winds and shocks also has implications for the measurement of chemical abundances in AGN \citep[e.g.,][]{Dors2021}. Here, inferring underlying physics, geometry, and outflow strength from a single-aperture spectrum demonstrates the power and effectiveness of {\tt OutLines} in the context of AGN feedback.

\section{Conclusion}\label{sec:conclusion}

Here, we present a novel method for modeling emission and absorption lines arising from astrophysical winds, bubbles, and outflows. Our method has many advantages over previous approaches, such as multi-Gaussian and Gauss-Hermite models or velocity quantiles, in that it is physically motivated but also agnostic to underlying conditions or assumptions about the driving mechanism(s). Moreover, our model is readily applicable to absorption and emission line profiles. Improvements over a previous implementation of this method include i) multiple approximations to the velocity field from the wind equations of motion: a velocity power law, an acceleration power law, and the CAK theory $\beta$ law; ii) a variety of density profiles expressing continuous (power law, double power law exponential) and single-episode (normal, log normal, fast-rise exponential decay) scenarios as well as a composite of many expanding shells or bubbles and iii) different geometries (spherical, filled cones, and hollow cones, with cones accompanied by an optional disk). All these scenarios are consistent with conditions observed in stars, nebulae, and galaxies.

We have made this modeling approach publicly accessible through the open-source software {\tt OutLines} in {\tt python}, available via {\tt pip} and {\tt GitHub} facilities. To date, {\tt OutLines} is the only line profile code with such extensive and interchangeable geometries, density profiles, and velocity fields, making it by far the most comprehensive and versatile tool available for the study of outflows, winds, and bubbles. Furthermore, it is the only code for which the geometry inferred from integrated spectral line profiles has been verified through comparisons with spatially resolved IFU observations.

We illustrate the usefulness of the {\tt OutLines} software with application to an \ion{H}{ii} region, a super star cluster, a starburst galaxy, and an AGN, deriving properties based on the line profile models. Furthermore, we demonstrate how the results from modeling with {\tt OutLines} provide insight into the underlying physics of winds, bubbles, and outflows, obtaining the following results specifically made possible through our custom line profiles:
\begin{itemize}
    \item Young ($<4$ Myr) \ion{H}{ii} regions undergoing expansion can exhibit two orders of magnitude in gas compression due to wind-triggered shocks even while optically thick radiation pressure is the main driving mechanism of the bulk motion of the gas. Thus, while shocks may dominate the radio continuum, it is radiation pressure from stars which primarily shapes the evolution of young \ion{H}{ii} regions.
    \item Suppression of optically thin radiative feedback occurs in young ($<2$ Myr) super star clusters in dwarf galaxies like the Green Peas indicates that the most recent stellar populations cannot mechanically evacuate optically thin channels for Lyman continuum photons to escape. Mechanical feedback from previous episodes of star formation may be necessary to shape the gas geometry for subsequent populations to leak ionizing radiation. Thus it is ionizing feedback from the most recent stellar populations which produces the density-bounded conditions in low gas density channels through which Lyman continuum photons escape.
    \item Expanding superbubbles produced by core-collapse supernovae can, in succession, accumulate up to produce an outflow as the shells relax or dissipate into a more continuous flow; however, optically thick radiation pressure is necessary for these shells to accelerate to the terminal velocity before blowout and formation of bidirectional cones extended beyond the galactic disk. The formation of outflows from systems of metal-rich shells indicates that radiative driving is a key mechanism in the baryon cycle which shapes the scaling relations of galaxies across cosmic time.
    \item Accretion disk winds of a Seyfert galaxy are sufficiently fast to trigger substantial shocks in the central kiloparsec. The resulting hot fluid drives an outflow via pressure generated from the hot gas itself or from cosmic rays accelerated by the shock. The outflow consists of ``cold'' clouds entrained in the periphery of hollow cones and has the potential to cause inside-out quenching as it impacts the host galaxy.
\end{itemize}

This broad applicability and deep insight into the physics of outwardly moving gas demonstrates that {\tt OutLines} is powerful tool which will prove to be useful across the astronomical community. Future work will involve application to massively multiplexed fiber spectroscopic surveys like WEAVE-LOFAR and 4MOST/WAVES for statistical assessments and to hydrodynamical simulations to explore the underlying physics.

\section*{Acknowledgements}

We thank the beta testers, including K. Z. Arellano Cordova, K. Duncan, A. Hall, and T. Horn, whose efforts helped to optimize the use of this code. We thank K. Z. Arellano Cordova, C. Esteban, and J. Garcia-Rojas for providing access to Knot A extractions from their reduced MEGARA data. We thank K. Parker for sharing results from the recently accepted \citet{Parker2025}. We thank K. Duncan for insight and comments regarding this manuscript. We thank A. Rankine and A. Lawrence for suggestions regarding the relativistic corrections. We also thank M. Divakara and G. Romero-Cruz for insightful discussions about outflow geometry at the National Institute of Astrophysics, Optics and Electronics (INAOE) and J. Chisholm, C. Martin, and N. Murray for insightful discussions about winds, massive stars, feedback, and {\it HST}/COS at the Aspen Center for Physics.

This work was performed in part at the Aspen Center for Physics, which is supported by a grant from the Simons Foundation (1161654, Troyer) and by National Science Foundation grant PHY-2210452.

Support for \emph{HST} GO programs 12583 and 15840 was provided by NASA through a grant from the STScI, which is operated by the Association of Universities for Research in Astronomy, Inc., under NASA contract NAS 5–26555.

Facilities: {\it HST}, {\it GTC}, {\it SSO}

Software, repositories, and other digital resources:
{
{\tt corner} \citep{corner},
\href{https://matplotlib-curly-brace.readthedocs.io/en/}{\tt curlyBrace},
{\tt emcee} \citep{emcee},
{\tt matplotlib} \citep{matplotlib},
\href{https://mast.stsci.edu/portal/Mashup/Clients/Mast/Portal.html}{\tt MAST}, 
\href{https://www.nist.gov/pml/atomic-spectra-database}{\tt NIST/ASD} \citep{NIST_ASD},
{\tt numpy} \citep{numpy},
\href{https://github.com/sflury/OutLines}{\tt OutLines} (this work, see also \href{https://doi.org/10.5281/zenodo.11238265}{10.5281/zenodo.11238265}),
{\tt PyNeb} \citep{2015A&A...573A..42L},
\href{https://www.mso.anu.edu.au/S7/Data_Release_2/}{\tt S7 DR2} \citep{S7DR2},
{\tt scipy} \citep{scipy},
\href{https://www.github.com/sflury/vygrboi}{\tt vygrboi}
}

AI use: No language learning model or otherwise generative artificial intelligence was used in this work.

\section*{Data Availability}

The data underlying this article for NGC 5471 were provided by K. Z. Arellano Cordova, C. Esteban, and J. Garcia-Rojas by permission. Data will be shared on reasonable request to the corresponding author with permission of the above owners. The data underlying this article for J1044+4012 and Mrk 1486 are publicly available from \href{https://mast.stsci.edu/portal/Mashup/Clients/Mast/Portal.html}{\tt MAST}. The data underlying this article for NGC 2992 are publicly available from \href{https://www.mso.anu.edu.au/S7/Data_Release_2/}{\tt S7 DR2}.

The {\tt OutLines} code presented here is publicly available on \href{https://github.com/sflury/OutLines}{github.com/sflury/OutLines} under DOI \href{https://doi.org/10.5281/zenodo.11238265}{10.5281/zenodo.11238265} or from \href{https://pypi.org/project/SpecOutLines/}{PyPI} via {\tt pip install SpecOutLines}.

\bibliographystyle{mnras}
\bibliography{bib,flury23,soft}



\appendix

\section{Geometry of Directed Outflows}\label{apx:projections}

To determine the integral $2\pi$\lb\ over the azimuthal coordinate term in d$\Omega$, for a directed conical outflow, we must first project the spherical cap of a cone onto the plane of the sky. Then, we must subsequently compute the fraction \lb\ of a given deprojected velocity ($w$) subtended by the cap. With this formalism in hand, we demonstrate incorporation of additional geometric components in the form of cavities and obstructing disks.

\subsection{Cones As A Spherical Cap Projection}

\begin{figure}
    \centering
    \includegraphics[width=\linewidth]{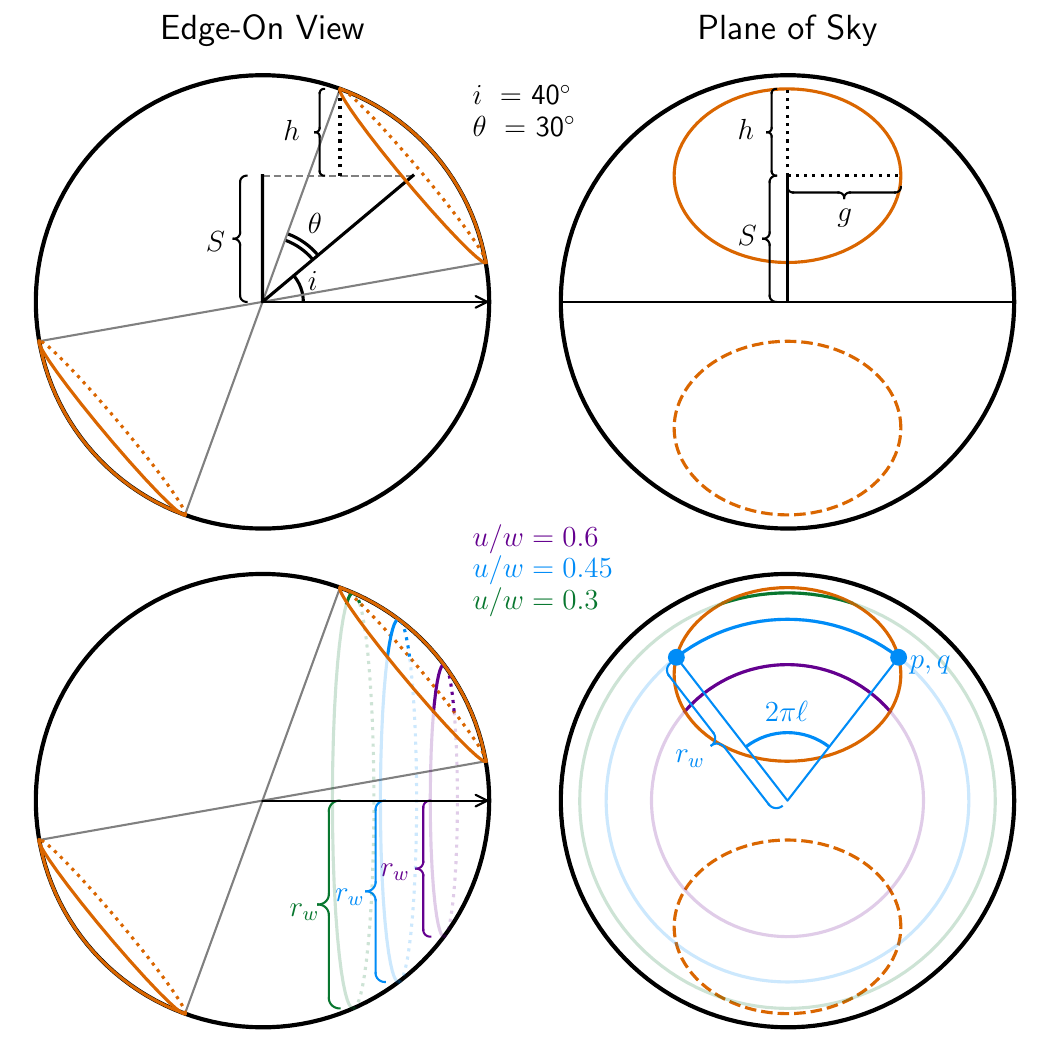}
    \caption{A cone geometry in the case of $i=40^\circ$, $\theta=30^\circ$. Left panels illustrate the cones seen edge-on, i.e., transverse to the line of sight. Right panels illustrate the cones seen face-on, i.e., projected onto the plane of the sky. Solid ellipses indicates the cone projected anterior to source relative to the observe while dashed ellipses indicate the posterior projection. Top panels illustrate the geometric terms $S$, $g$, and $h$ which translate the opening angle $\theta$ and inclination $i$ of the cones into a projection onto the sky. Bottom panels illustrate how an observed velocity $u$ de-projects into an azimuthal band with radius $r_{w}$ with unique azimuthal coverage $\pi\ell$ within the cone defined by the intersection points $p,q$ of the velocity band with the projected cone. High $u/w$ (low $r_{w}$) corresponds to wings of a line profile far from the systemic velocity while low $u/w$ (high $r_{w}$) corresponds to the core of the line profile close to the systemic velocity.}
    \label{fig:cone_geom}
\end{figure}

An outflowing cone on the plane of the sky appears as a the projected spherical cap with axes $g$ and $h$ and is offset from the projected outflow source by some distance $S$. These quantities can be expressed in terms of the inclination and opening angle of the ouflow center such that
\begin{align}
    S &= \sin{i}\cos{\theta} \\
    g &= \sin{\theta} \\
    h &= \cos{i}\sin{\theta}.
\end{align}
We illustrate a single case of such a projection in Figure \ref{fig:cone_geom}. 
In the case where $i+\theta>90^\circ$, the projection is no longer an ellipse but rather an ellipse plus a lune to account for the projection of the ``top'' of the spherical cap and an additional lune to the opposing side to account for the projection of the second cone. We visualize such a scenario in the last two panels of Figure \ref{fig:cone-proj}.

\subsection{Cone Inscription of Deprojected Velocities}

The fractional arc length \lb\ of the deprojected velocity inscribed by the cone can be determined by solving the system of linear equations for a circle and the projected spherical cap area to determine where they intersect, which in turn determine the bounds on the integral(s) over the azimuthal coordinate of the solid angle. The azimuthal ``ring'' of gas with velocity $w$ has a radius $r_{\rm w}$ defined by the sine of the 
angle $\varrho$ between the observed and deprojected velocities $u$ and $w$, respectively. 
Relativistic aberration (i.e., beaming effect) is invariant across the azimuthal angle, ergo a classical treatment in the source frame is still accurate for deriving the geometry of the projected cone.
The apparent deprojection radius $r_{\rm w}$ for each azimuthal ring is thus
\begin{equation}
 r_{\rm w}=\sqrt{1-\left(\frac{u}{w}\right)^2},
\end{equation}
which we show in Figure \ref{fig:cone_geom}.
Solving the system of equations of the projected cap and the deprojected velocity ring gives intersection points of ($p$, $q$) such that
\begin{equation}
    p = \pm\sqrt{r_{\rm w}^2-q^2}
\end{equation}
and, from substituting $p$ and employing the quadratic equation,
\begin{equation}
    q = \frac{-B\pm\sqrt{B^2-4AC}}{2A}
\end{equation}
where the coefficents are
\begin{align}
    A &= \left(\frac{g}{h}\right)^2-1 \\
    B &= -2S\left(\frac{g}{h}\right)^2 \\
    C &= r_{\rm w}^2+\left(\frac{gS}{h}\right)^2-g^2.
\end{align}
To reduce computation time, $A$ and $B$ are calculated outside of the integral as only $C$ depends on the de-projected velocity $w$.
The intersection points ($p$, $q$) given by $A$, $B$, and $C$ yield the arc of the enclosed deprojected velocity such that
\begin{equation}
    \pi\ell_{\rm w} = 
\begin{cases}
    \arccos{\frac{q}{r_{\rm w}}}&q>0\\
    \pi-\arccos{\frac{q}{r_{\rm w}}}&q<0
\end{cases}
\end{equation}
which notably avoids the additional step of calculating $p$. We illustrate such a solution for $\pi\ell$, including the velocity dependence, in Figure \ref{fig:cone_geom}.
If multiple real solutions for ($p$, $q$) exist due to the lune from the projection of the second cone into the plane, then {\tt OutLines} simply sums the two arcs.

\begin{figure}
    \centering
    \includegraphics[width=\columnwidth]{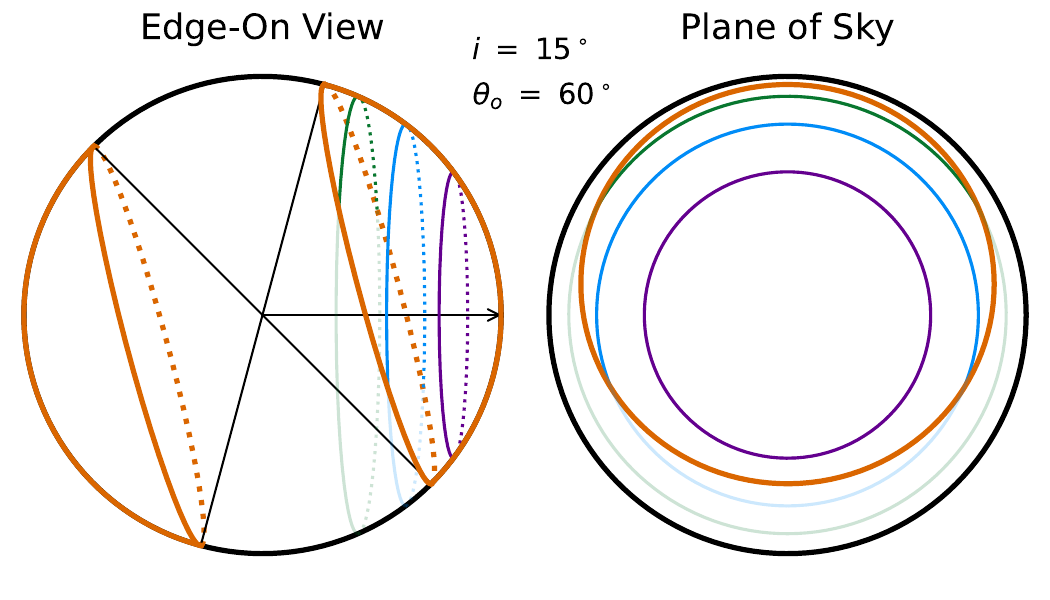}
    \includegraphics[width=\columnwidth]{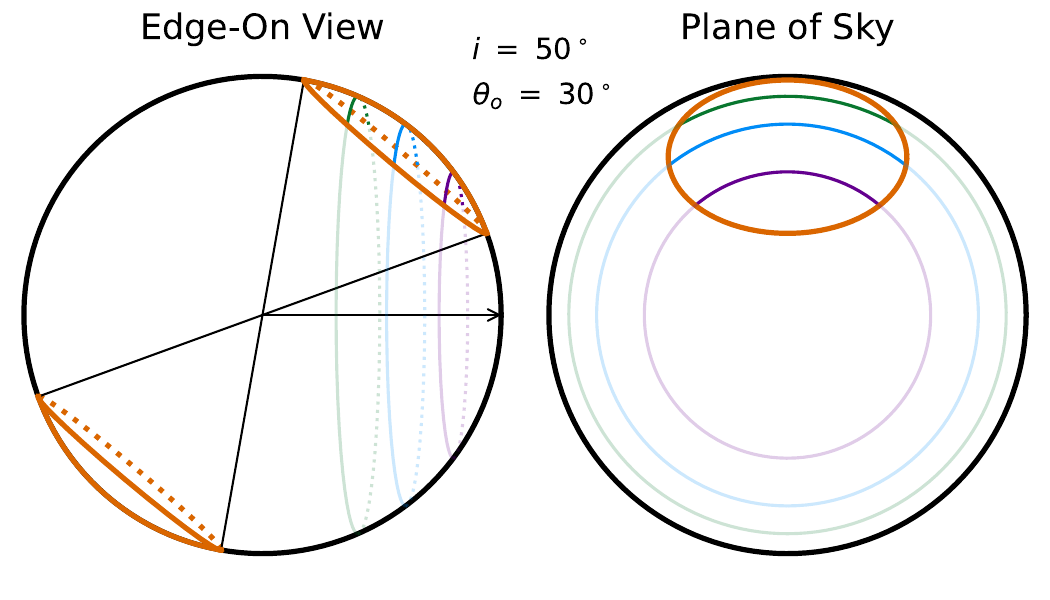}
    \includegraphics[width=\columnwidth]{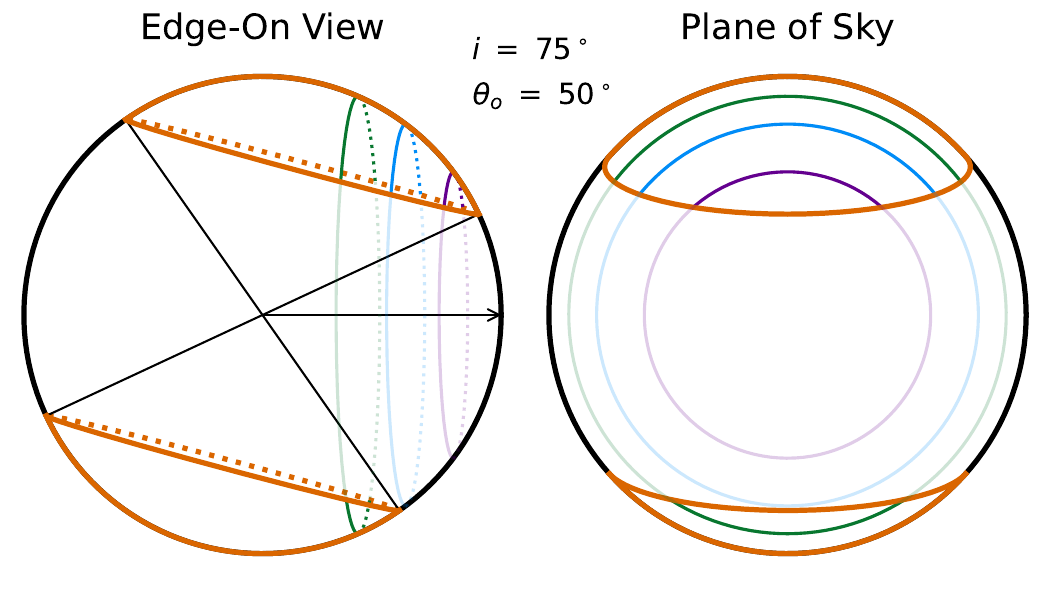}
    \includegraphics[width=\columnwidth]{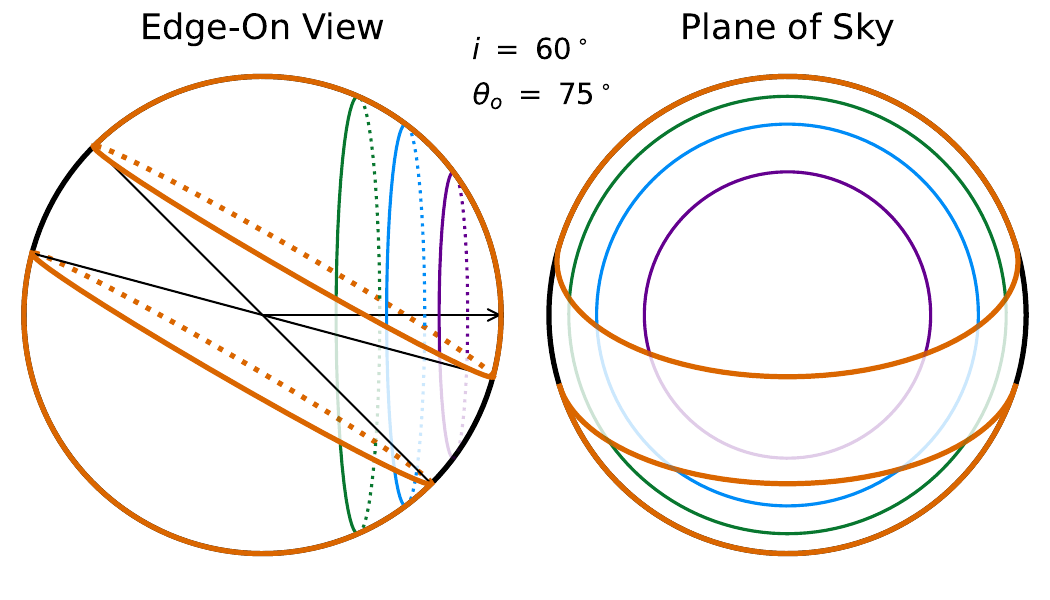}
    \caption{Following Figure \ref{fig:cone_disk_profiles}, we illustrate bidirectional cone geometries (left of each panel) projected onto the plane of the sky (right of each panel) for various combinations of $i$ and $\theta_o$ (orange). Circles indicate deprojected velocities of $u/w=$0.4 (green), 0.6 (blue), and 0.8 (purple). Opaque segments indicate where a deprojected velocity is inscribed by the projected spherical cap. Arrow in the left of each panel indicates direction of the projections onto the plane of the sky (i.e., the line of sight).}
    \label{fig:cone-proj}
\end{figure}

We show example projections in Figure \ref{fig:cone-proj} for several values of $u/w$, $i$, and $\theta$. As a tool, we also include in {\tt OutLines} a module to produce such visualizations for any given combination of $i$ and $\theta$. Calling {\tt OutLines.PlotConeProjection(i,thetaO)} will draw the geometry and projection as shown in Figure \ref{fig:cone-proj} for any inclination and opening angle.

\subsection{Cavity Effects}

\begin{figure}
    \centering
    \includegraphics[width=\columnwidth]{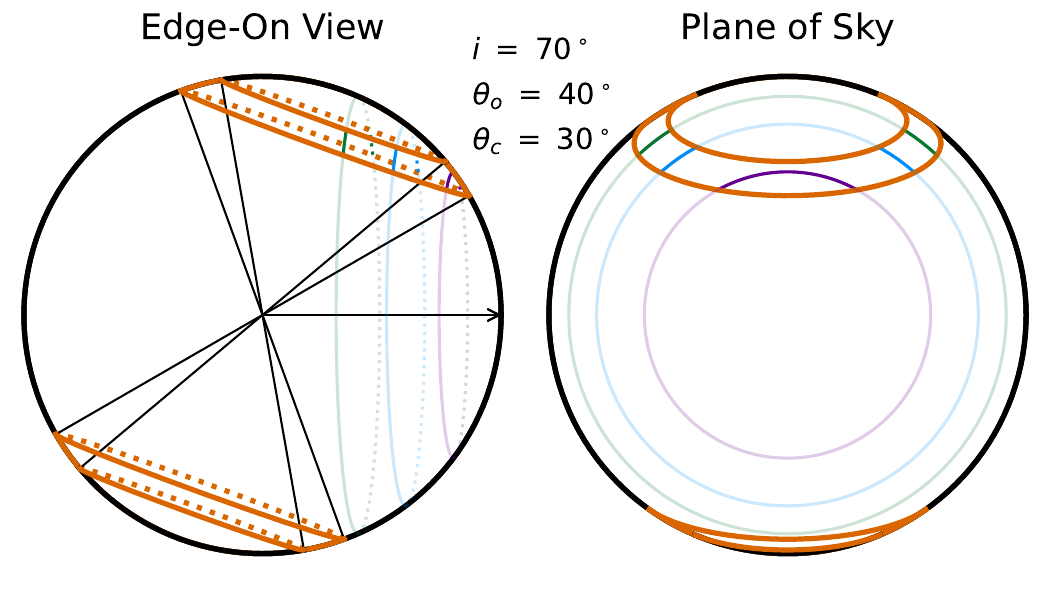}
    \caption{Same as Figure \ref{fig:cone-proj} but for directed cones with an inner cavity.}
    \label{fig:hollow-cone-proj}
\end{figure}

As discussed in \S\ref{sec:geometry}, some outflows exhibit a bidirectional cones with cavities, resulting in a hollow cone geometry. The fractional arc length $\ell_{\rm cavity}$ lost due to the cavity can be calculated in the same manner as \lb\ and subtracted, which we show in Figure \ref{fig:hollow-cone-proj}. Calling {\tt OutLines.PlotConeProjection(i,thetaO,thetaC)} will illustrate projections similar to those of the filled cones but for the hollow cone geometry.

\subsection{Obstruction by a Disk}

In \S\ref{sec:geometry}, we present the possible inclusion of a disk which encircles the source of the outflow and is inclined orthogonally to a cone or hemisphere axis. This disk obstructs the posterior cone or hemisphere but may affect either the red-shifted or both wings of the line profile. {\tt OutLines} projects the disk onto the line of sight and then determines the length of each deprojected velocity which is still observed after accounting for the disk effect. Since the disk is centered at the source of the outflow and assumed to be orthogonal to the outflow cone, the projected disk is simply an ellipse with semi-major and semi-minor axes of ($w[x_d|\beta],w[x_d|\beta]\cos i$). Thus, the intersection points ($p_d$,$q_d$) between the deprojected velocity and the disk can be written as
\begin{equation}
    p_d = \pm\sqrt{r_{\rm w}^2-q_d^2}
\end{equation}
and
\begin{equation}
    q_d = \pm\sqrt{\frac{r_{\rm w}^2-w^2[x_d|\beta]}{\sec^2 i-1}}.
\end{equation}
Rather than compute \lb\ and $\ell_{\rm disk}$ separately, which would needlessly incur a computational overhead, we can instead simply compare $q$ and $q_d$ to determine the final \lb.
For the posterior cone or hemisphere, \lb\ is given by
\begin{equation}
    \pi\ell_{\rm w} = 
\begin{cases}
    \arccos{\frac{q}{r_{\rm w}}}&q>q_d\\
    \arccos{\frac{q_d}{r_{\rm w}}}&-q_d<q<q_d\\
    2\arccos{\frac{q_d}{r_{\rm w}}}-\arccos{\frac{q}{r_{\rm w}}}&q<-q_d\\
\end{cases}.
\end{equation}
If $q$ has multiple solutions due to the projected lune from the anterior cone, {\tt OutLines} will add to \lb\ the first case evaluated using the second solution for $q$ with the imposition that $q>q_d$ since the anterior cone is not obstructed by the disk. Note that, in the case of $i+\theta>90^\circ$, the red component will never be fully obstructed.
For the anterior cone, \lb\ is given by
\begin{equation}
    \pi\ell_{\rm w} = 
\begin{cases}
    \arccos{\frac{q}{r_{\rm w}}}&q>0,\ q_1>q_d\\
    \arccos{\frac{q_d}{r_{\rm w}}}&q>0,\ q_1<q_d\\
    \pi-\arccos{\frac{q}{r_{\rm w}}}&q<0\\
\end{cases}
\end{equation}
where the $q_1$ criterion only applies in the case of a lune projected from the posterior cone, which can be obstructed by the disk. Note that, in the case of $i+\theta>90^\circ$, the blue component can be partially obstructed, primarily affecting the line core. Under this formalism, $\ell_{\rm cavity}$ can still be subtracted to determine the hollow cone geometry as both \lb\ and $\ell_{\rm cavity}$ will account for any disk effects on the deprojected velocity apriori.

\section{Implementation of OutLines}\label{apx:OutLines}

\begin{table*}
    \centering
    \caption{{\it Left}: the {\tt Profile} object classes in {\tt OutLines} with which users can generate model profiles for different spectral line types. {\it Right}: keyword settings for the {\tt Profile} classes to customize models. {\tt Disk} not implemented for spherical geometries. {\tt Pulse} only applicable ensemble density profiles.}
    \begin{tabular}{l p{1.2in} c | p{0.8in} p{0.5in} p{0.5in}}
    class & use & atomic data & keywords & options & default \\
    \hline
    \hline
    {\tt Nebular}     & nebular emission lines & $\lambda_0$ & \multirow{ 6}{1.4in}[0pt]{{\tt Geometry}\newline {\tt VelocityField}\newline {\tt DensityProfile}\newline {\tt AddStatic}\newline {\tt FromRest}\newline {\tt Disk}\newline {\tt Aperture}\newline {\tt Pulse} } & \multirow{ 6}{1.4in}[0pt]{Table \ref{tab:geom_terms}\newline Table \ref{tab:velocity_terms}\newline Table \ref{tab:dens_terms}\newline {\tt boolean}\newline {\tt boolean}\newline {\tt boolean}\newline {\tt boolean}\newline Table \ref{tab:dens_terms}\newline Bubbles/Shells} & \multirow{ 6}{1.4in}[0pt]{\tt Spherical\newline BetaCAK\newline PowerLaw\newline False\newline True\newline False\newline False\newline `Normal' } \\
    {\tt Resonant} & resonant emission lines & $\lambda_0,\ f,\ p_r$ &  \\
    {\tt Fluorescent} & fluorescent emission lines & $\lambda_0,\ f,\ p_r,\ p_f$ &  \\
    {\tt Absorption} & resonant absorption lines & $\lambda_0,\ f$ &  \\ &&& \\ &&& \\ &&& \\
    \end{tabular}
    \label{tab:outline}
\vspace{6pt}
\end{table*}

\subsection{Installation}

Installing {\tt OutLines} can be done in the terminal command line using {\tt pip} as shown below.
\begin{lstlisting}[language=bash]
pip3 install SpecOutLines
\end{lstlisting}

Alternatively, the GitHub repository can be cloned as below
\begin{lstlisting}[language=bash]
git clone https://github.com/sflury/OutLines.git
\end{lstlisting}

Dependencies include {\tt python >=3.12.9}, which may require updating of python globally or creating, say, an {\tt anaconda} environment. Other dependencies, including {\tt numpy}, {\tt scipy}, and {\tt matplotlib} will be automatically managed by {\tt pip}, including within {\tt conda} environments.

\subsection{ The {\tt Profile} Classes}

The primary front-end user interaction with {\tt OutLines} is through the various {\tt Profile} object classes, which we list in Table \ref{tab:outline}. Calling one of these classes instantiates an object containing the line profile model, parameter guesses, and parameter bounds. When calling a class, users need only provide the relevant atomic data and any optional keyword arguments specifying the velocity field, geometry, and density profile to customize the model (see Table \ref{tab:outline}). The terms and related keyword arguments for the profile model are listed in Table \ref{tab:geom_terms} for the geometry, Table \ref{tab:velocity_terms} for the velocity field, and Table \ref{tab:dens_terms} for the density profile, with the {\tt OutLines} defaults indicated by a \textsuperscript{\textdagger}. { In some cases, users may want to consider a velocity field or density profile not included in {\tt OutLines}. In such cases, if a user passes a function to either keyword instead of a string, {\tt OutLines} will implement the user-defined field or profile.}

\begin{table}
\centering
\caption{Geometry keyword options and related required arguments for {\tt OutLines}. $^\dagger$ indicates the {\tt OutLines} default. 
\label{tab:geom_terms}}
\begin{tabular}{l c l c}
 {\tt Geometry}    & terms & assumptions\\
 \hline
 \hline\\[-7.5pt]
{\tt `Spherical'}~$^\dagger$ & none & $i=0$, $\theta_c=0^\circ$, $\theta_o=90^\circ$\\
{\tt `Hemisphere'} & $i$ & $\theta_c=0^\circ$, $\theta_o=90^\circ$\\
{\tt `FilledCones'} & $i$, $\theta_o$ & $\theta_c=0^\circ$\\
{\tt `HollowConesFixedCavity'} & {$i$, $\theta_o$} & {$\theta_c=\theta_o-10^\circ $}\\

{\tt `HollowCones'} & \multirow{2}{*}{$i$, $\theta_c$, $\theta_o$} & \multirow{2}{*}{none}\\
{\tt `OpenCones'} & \\
\end{tabular}
\end{table}

\begin{table}
\centering
\caption{Velocity field keyword options for {\tt OutLines}, each paramaterized by an exponent $\beta$. $^\dagger$ indicates the {\tt OutLines} default.\label{tab:velocity_terms}}
\begin{tabular}{l c p{4.5cm}}
{\tt VelocityField} & equation & references  \\
\hline 
\hline\\[-7.5pt]
{\tt `BetaCAK'}~$^\dagger$ & \ref{eqn:betaCAK} & {\citet{1975ApJ...195..157C,Castor1979}} \\
{\tt `AccPlaw'} & \ref{eqn:accplaw} & {\citet{2010ApJ...717..289S}} \\
{\tt `VelPlaw'} & \ref{eqn:velplaw} & {\citet{FaucherGiguere2012}} \\
\end{tabular}
\end{table}

\begin{table}
\centering
\caption{Density profile keyword options and related required arguments for {\tt OutLines}. $^\dagger$ indicates the {\tt OutLines} default.\label{tab:dens_terms}}
\begin{tabular}{l c c}
{\tt DensityProfile}     &  terms & equation\\
\hline 
\hline\\[-7.5pt]
\multicolumn{3}{c}{Winds} \\[1.5pt]
\hline
{\tt `PowerLaw'}~$^\dagger$ & $\alpha$ & \ref{eqn:denplaw} \\
{\tt `Exponential'} & $\gamma$ & \ref{eqn:expon}  \\
{\tt `DoublePowerLaw'} & $\alpha_1$, $\alpha_2$, $x_1$ & \ref{eqn:dplaw} \\
\hline\\[-7.5pt]
\multicolumn{3}{c}{Bubbles / Shells} \\[1.5pt]
\hline
{\tt `Normal'} & $x_1$, $\sigma_x$ & \ref{eqn:norm} \\
{\tt `LogNormal'} & $\log x_1$, $\log\sigma_x$ & \ref{eqn:lognorm} \\
{\tt `DLogic`} & $\log x_1$, $k$ & \ref{eqn:dlogic} \\
{\tt `Shell'} & $x_1$, $\sigma_x$  & \ref{eqn:shell} \\
{\tt `FRED'} & $x_1$, $\tau_1$, $\tau_2$  & \ref{eqn:fred} \\
\hline\\[-7.5pt]
\multicolumn{3}{c}{Ensembles} \\[1.5pt]
\hline
{\tt `Pulses'} & $x_1$, $\sigma_x$, $x_k$ & \ref{eqn:pulses} \\
{\tt `DampedPulses'} & $\gamma$, $x_1$, $\sigma_x$, $x_k$ & \ref{eqn:dampedpulses}
\end{tabular}
\end{table}

Users will primarily use {\tt OutLines} by instantiating one of the several {\tt Profile} classes: {\tt Nebular}, {\tt Resonant}, {\tt Fluorescent}, and {\tt Absorption}. The object-bound methods of the {\tt Profile} classes can be used to obtain or update parameter values and to generate line profiles. We provide examples of using these classes to generate line profiles in Appendix \ref{apx:examples}. To obtain line profiles computed by {\tt OutLines}, users can call the {\tt get\_profile()} bound method for either class, which requires only an array of wavelengths and computes a line profile for the parameters stored in the object. Alternatively, the line profile function is directly accessible through the {\tt Profile} attribute, which requires a wavelength array just as {\tt get\_profile()} but also takes the full set of parameters. For purposes of modeling observations, users may prefer to use the {\tt Profile} function where the input parameters may be readily varied without updating the corresponding attribute in the object.

The {\tt get\_params()} and {\tt get\_bounds()} methods will pass the relevant values from the object to the user. The {\tt print\_settings()}, {\tt print\_params()}, and {\tt print\_bounds()} will print the selected profile settings (velocity field, geometry, density profile, and optional static line component), the corresponding current parameters, and the associated boundaries. The {\tt update\_params()} and {\tt update\_bounds()} methods will allow users to update any of the profile parameters and their related bounds. All {\tt Profile} classes can each support multiple lines, something we encourage users to consider.

\subsection{The {\tt Properties} Class}

After modeling a wind or outflow from observed spectral lines, users will want to derive important diagnostic properties to determine what drives the wind, how, and to what extent the outflow impacts its environment. Because the relevant information is already contained in the {\tt Profile} classes, {\tt Properties} is designed to inherit one of the {\tt Profile} classes and will automatically compute relevant properties defined in \ref{sec:properties}, with the caveat that $n_0$ and $R_0$ remain unknown. 

Users can access these results directly from the object class once instantiated, either through the stored dictionary or by printing to the command line via the bound method {\tt print\_propts()}. As with the {\tt Emisison} and {\tt Absorption} classes, {\tt Properties} also has the bound method {\tt update\_params()}, which behaves similar to the bound method in the {\tt Profile} classes by allowing users to change parameters associated with the line profile.

\begin{table}
    \centering
    \caption{Outflow properties computed by the {\tt OutLines.Properties} object class. Property names listed here correspond to the dictionary keys in {\tt props} attribute and the terminal output by {\tt print\_props} method.\label{tab:properties}}
    \begin{tabular}{l c l}
    name     & equation &property\\
    \hline 
    \hline
    {\tt x.out}     &  & $x_{out}$ characteristic radius\\
    {\tt v.out}     & \ref{eqn:vout} & $v_{out}$ characteristic radius\\
    {\tt Rcal}     & \ref{eqn:Rcal} & $\mathcal{R}$ relative column density\\
    {\tt Mdot}     & \ref{eqn:MdotNR2} & $\dot{M}/n_0R_0^2$ mass outflow rate\\
    {\tt pdot}     & \ref{eqn:pdotNR2} & $\dot{p}/n_0R_0^2$ momentum flux\\
    {\tt Edot}     & \ref{eqn:EdotNR2} & $\dot{E}/n_0R_0^2$ mechanical luminosity
    \end{tabular}
\end{table}

\subsection{Visualization}

To assist with interpretation of outflow properties and geometry, {\tt OutLines} contains several functions to generate figures like the ones presented here. To produce isocontours as in Figure \ref{fig:isocontours}, we provide the {\tt PlotIsoContours} function. To produce the edge-on geometric cartoon as in Figure \ref{fig:cone_geometry}, we provide the {\tt PlotGeometry} function. To produce the cone and velocity projection geometry for filled cones as in Figure \ref{fig:cone_geom}, we provide the {\tt PlotProjGeom} function. To produce the velocity projections for filled and hollow cones as in Figures \ref{fig:cone-proj} and \ref{fig:hollow-cone-proj}, we provide the {\tt PlotConeProj} function.

\subsection{Numerical Methods}

Some equations require root-finding where analytic solutions are not possible, including the ``source-effect'' and aperture velocity limits (Equations \ref{eqn:w0source} and \ref{eqn:wAper}, respectively) and the characteristic outflow radius $x_{out}$ where the momentum density is maximized. As in \citet{Flury2023}, we employ Brent's method implemented by {\tt scipy.optimize} to solve for the root in each case.

The initial profile model of \citet{Flury2023} solved the integral for the emission line profile (Equation \ref{eqn:phi_ems}) using the adaptive Romberg method, which included Richardson extrapolation under the Bulirsch-Stoer method. As implemented by the now-deprecated {\tt scipy.integrate.romberg}, this method can be costly and in later {\tt scipy} versions has loss of accuracy in non-spherical geometries due to poor sampling of the integrand. { For improvements to both speed and accuracy, we now adopt Gauss-Legendre quadrature with a fixed number of nodes to solve the integrals for $\phi_\lambda$. We implement quadrature integration following the {\tt scipy.integrate.fixed\_quad}; however, for speed, we have pre-computed the nodes and weights using {\tt scipy.special.roots\_legendre}, which takes $\approx1\rm\:ms$ for each call to {\tt scipy.integrate.fixed\_quad}. We find at least 16 nodes necessary to obtain reliable profile solutions in many cases but conservatively assume 96 nodes for optimal accuracy across the profile parameter spaces without significant computational expense. For bubbles, we restrict velocity bounds where possible to improve sampling of the density profile. Future work may include adopting Gauss-Kronrod integration for improved accuracy and precision, e.g., following \citet{laurie1997}.}

To normalize the line profile, we use a linear integration method to effectively handle cusps which may occur at peaks and troughs, typically near the line core. We implement this integration following the shoelace algorithm case of Green's theorem. Benchmarking against the {\tt scipy.integrate.trapezoid} implementation of trapezoid rule integration, we find comparable results with a marginally faster (by $<1\rm\:ms$) compute time.


For a single core on a conventional laptop computer, benchmarking indicates that {\tt OutLines} can calculate an emission line profile at $R=5,000$ in $6\rm\:ms$ for the most complex geometry (disk plus hollow cones), and that the compute time increases almost linearly with $R$ such that run times are $2.5\rm\:ms$ at $R=2,000$, $4\rm\:ms$ at $R=3,000$, $13\rm\:ms$ at $R=10,000$, $25\rm\:ms$ at $R=20,000$, and $65\rm\:ms$ at $R=50,000$. For the simplest geometry (spherical), the compute times are much faster ($<1\rm\:ms$ for $R\leq30,000$). Given reasonable guess parameters, iterative non-linear least squares fitting routines like the trust-reflective algorithm (see {\tt scipy.optimize.least\_squares}) perform rapidly, typically converging on a solution within 10 to 30 s depending on the data and the assumed model. For 25 walkers, a typical MCMC sampler (here benchmarked using {\tt emcee}, \citealt{emcee}) takes anywhere from $0.1$ to $2$ s per step to run on a conventional laptop computer when fitting an {\tt OutLines} model to an observed line profile, even when accounting for complex geometries and LSF effects. The addition of convolution with a line spread function does increase the computation time; however, the extent to which this occurs depends on the resolution, size of the data set, and the implementation of the convolution. 

\section{Example Line Profiles with {\tt OutLines}}\label{apx:examples}

Below, we provide example usages of the {\tt Absorption}, {\tt Nebular}, and {\tt Fluorescent} line profile object classes in {\tt OutLines}. We demonstrate how to select different options for the line profile model, update model parameters and bounds associated with the line profile, display documentation, print various model settings and parameters, and extract computed line profiles for visualization. We also demonstrate how to compute outflow properties from the {\tt Properties} class and print the results to the command line.

\subsection{\ion{C}{ii} $\lambda1334$}

Example instantiations of {\tt Absorption}, {\tt Resonant}, and {\tt Fluorescent} objects from the {\tt Profile} class using atomic data from \citet{Morton2003}. Here, the object classes are instantiated, documentation, setting, and parameters are printed to the terminal, and the column density and terminal velocity are updated.

\begin{lstlisting}[language=Python]
# import the OutLines package
import OutLines as OL
# instantiate the absorption line 
# Profile object class
modelAbs = OL.Absorption(1334.519,0.128)
# print documentation of the model object
modelAbs.docs()
# print model settings
modelAbs.print_settings()
# print the current parameters
model.print_params()
# instantiate the resonant and fluorescent 
# Profile object classes
modelRes = OL.Resonant(1334.532,0.128,0.456)
modelFlu = OL.Fluorescent(1335.663,0.115,0.456,0.544)
# updates for all three objects
for model in [modelAbs,modelRes,modelFlu]:
    # change the log column density to 1e15 cm^-2
    model.update_params('LogColumnOutflow1',15)
    # change the velocity to 500 km s^-1
    model.update_params('TerminalVelocity', 500.00)
\end{lstlisting}

\subsection{[\ion{O}{iii}] $\lambda4959,5007$}\label{apx:o3_examp}

\begin{figure}
    \centering
    \includegraphics[width=\linewidth]{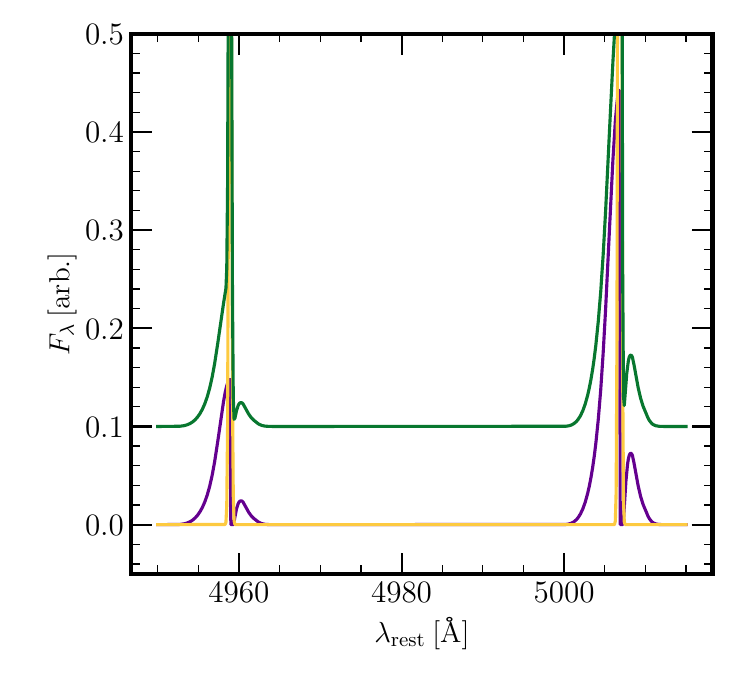}
    \caption{Example [\ion{O}{iii}] $\lambda\lambda4959,5007$ line doublet profile produced using {\tt OutLines} as shown in Appendix \ref{apx:o3_examp}.}
    \label{fig:o3_examp}
\end{figure}

Example insantiation of the {\tt Nebular} object from the {\tt Profile} class in {\tt OutLines} using atomic data from {\tt NIST} \citep{NIST_ASD} and \citet{Storey2000}. Here, we illustrate how to customize the model with keyword arguments passed to the object class on instantiation, updating parameters, printing setting and parameters, computing and printing outflow properties, and visualizing a line profile as in Figure \ref{fig:o3_examp}.

\begin{lstlisting}[language=Python]
# import packages
import OutLines as OL
from numpy import linspace
import matplotlib.pyplot as plt
# instantiate the nebular line
# profile object
kwargs = dict(Geometry='HollowCones',AddStatic=True,\ VelocityField='AccPlaw',Disk=True)
model = OL.Nebular([4958.911,5006.843],**kwargs)
# update some parameters
model.update_params(['TerminalVelocity'],[500,30,15,45])
model.update_params(['OpeningAngle','CavityAngle',\ 'Inclination'],[30,15,45])
model.update_params(['FluxOutflow1','FluxStatic1'],\ [1/2.98,1/2.98]) # Storey & Zeppen 2000
# print model settings and current parameters
model.print_settings()
model.print_params()
# compute outflow properties
props = OL.Properties(model)
# print outflow properties
props.print_props()
# plot the profile
wave = linspace(4950,5015,651)
plt.plot(wave,model.get_outflow(wave),label='outflow')
plt.plot(wave,model.get_static(wave),label='static')
plt.plot(wave,model.get_profile(wave)+0.1,label='total')
plt.show()
\end{lstlisting}

\section{Example Modeling}\label{apx:modeling}

Below, we provide examples for modeling absorption and nebular emission line features using line profiles computed by {\tt OutLines}. We indicate suggested methods for obtaining best-guess parameters and demonstrate how to use {\tt OutLines} in tandem with MCMC samplers using the initial value generator and logarithmic posterior probability contained in the object instance of the profile class.

\subsection{\ion{Si}{ii} $\lambda1190,1193$ Absorption Doublet}

\begin{lstlisting}[language=Python]
# import the OutLines package
import OutLines as OL
# MC sampler for posterior
import emcee
# wavelength, continuum-subtracted flux, and uncertainty
wave,flux,flux_error = data
# set up OutLines model using NIST vacuum data
kwargs = dict(DensityProfile='DampedPulses',\ AddStatic=True)
model = OL.Absorption([1190.416,1193.290],[0.277,0.575],\ **kwargs)
# initialize walkers and sampler and run the sampler
init,nwalk = model.init_params()
# run the MCMC sampler
sampler = emcee.EnsembleSampler(nwalk,model.npar,\ model.log_probability,args=(wave,flux,flux_error)) 
sampler.run_mcmc(init,1200, progress=True)
samples = sampler.get_chain(discard=200,flat=True)
# extract parameters from posterior sample
for i,param in enumerate(model.get_param_names()):
    value = nanquantile(samples[:,i],0.5)
    model.update_params(param,value)
# print line profile parameters to terminal
model.print_params()
\end{lstlisting}

\subsection{[\ion{S}{ii}] $\lambda6716,6731$ Emission Doublet}

\begin{lstlisting}[language=Python]
# import the OutLines package
import OutLines as OL
# scipy nonlinear least squares for initial guess
from scipy.optimize import curve_fit
# MC sampler for posterior
import emcee
# wavelength, continuum-subtracted flux, and uncertainty
wave,flux,flux_error = data
# set up OutLines model using NIST air wavelengths
mod_kwargs = dict(AddStatic=True)
model = OL.Nebular([6716.440,6730.815],**mod_kwargs)
# simple trust-reflective fit for initial guess
tr_kwargs = dict(p0=model.get_params(),\ bounds=model.get_bounds(),sigma=flux_error)
p0,c0 = curve_fit(model.Profile,wave,flux,**tr_kwargs)
model.update_params(model.get_param_names(),p0)
# initialize walkers and sampler and run the sampler
init,nwalk = model.init_params()
# run the MCMC sampler
sampler = emcee.EnsembleSampler(nwalk,model.npar,\ model.log_probability,args=(wave,flux,flux_error)) 
sampler.run_mcmc(init,1200, progress=True)
samples = sampler.get_chain(discard=200,flat=True)
# extract parameters from posterior sample
for i,param in enumerate(model.get_param_names()):
    value = nanquantile(samples[:,i],0.5)
    model.update_params(param,value)
# print line profile parameters to terminal
model.print_params()
# obtain outflow properties
props = OL.Properties(model)
# print outflow properties to terminal
props.print_props()
\end{lstlisting}

\subsection{[\ion{O}{iii}] $\lambda4959,5007$ Emission Doublet for the Seyfert 2 in NGC 2992}

\begin{lstlisting}[language=Python]
# import the OutLines package
import OutLines as OL
# MethodType for changing prior
from types import MethodType
# scipy nonlinear least squares for initial guess
from scipy.optimize import curve_fit
# scipy convolution function for speed
from scipy.signal import convolve
# MC sampler for posterior
import emcee
# log prior -- uniform bounded
def log_prior(self,theta):
    tlower,tupper = self.get_bounds()
    tbound = list(map(self.__fun_bound__,theta,tlower,tupper))
    # enforce theta_o - theta_c >= 5 degrees
    # suggest atomic data to within 5%
    if all(tbound) and theta[8]-theta[9] > 0.087:
        return -0.5*((theta[4]/theta[3]-2.98)/0.05)**2 \
               -0.5*((theta[6]/theta[5]-2.98)/0.05)**2
    else:
        return -inf
# log likelihood for chi squared
# including convolution of line profile with LSF
# assuming the LSF kernel `kern` is defined
def log_likelihood(self,theta,x,y,yerr):
    ybar = convolve(model.Profile(x,*theta)+cont[ind],kern,'same')
    ln_prb = -((y-ybar)/yerr)**2 - 2*log(yerr)
    return nansum(ln_prb)
# instantiate the nebular line
# profile object using
# rest wavelengths from NIST
kwargs = dict(Geometry='HollowCones',AddStatic=True,\ VelocityField='AccPlaw',Disk=True)
model = OL.Nebular([4958.911,5006.843],**kwargs)
# update some parameters
model.update_params(['TerminalVelocity'],[500,30,15,45])
model.update_params(['OpeningAngle','CavityAngle',\ 'Inclination'],[60,30,60])
model.update_params(['FluxOutflow1','FluxStatic1'],\ [1/2.98,1/2.98]) # Storey & Zeppen 2000
# simple trust-reflective fit for initial guess
tr_kwargs = dict(p0=model.get_params(),\ bounds=model.get_bounds(),sigma=flux_error)
p0,c0 = curve_fit(model.Profile,wave,flux,**tr_kwargs)
model.update_params(model.get_param_names(),p0)
# initialize walkers and sampler and run the sampler
init,nwalk = model.init_params()
# use the custom prior and likelihood
model.log_prior = MethodType(log_prior,model)
model.log_likelihood = MethodType(log_likelihood,model)
# run the MCMC sampler
sampler = emcee.EnsembleSampler(nwalk,model.npar,\ model.log_probability,args=(wave,flux,flux_error)) 
sampler.run_mcmc(init,1200, progress=True)
samples = sampler.get_chain(discard=200,flat=True)
# extract parameters from posterior sample
for i,param in enumerate(model.get_param_names()):
    value = nanquantile(samples[:,i],0.5)
    model.update_params(param,value)
# print line profile parameters to terminal
model.print_params()
# obtain outflow properties
props = OL.Properties(model)
# print outflow properties to terminal
props.print_props()
\end{lstlisting}

\section{Posterior Sampling}\label{apx:corners}

\begin{figure}
    \centering
    \includegraphics[width=\linewidth]{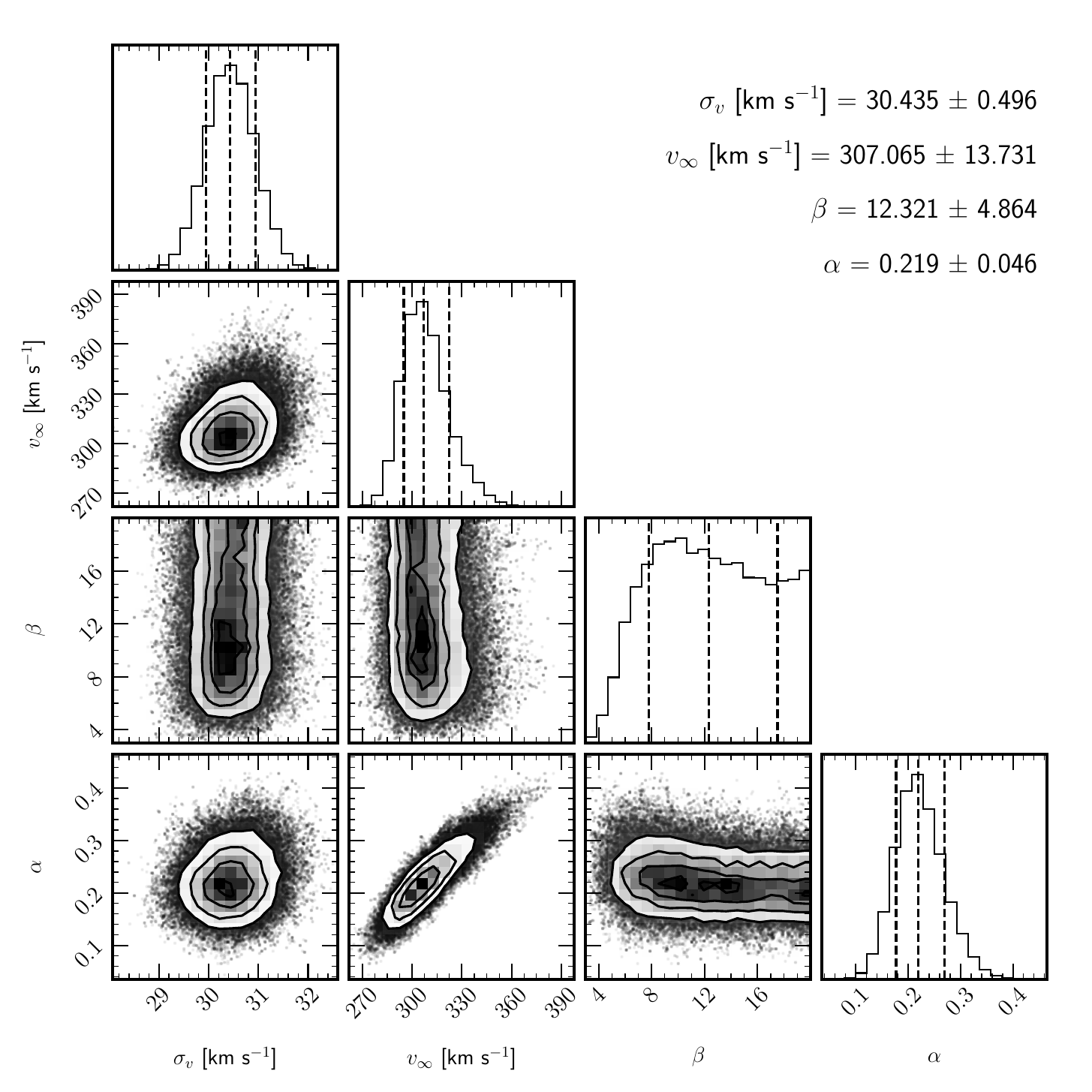}
    \caption{Posterior sampling of {\tt OutLines} line profile parameters for [\ion{S}{ii}]$\lambda6716,6731$ in giant extragalactic \ion{H}{ii} region NGC 5471 Knot A. The parameters are CAK theory velocity field $\beta$, terminal velocity $v_\infty$, and power law density profile $\alpha$.}
    \label{fig:ngc5471_corner}
\end{figure}

\begin{figure}
    \centering
    \includegraphics[width=\linewidth]{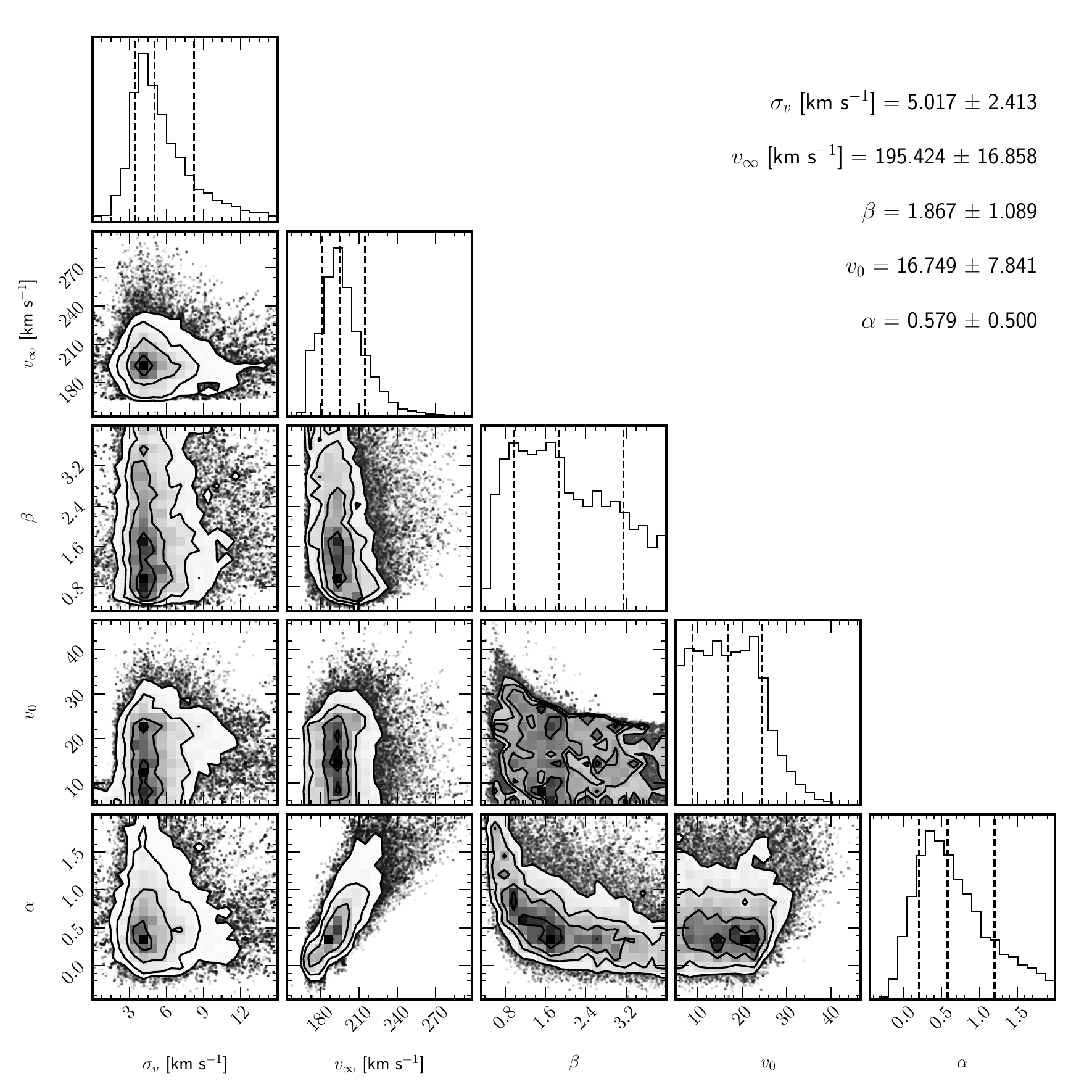}
    \caption{Posterior sampling of {\tt OutLines} line profile parameters for \ion{C}{ii}$\lambda1334$ in the super star cluster of Green Pea J1044+0353. The parameters are CAK theory velocity field $\beta$, terminal velocity $v_\infty$, and power law density profile $\alpha$.}
    \label{fig:j1044_corner}
\end{figure}

\begin{figure*}
    \centering
    \includegraphics[width=\linewidth]{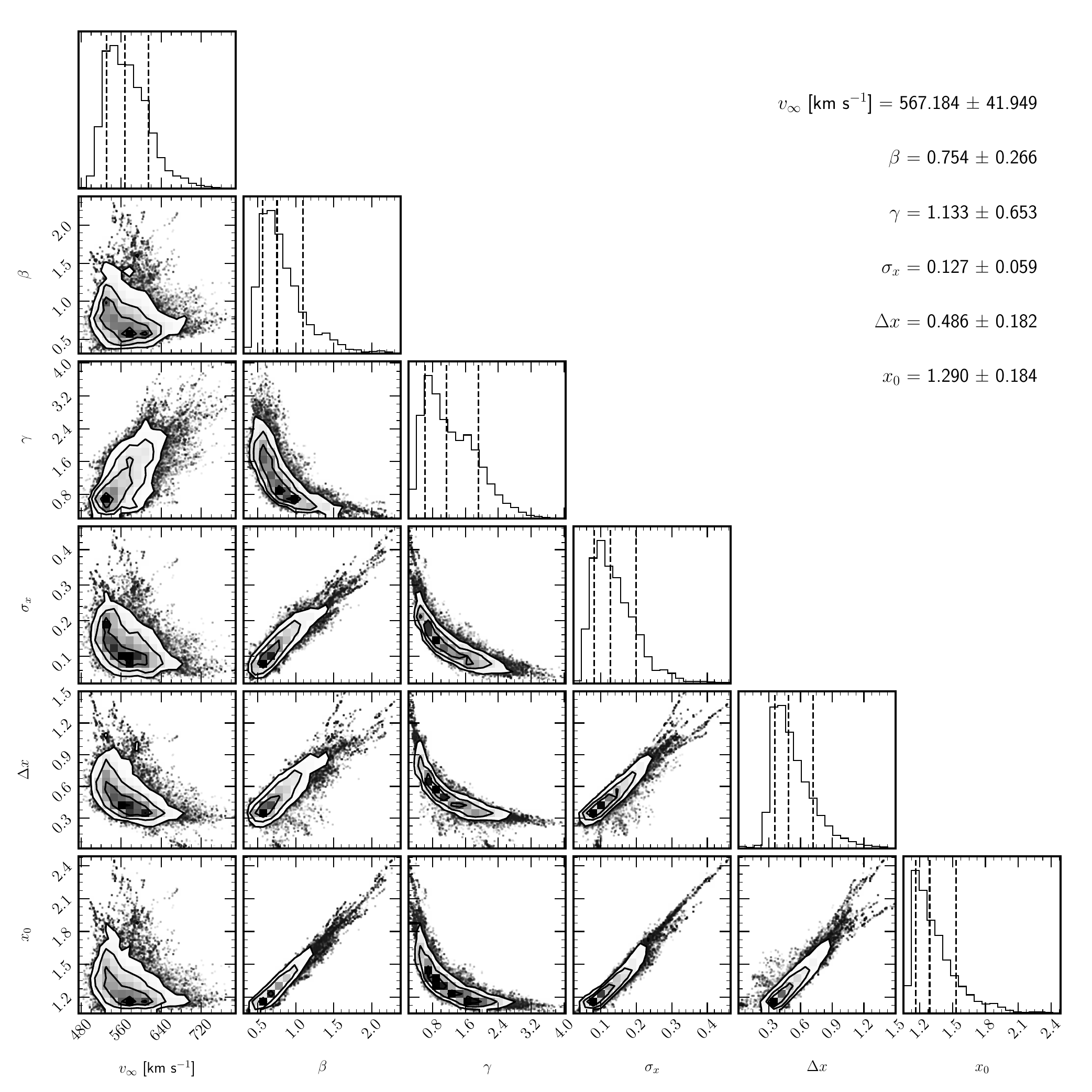}
    \caption{Posterior sampling of {\tt OutLines} line profile parameters for \ion{Si}{ii}$\lambda1190,1193$ in starburst galaxy Mrk 1486. The parameters are CAK theory velocity field $\beta$, terminal velocity $v_\infty$, and damped pulses with minimum radius $x_1$, shell thickness $\sigma_x$, and interval distribution $x_k$. $\alpha$.}
    \label{fig:mrk1486_corner}
\end{figure*}

\begin{figure*}
    \centering
    \includegraphics[width=\linewidth]{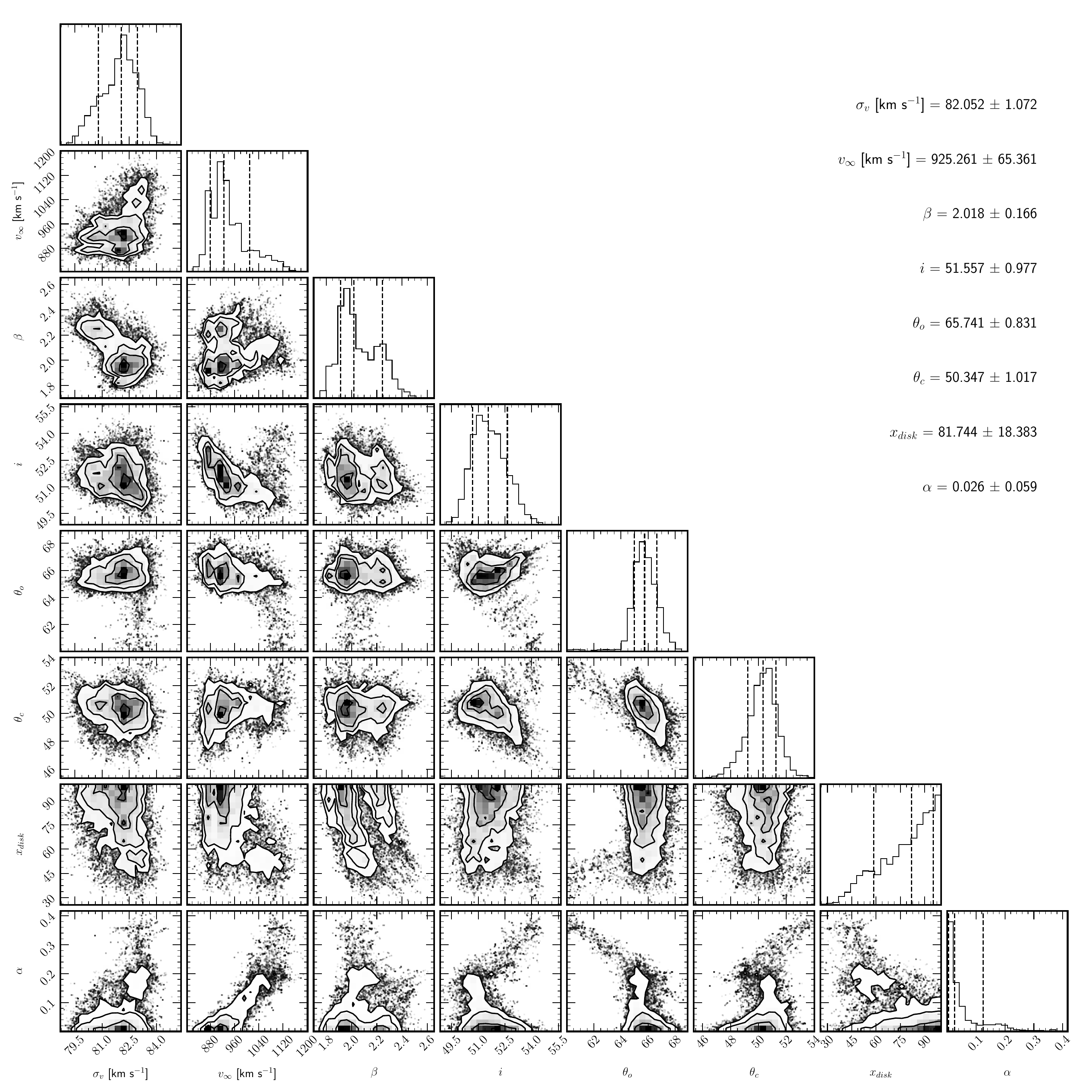}
    \caption{Posterior sampling of {\tt OutLines} line profile parameters for [\ion{O}{iii}]$\lambda4959,5007$ in the Seyfert 2 nucleus of NGC 2992. The parameters are acceleration power law velocity field $\beta$, terminal velocity $v_\infty$, power law density profile $\alpha$, and the hollow cone plus disk geometry components of inclination $i$, opening angle $\theta_o$, cavity angle $\theta_c$, and disk radius $x_d$.}
    \label{fig:ngc2992_corner}
\end{figure*}

All posterior samples for each of the {\tt OutLines} modeling cases were obtained using {\tt emcee} \citep{emcee} and visualized with figures produced by {\tt corner} \citep{corner} and subsequently adapted. Results are shown for NGC 5471 Knot A in Figure \ref{fig:ngc5471_corner}, the super star cluster of Green Pea J1044+0353 in Figure \ref{fig:j1044_corner}, Mrk 1486 in \ref{fig:mrk1486_corner}, and the Seyfert 2 nucleus of NGC 2992 in \ref{fig:ngc2992_corner}.

\bsp	
\label{lastpage}


\end{document}